\documentclass[prd,twocolumn,nofootinbib,showpacs,superscriptaddress]{revtex4-1}

\usepackage{amsfonts}
\usepackage{amsmath}
\usepackage{amssymb}
\usepackage{bm}
\usepackage{dcolumn}
\usepackage{epsfig}
\usepackage{graphicx}
\usepackage{graphics}
\usepackage[latin1]{inputenc}
\usepackage{latexsym}
\usepackage{rotating}
\usepackage{hyperref}
\usepackage[usenames]{color}
\usepackage{float}
\usepackage{xspace} 
\usepackage{mathrsfs}
\usepackage{subfigure}
\usepackage{multirow}

\newcommand {\Msun}{\ensuremath{M_{\odot}}}

\newcommand{\eeq}{\end{equation}}
\newcommand{\bea}{\begin{eqnarray}}

\def\ltsima{$\; \buildrel < \over \sim \;$}
\def\simlt{\lower.5ex\hbox{\ltsima}}
\def\gtsima{$\; \buildrel > \over \sim \;$}
\def\simgt{\lower.5ex\hbox{\gtsima}}

\newcommand{\mc}{\ensuremath{{\cal{M}}_C}}

\def\lesssim{\mathrel{\hbox{\rlap{\hbox{\lower4pt\hbox{$\sim$}}}\hbox{$<$}}}}
\def\gtrsim{\mathrel{\hbox{\rlap{\hbox{\lower4pt\hbox{$\sim$}}}\hbox{$>$}}}}
\def\alt{\mathrel{\hbox{\rlap{\hbox{\lower4pt\hbox{$\sim$}}}\hbox{$<$}}}}
\def\agt{\mathrel{\hbox{\rlap{\hbox{\lower4pt\hbox{$\sim$}}}\hbox{$>$}}}}
\def\prd{{\it Phys. Rev.} D~}

\def\apjl{{\it Astrophys. J.} Lett~}
\def\apj{{\it Astrophys. J.}}
\def\Msun{M_\odot}
\def\PRD{{\it Phys. Rev.} D~}
\def\CQG{{\it Class. Quantum Grav.}}

\def\mnras{{\it MNRAS}}

\def\gta{\ifmmode {\mathbin{\lower 3pt\hbox   
    {$\,\rlap{\raise 5pt\hbox{$\char'076$}}\mathchar"7218\,$}}}
    \else {${\mathbin{\lower 3pt\hbox
    {$\rlap{\raise 5pt\hbox{$\char'076$}}\mathchar"7218\,$}}}
    $}\fi}
\def\lta{\ifmmode {\,\mathbin{\lower 3pt\hbox   
    {$\,\rlap{\raise 5pt\hbox{$\char'074$}}\mathchar"7218\,$}}}
    \else {${\mathbin{\lower 3pt\hbox
    {$\rlap{\raise 5pt\hbox{$\char'074$}}\mathchar"7218\,$}}}
    $}\fi}

\newcommand{\SU}{\affiliation{Department of Physics, Syracuse University, Syracuse, NY 13244, USA}}
\newcommand{\WVU}{\affiliation{Department of Physics, West Virginia University, White Hall, Morgantown, WV 26506, USA}}

\newcommand{\WM}{\affiliation{Center for Gravitation and Cosmology, University of Wisconsin-Milwaukee, Milwaukee, WI 53211, USA}}
\newcommand{\MU}{\affiliation{Department of Physics, Montana State University, Bozeman, MT 59717, USA}}
\newcommand{\RIT}{\affiliation{Center for Computational Relativity and Gravitation, School of Mathematical Sciences, and School of Physics and Astronomy, Rochester Institute of Technology, Rochester, NY 14623, USA}}
\newcommand{\CITA}{\affiliation{Canadian Institute for Theoretical Astrophysics, 60 St. George Street, 
University of Toronto,  Toronto, ON M5S 3H8, Canada}}

\begin{document}

\title{Accurate and efficient waveforms for compact binaries on eccentric orbits}
\author{E. A. Huerta}
\email{elhuertaescudero@mail.wvu.edu}
\WVU%
\author{Prayush Kumar}
\SU\CITA%
\author{Sean T. McWilliams}
\WVU%
\author{Richard O'Shaughnessy}
\WM\RIT%
\author{Nicol\'as Yunes}
\MU%


\begin{abstract}
Compact binaries that emit gravitational waves in the sensitivity band of ground-based detectors can have non-negligible
eccentricities just prior to merger, depending on the formation scenario. We develop a purely analytic, frequency-domain
model for gravitational waves emitted by compact binaries on orbits with small eccentricity, which reduces to the
quasi-circular post-Newtonian approximant {\texttt{TaylorF2}} at zero eccentricity and to the post-circular
approximation of Yunes et al. (2009)  at small eccentricity. Our model uses a spectral approximation to the (post-Newtonian) Kepler problem to model the orbital phase as a function of frequency, accounting for eccentricity effects up to ${\cal{O}}(e^8)$ at each post-Newtonian order. 
Our approach accurately reproduces an alternative time-domain eccentric waveform model  for $e\in [0, 0.4]$ and binaries with total mass $\lesssim 12 \Msun$.  As an application, we evaluate the signal amplitude that eccentric binaries produce in different networks of existing and forthcoming gravitational waves detectors. Assuming a population of eccentric systems containing black holes and neutron stars that is uniformly distributed in co-moving volume, we estimate that second generation detectors like Advanced LIGO could detect approximately 0.1--10 events per year out to redshift \(z\sim 0.2\), while an array of Einstein Telescope detectors could detect hundreds of events per year to redshift \(z \sim 2.3\).
\end{abstract}

\maketitle

\section{Introduction}

Numerical studies suggest that, depending on the metallicity and chemical composition, the end point of stellar evolution for most massive stars above \(\gtrsim7\Msun\) will be either a stellar mass black hole (BH) or a neutron star (NS)~\cite{Pols:1998}. Binary systems of stellar-mass compact objects are the most promising sources of gravitational waves (GWs) for second generation ground-based interferometric detectors, such as Advanced LIGO (aLIGO), Virgo, LIGO India, and Kagra~~\cite{aLIGO,virgo,kagra}, as well as for planned third generation detectors like the Einstein Telescope (ET)~\cite{Freise:2009}. Most previous analyses have assumed that these compact binaries will be in nearly quasi-circular orbits by the time they reach the sensitive frequencies of these detectors (i.e.,~orbital frequencies greater than a few Hz). Such assumptions were made because GW emission rapidly reduces the eccentricity of a binary system~\cite{Peters:1964}, and thus most astrophysical binaries that were formed at large separations and low frequencies will circularize well before their signal enters the detector's sensitive band. 

Many recent investigations, however, now suggest that there may exist several plausible, observationally-unconstrained astrophysical mechanisms through which GWs emitted by compact binaries with significant eccentricity may persist into the detector's sensitivity band \cite{antonini,Samsing:2014,Thompson:2011,East:2013}. Dense stellar environments in galactic nuclei can facilitate frequent interactions, enabling direct dynamical capture \cite{Leary:2009,Kocsis:2012,Tsang:2013,East:2013, Wen:2003,east:2012a,Pooley:2003} into high-eccentricity orbits via single-single, binary-binary and binary-single processes \cite{Samsing:2014}.  A high density of compact objects in the galactic center cusp is expected on theoretical grounds, from mass segregation of individual objects \cite{Freitag:2006A,Anton:2012} and satellite stellar systems (e.g., clusters and small galaxies) \cite{Anton:2012,Antonini:2014}. The compact object density, and hence the event rate, remains highly uncertain, but~\cite{Hopman:2006,Antonini:2014} suggest $\gtrsim 10^3$ BHs in the central $0.1$ pc of our galaxy, consistent with the high event rates quoted in~\cite{Leary:2009,Kocsis:2012}.

Globular clusters also provide an environment for stars and binaries to interact and form eccentric binaries.  While typical interactions occur at large separations, binary-single and binary-binary interactions can produce close encounters that form high-eccentricity systems even for relatively low stellar velocities \cite{Samsing:2014}.   Additionally, these environments naturally form hierarchical triples, whose secular interactions (the Kozai-Lidov mechanism~\cite{Kozai:1962,Lidov:1962}) can drive the inner binary to high eccentricity \cite{Anton:2014,Wen:2003,Aarseth:2012,Miller:2002,Thompson:2011}. This effect also naturally occurs in binaries bound to supermassive BHs, and could therefore yield eccentric systems in galactic centers as well \cite{antonini}. While relatively few BHs are known to exist in globular clusters, recently several accreting BH candidates have been found in extragalactic globular clusters~\cite{Maccarone:2007,Irwing:2010}, and more recently in the Milky Way globular cluster M22~\cite{Strader:2012}. Recent numerical calculations also lend support to the idea that a significant population of BHs will persist in clusters for the full cluster lifetime \cite{Sippel:2013,Morscher:2013}. 

Another potential mechanism for forming eccentric ultra-compact binaries is tidal capture. NSs can absorb orbital energy during close passages, i.e.,~capture via closest approach for pericenter distances $r_p \simeq {\cal{O}}(10-100\, {\rm{km}})(M/10 M_{\odot})$~\cite{Lee:2010}.  The tidal capture event rate is very sensitive to assumptions made about the retention rate of NSs~\cite{Murphy:2011}, the fraction of NSs in the core of the globular cluster~\cite{Pfahl:2002}, BH ejection~\cite{Oleary:2006,Kulkarni:1993,nat,Portegies:2000,Aarseth:2012} and the tidal capture cross section used.

\subsection{Previous Work}

All of the above scenarios are quite uncertain; hence, even a null GW result
will significantly constrain previously inaccessible astrophysics.  A null result could also arise from selection biases
against eccentric binaries  
-- for example, due to 
systematic errors in the waveform modeling if, for example, one tried to extract eccentric signals with quasi-circular
waveforms. 
Selection biases are estimated by characterizing the ``effectualness'' of model A for finding members of model B: a
candidate signal from B is compared with a dense, discrete, complete
sample of  A.   
Using leading-order post-Newtonian (PN) expansions, i.e., an expansion in small velocities and weak fields,
for eccentric binary systems, Ref.~\cite{Martel:1999} concluded that quasi-circular templates could effectively detect
low-eccentricity compact binary sources: nonspinning searches were ``effectual'' for eccentric binaries.    Subsequent studies quantified  the
selection bias against eccentric binaries \cite{Cokelaer:2009,Brown:2010,Huerta:2013a}.    For example, in initial LIGO \cite{Cokelaer:2009},  targeting binary NSs with quasi-circular templates would lead to a  detection loss \(\gtrsim 10\%\) for
binaries with eccentricities \(e\gtrsim 0.05\) at a Keplerian mean orbital frequency of 20 Hz.
By contrast, in this paper we perform  ``faithfulness'' studies, demonstrating that eccentricity has a distinguishable
impact by comparing GWs from otherwise identical binaries with eccentricity to those without. 

The studies mentioned above motivated the development of accurate eccentric waveform models to target compact binaries on eccentric orbits. Several authors over the years have generated time-domain eccentric waveforms, including kludge waveforms typically applied at high mass-ratio~\cite{kludge,improved,gla,Sunda:2008,Ganz:2007,MMT,mino,Mino:1997}, precessing time-domain signals~\cite{Berti:2006C,CampLou:2009,cutler,NCornish:2010,Key:2011J,Levin:2011C,MikKoc:2012}, and finally relatively high-accuracy time-domain PN models~\cite{ArunBlanchet:2008}. Though useful for qualitative work, these calculations faced some limitations.  First, PN approximations become inaccurate when the orbital velocity becomes large enough, which can occur during pericenter passages for certain orbits, but always occurs close to merger. In such instances, numerical relativity was required to build confidence in the associated dynamical evolution and waveform models~\cite{Gold:2013,Hinder:2010,Tania:2013}. Second, the construction of such models is computationally expensive because of the need to solve ordinary differential equations in the time-domain with a very fine and constant discretization, so that aliasing and Nyquist noise is under control when computing the discrete Fourier transform (DFT) of the GW response function.  

These limitations have been recently addressed in two different waveform models: the \emph{\(x\)-model} of~\cite{Hinder:2010} and the post-circular (PC) model of~\cite{Yunes:2009}. The former is a time-domain model with conservative orbital dynamics accurate to 3 PN\footnote{A term of Nth PN order is one that is proportional to $(v/c)^{2N}$, where $v$ is the orbital velocity~\cite{Blanchet:2006}.} order and radiation-reaction accurate to 2 PN order. The \(x\)-model has been validated against one numerical relativity simulation of an equal-mass BH-BH with initial eccentricity \(e = 0.1\), 21 GW cycles before merger~\cite{Hinder:2010}. The \(x\)-model, however, is quite computationally expensive (mainly because of the need to solve for the orbital evolution in the time-domain and then to Fourier transform the resulting response function to the frequency domain)~\cite{Huerta:2013a} and not sufficiently accurate to model low-mass binary inspirals~\cite{boyle}. 

The PC model is a frequency-domain approach, where the conservative and dissipative orbital dynamics are treated in the PN approximation, but further expanded in a small eccentricity approximation through an analytic, high-order spectral decomposition~\cite{Yunes:2009}. The frequency-domain response function is then computed through the stationary-phase approximation (SPA)~\cite{Bender:1999,Yunes:2009}. Although in principle this model can be implemented to arbitrary PN order, only the leading PN order terms were included explicitly in \cite{Yunes:2009}, while 1 PN corrections~\cite{Tessmer:2010sh}, 2 PN corrections~\cite{Tessmer:2010ii} and 3 PN corrections~\cite{Arun:2008,ArunBlanchet:2008} are now available to extend~\cite{Yunes:2009}. 

\subsection{Executive Summary}

In this work we develop the enhanced post-circular (EPC) model. This model is an extension of the PC analysis in~\cite{Yunes:2009}, designed to reproduce in the zero-eccentricity limit the {\texttt{TaylorF2}} model at 3.5 PN order and to reproduce in the small eccentricity limit the PC model to leading (Newtonian) order. The {\texttt{TaylorF2}} model is a waveform family constructed from the PN approximation for non-spinning, quasi-circular binaries directly in the frequency-domain, using the SPA. Furthermore, as shown in~\cite{Prayush:2013a,pnbuo}, {\texttt{TaylorF2}} 3.5 PN is accurate and computationally efficient to construct effectual searches of quasi-circular binary systems with total mass \(\lesssim12\Msun\)~\cite{Prayush:2013a,pnbuo}. It is worth emphasizing that we could also try to reproduce the evolution of the time domain PN-based approximant  {\texttt{TaylorT4}}  3.5 PN in the quasi-circular limit, since this model provides an accurate representation of the evolution of comparable mass quasi-circular binaries~\cite{boyle}. Nonetheless, anticipating that the matched-filtering in future GW searches will be carried out in the frequency domain, we provide a waveform family that is directly applicable in this framework. 
 
The EPC model is not a consistent PN expansion to 3.5 PN order of the PC model. Instead, it adds 3.5 PN order corrections through a mapping between the {\texttt{TaylorF2}} and the PC models. Such a mapping will necessarily neglect high-order eccentricity-dependent terms at first and higher PN order. We will show, however, that the EPC model is remarkably simple and sufficient for data analysis explorations. For simplicity and consistency with prior work, we adopt the restricted PN (``quadrupole'') approximation, wherein we include the aforementioned PN phase corrections, but we neglect PN amplitude corrections. Recent work suggests that amplitude corrections may play an important role in detection~\cite{VanDen:2007,Arun:2007,Trias:2008,Sintes:2000,Sintes:2000L,Moore:2002, Hellings:2003,VanDen:2007a,Arun:2007a,Porter:2008,Arun:2009}, but we defer such an analysis to future work. 

We compare and validate the EPC model against other waveforms commonly employed in the literature~\cite{Hinder:2010,Huerta:2013a}. In particular, we show analytically and numerically that the EPC reduces to the {\texttt{TaylorF2}} model at 3.5 PN order in the limit of zero eccentricity. We also compare the EPC model numerically to other eccentric waveform families for comparable mass binaries~\cite{Hinder:2010,Huerta:2013a} using data analysis measures, focusing particularly on the \(x\)-model~\cite{Hinder:2010}. We find that the EPC model agrees better with the {\texttt{TaylorF2}} 3.5 PN model at small eccentricity than the \(x\)-model, primarily because the \(x\)-model only includes the dissipative dynamics to 2 PN order. We also find that the EPC model loses accuracy at a slow rate as the eccentricity increases, remaining accurate relative to the \(x\)-model up to eccentricities of $0.4$.  

Once validated, we use the EPC model to study the importance of eccentricity in GW searches with second- and third-generation detectors. We confirm that eccentricity corrections increase the in-band GW signal strength for a fixed mass system, thereby increasing the distances to which the system could be detected. Similarly, the presence of eccentricity increases the range of masses that are accessible to the detectors. We also use the EPC model to estimate the cosmological range to which eccentric inspirals of NSBH and NSNS binaries could be observed. Such binaries have been proposed as possible progenitors of short gamma-ray bursts (SGRBs)~\cite{Paczynski:1986,Eichler:1989,Bart:2005,grindlay,Troja:2010}. Following~\cite{Oleary:2007}, and assuming that one to a few eccentric binaries form per young massive star cluster over its lifetime, we estimate that aLIGO could observe approximately 0.1--10 events per year out to \(z\sim 0.2\), while an array of ET detectors could observe from 60--7900 events per year out to \(z \sim2.3\).

This paper is organized as follows. 
In Section~\ref{sec:time-domain-wf} we summarize the construction of time-domain and frequency-domain waveform models.
Section~\ref{sec:EPC} presents and develops the EPC model.
Section~\ref{sec:astro-cons} studies the astrophysical consequences of doing data analysis with the EPC model.
Section~\ref{sec:conclusions} concludes and points to future work. 
Henceforth, we use geometric units with $G=c=1$. We follow the conventions in~\cite{mtw} and~\cite{Yunes:2009}. 

\section{Eccentric Waveform Models}
\label{sec:time-domain-wf}

In this section, we review the basics on how to construct time-domain and frequency-domain eccentric waveform models. We concentrate on the time-domain \(x\)-model proposed in Ref.~\cite{Hinder:2010} and the frequency-domain PC model proposed in Ref.~\cite{Yunes:2009}. Ultimately, detection and parameter estimation is usually carried out in the frequency-domain. By time-domain waveform model, we here mean a waveform that is constructed (usually numerically) in the time-domain and then is DFTed to obtain a frequency representation. By contrast, a frequency-domain model is constructed (usually analytically) directly in the Fourier domain. 

\subsection{The \(x\)-model}
\label{xmodel}

The \(x\)-model~\cite{Hinder:2010} is a parameter-free, time-domain PN-based waveform family. The binary orbit is given in the Keplerian parameterization to 3 PN order, which can be written as
\begin{align}
\frac{r}{M} &= \frac{1-e_t\cos u}{x} + \sum_{j=1}^{j=3} \delta r_{j} \; x^{j-1}\,, 
\\
\ell &= u - e_t\sin u + \sum_{j=2}^{j=3} \delta \ell_{j} \; x^{j}\,,
\end{align}
while its conservative evolution is given also to 3 PN order by~\cite{Hinder:2010}: 
\begin{align}
\label{eq:xmodel:conservative}
M {\dot{\phi} \choose \dot{\ell}} &= \sum_{j=0}^{j=3} {\delta\dot{\phi}_{j} \choose \delta\dot{\ell}_{j}}  x^{3/2+j}\,.
\end{align}
In these equations, $r$ and $\phi$ are the magnitude of the relative separation vector and the relative orbital phase, $e_{t}$ is the so-called \emph{temporal} eccentricity, defined by the change in mean anomaly $\ell$ when the eccentric anomaly $u$ changes by one full cycle to leading PN order, $x=\left(M\, \dot{\ell} \right)^{2/3} \ll 1$ is the PN expansion parameter, with $\dot{\ell} = n$ the mean Keplerian orbital frequency, $M=m_{1}+m_{2}$ is the total mass, and $n$ the mean motion, and the PN coefficients $(\delta r_{j},\delta \ell_{j},\delta\dot{\phi}_{j},\delta\dot{\ell}_{j})$ can be found in~\cite{Hinder:2010} and references therein. The above orbital evolution is \emph{conservative} in that energy and (z-component of) angular momentum are conserved quantities, and the equations can thus be derived from a given PN Hamiltonian~\cite{Gopakumar:2005b}.

The true inspiral evolution, however, is not conservative because GWs carry energy and angular momentum away from the binary. The loss of energy and angular momentum can be mapped to a change in the PN expansion parameter $x$ and the eccentricity $e_{t}$, which are no longer conserved, but rather evolved according to the 2 PN equations 
\begin{align}
M {\dot{x} \choose \dot{e}_{t}} &= \sum_{j=0}^{4} {\delta\dot{x}_{j} \choose \delta \dot{e}_{j}} x^{5 + j/2}\,,
\end{align}
where the dissipation coefficients $(\delta \dot{x}_{j}, \delta \dot{e}_{j})$ can be found for example in~\cite{Hinder:2010} and references therein.

The above equations define the orbital evolution in the $x$-model, which is solved numerically in the time-domain. This evolution has been validated against one numerical relativity simulation of an equal-mass, BH-BH binary with initial eccentricity of $0.1$ 21 GW cycles before merger~\cite{Hinder:2010}. 
Once the orbital evolution has been obtained, one can obtain the time-domain GW response function to leading (mass-quadrupole or Newtonian) order~\cite{w1}, which one then DFTs to obtain frequency-domain templates for data analysis studies.  

The $x$-model reduces to some well-studied template families used in GW data analysis. For example, in the limit of zero eccentricity, the orbital phase in the \(x\)-model reduces to the {\texttt{TaylorT4}} PN model at 2 PN order~\cite{Brown:2010}. In fact, the {\texttt{TaylorT4}}  2 PN differential equations that define the orbital evolution are the same as those of the $x$-model in the zero-eccentricity limit by construction. However, the amplitude of the \(x\)-model differs from 2 PN  {\texttt{TaylorT4}} in that \(M\,\dot{\ell}\) in Eq.~\eqref{eq:xmodel:conservative} introduces an additional amplitude contribution~\cite{Brown:2010}.

The $x$-model, however, deviates from some other well-studied template families. One example is the {\texttt{TaylorF2}} templates, a family constructed to model GWs from the quasi-circular inspiral of non-spinning compact binaries. This template family is defined directly in the frequency-domain through the SPA via
\begin{eqnarray}
\label{strain}
\tilde{h}(f) &=& A f^{-7/6}\,e^{i \Psi_{\rm F2}(f)}\,,\\
\label{phasef2}
\Psi_{\rm F2}(v)&=&2\,\pi\, f \, t_c - 2\phi_c -\frac{\pi}{4} + \Psi_{\rm PN}(v)\,,
\end{eqnarray}
where the PN phase is defined as
\begin{equation}
\Psi_{\rm PN}(v) = \frac{3}{128\, \eta\, v^5}\sum^{i=7}_{i=0}a_n v^n\,,
\end{equation}
and where \(A\propto  \mc^{5/6}/D_L\), with the chirp mass, \(\mc= M\,\eta^{3/5}\), the symmetric mass ratio \(\eta = m_1 m_2/M^2\), and the luminosity distance \(D_L\), while \(v=\left(\pi\, M f\right)^{1/3}\) is the orbital velocity of the binary. The {\texttt{TaylorF2}} waveform phase we will use throughout this article, Eq.~\eqref{phasef2}, includes PN corrections up to 3.5 PN order. The corresponding \(a_n\) coefficients in Eq.~\eqref{phasef2} at this PN order can be found in~\cite{pnbuo}.

The $x$-model captures all critical features that eccentricity introduces to non-spinning binary physics, both on the dynamics and on the waveform, to high PN order, i.e., to 3 PN order in the conservative dynamics and to 2 PN order in the dissipative dynamics.  
First and foremost, eccentric binaries precess, and the $x$-model captures this well at high PN order.
Second, eccentricity shortens the duration of the orbit and hence of the waveform, compared to circular binaries starting at the same mean orbital frequency.  
Third, binaries with eccentricity have complicated, highly modulated waveforms, which is also captured in the $x$-model to high PN order.  

\begin{figure}[ht]
\centerline{
\includegraphics[height=0.35\textwidth,  clip]{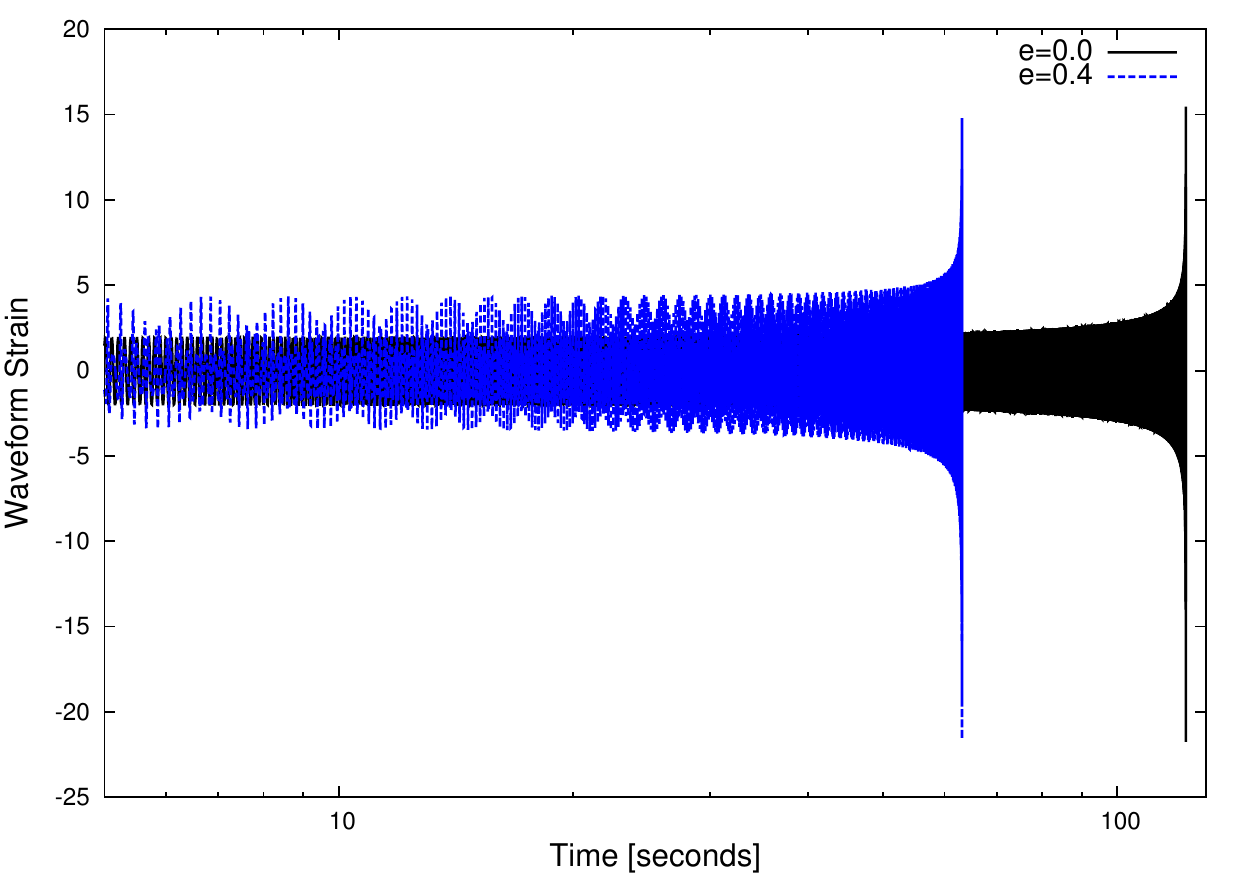}
}
\caption{(Color Online) Sample waveforms for an equal mass binary system with total mass \(M= 10\Msun\) starting at a Keplerian mean orbital frequency \(f_K = 5\) Hz. The figure uses a logarithmic scale in the time axis to clearly exhibit the structure of the eccentric waveform at low frequencies.}
\label{length}
\end{figure}
As a concrete example, Fig.~\ref{length} shows the waveforms predicted by the $x$-model for a circular ($e=0$) and eccentric ($e=0.4$)  $(5\Msun,\,5\Msun)$ binary. Observe the amplitude modulations present in the eccentric waveform and the fact that the eccentric inspiral is noticeably shorter. The latter is driven both by long-term effects at low frequencies and by waveform termination. In the figure, the orbital evolution terminates when the system reaches the innermost stable circular orbit (ISCO) for a test particle in an eccentric orbit around a Schwarzschild BH, i.e., \(r_{\rm{ISCO}} = 6\,M + 2\,e_{\rm{ISCO}}\)~\cite{Cutler:1994}, where $e_{\rm{ISCO}}$ stands for the eccentricity at the ISCO.  

\subsection{Limitations of the $x$-model}

Even though the \(x\)-model is capable of reproducing the main features of the eccentric numerical simulation used to calibrate it, the model does have some limitations, which we list below:
\begin{itemize}
\item {\emph{Computational expense}}. The \(x\)-model requires the numerical solution of the orbital evolution equations in the time-domain at a \emph{small and constant} discretization so that a DFT of the GW response can be accurately computed~\cite{Huerta:2013a}.   
\item {\emph{PN accuracy}}. Although the $x$-model reduces to the 2 PN {\texttt{TaylorT4}} approximant when \(e_{0}\rightarrow0\), higher PN order models (e.g.~{\texttt{TaylorT4}} at 3.5 PN order) are needed to describe the dynamical evolution of low-mass, quasi-circular binaries at the level of accuracy required for GW data analysis~\cite{boyle}. 
\end{itemize}

Let us discuss some of these limitations in more detail. To do so, we employ some basic data analysis tools.  Given two signals $h_1$ and $h_2$, the noise-weighted inner product is defined as
 \begin{equation}
 \left(h_1\,|\,h_2\right) = 2\int^{f_{\rm{max}}}_{f_{\rm{min}}}\frac{\tilde{h}^{*}_1(f)\,\tilde{h}_2(f) + \tilde{h}_1(f)\,\tilde{h}^{*}_2 (f)}{S_n(f)} \rm{d} f\,,
 \label{inner}
 \end{equation}
and the normalized overlap is
\begin{equation}
 \left(\hat{h}_1\,|\,\hat{h}_2\right) = \frac{  \left(h_1\,|\,h_2\right)}{\sqrt{ \left(h_1\,|\,h_1\right)\, \left(h_2\,|\,h_2\right)}}\,,
 \label{nor_over}
 \end{equation}
where \(S_n(f)\) is the power spectral density of the detector noise, and \(\tilde{h}(f)\) is the Fourier transform of the signal. For the former, in this paper we always use the Zero Detuned High Power (ZDHP) spectral density for aLIGO~\cite{aLIGO} and the optical configurations B or D for the Einstein Telescope (ETB, ETD),  respectively~\cite{Freise:2009}. The lower limit in integration depends on the detector under consideration, namely \(f_{\rm{min}} = [1\,{\rm{Hz}}, 10\,{\rm{Hz}}]\), for the ET and aLIGO, respectively, and \(f_{\rm{max}} \) is the last frequency at which the waveform is sampled.  

Maximizing over the time of coalescence, \(t_c\), and the phase of coalesce, \(\phi_c\), one can compute the maximized overlap, \({\cal{O}}\left(h_1, \,h_2\right)\), between any two given signals, namely
 \begin{equation}
 {\cal{O}}\left(h_1, \,h_2\right) = \underset{t_c\,\phi_c}{\rm{max}}\,\left(\hat{h}_1 \, | \, \hat{h}_2 \, e^{i\left(2\pi f t_c -\phi_c\right)} \right)\,.
 \label{max_over}
 \end{equation}
This quantity is a data analysis measure of how similar two waveforms $h_{1}$ and $h_{2}$ are, without allowing for any biasing in the system parameters other than $t_{c}$ and $\phi_{c}$. When quoting overlaps, we will assume optimally oriented sources.

The $x$-model agrees quite well with the {\texttt{TaylorT4}} waveform family at 2 PN order, but it disagrees with higher PN order families. For example, in the limit of zero eccentricity, the overlap between the $x$-model and the {\texttt{TaylorT4}} model at 3.5 PN order is roughly $0.5$ and $0.7$ for binaries with component masses \((1.35\Msun, \,1.35\Msun)\)  and  \((6\Msun,\, 6\Msun)\) respectively. This is to be expected, since the $x$-model is built with 2 PN equations to describe the dissipative dynamics. Even if one compares the $x$-model to the {\texttt{TaylorT4}} model at 2 PN order, the overlap drops rapidly with increasing eccentricity, crossing the $0.97$ threshold at an initial eccentricity at a GW frequency of 10 Hz of approximately $0.02$ and $0.05$ for binaries with component masses \((1.35\Msun, \,1.35\Msun)\)  and  \((6\Msun,\, 6\Msun)\) respectively. 

\subsection{The post-Circular Approximation}
\label{sec:sub:PC}

One of the first attempts to develop a consistent eccentric waveform model in the frequency domain was through the PC approximation~\cite{Yunes:2009}; we refer the reader to Sections II and III of~\cite{Yunes:2009} for a careful description of the generalization of the SPA to eccentric orbits. The PC stipulates that one can expand all quantities assuming the eccentricity is small. In principle, one can keep an arbitrary number of eccentricity corrections, but in the work of~\cite{Yunes:2009}, terms up to ${\cal{O}}(e^{8})$ were kept. Expanding the time-domain response in this way, one can obtain an expression that is amenable to the SPA when computing the Fourier transform. The main result of the PC approach is the frequency-domain response function for eccentric binary inspirals~\cite{Yunes:2009}:
\begin{equation}
\label{PC-model}
\tilde{h}(f>0) ={\cal A} f^{-7/6} \sum_{\ell=1}^{10} \xi_{\ell} \left(\frac{\ell}{2}\right)^{2/3} e^{-i(\pi/4 + \Psi_{\ell})}\,,
\end{equation}
where ${\cal{A}}$ is a function of the component masses and the distance to the source only and $\xi_{\ell}$ are functions of the beam pattern functions and the eccentricity~\cite{Yunes:2009}. The Fourier phase to leading PN order is given by 
\begin{widetext}
\begin{eqnarray}
\label{phasepcexp}
\Psi_{\ell}& =& -\ell\,\phi_c +2\pi f t_c + \frac{3}{128 (M\pi f)^{5/3}}\left(\frac{\ell}{2}\right)^{8/3}\Bigg[1-\frac{2355}{1462}e^2_0 \chi^{-19/9} + \left( \frac{5222765}{998944}\chi^{-38/9} - \frac{2608555}{444448}\chi^{-19/9}\right)\,e^4_0 \\\nonumber&+& \left(  -\frac{75356125}{3326976} \chi^{-19/3}- \frac{1326481225}{101334144} \chi^{-19/9}+ \frac{17355248095}{455518464}\chi^{-38/9}\right)\,e^6_0 \\\nonumber &+& \left(-\frac{250408403375}{1011400704} \chi^{-19/3}+\frac{4537813337273}{39444627456}\chi^{-76/9}-\frac{6505217202575}{277250217984}\chi^{-19/9}+\frac{128274289063885}{830865678336}\chi^{-38/9} \right)\,e^8_0\\\nonumber &+& {\cal{O}}(e^{10}_0)\Bigg]\,,
 \end{eqnarray}
 \end{widetext}
where $\chi \equiv F/F_{0} = f/f_{0}$, with $F$ the orbital frequency, $f$ the GW frequency and subscript zero to indicate the initial values of these quantities at which the eccentricity $e_{0}$ is defined.  Such a model does contain amplitude corrections relative to its circular counterpart, because the $\xi_{\ell}$'s are functions of eccentricity, which  in turn is a function of the orbital frequency. 

The PC formalism is appealing but the waveforms presented above are not adequate for a data analysis study. This is simply because the expansions have been truncated at leading PN order. Therefore, by construction, the PC model does not reduce to the {\texttt{TaylorF2}} waveform family in the limit of zero eccentricity beyond leading PN order. The resulting overlaps are thus terrible. Having said that, the PC formalism is still promising because one could in principle include higher-order PN corrections, thus systematically improving the waveform family. 

\section{The Enhanced Post-Circular Waveform Model}
\label{sec:EPC}

Systematic, well-controlled PN expressions for time-domain elliptic orbits are available in the literature~\cite{Blanchet:1989M,Gopakumar:2002,Arun:2008,GopakumarandK:2006,Tessmer:2010sh,Tessmer:2010ii}. One could in principle use this work to extend the analysis of~\cite{Yunes:2009} to higher PN order. Doing so, however, becomes increasingly difficult with higher PN order, so much so that the analytic expressions become unwieldly and the resultant frequency-domain waveform becomes computationally expensive. 

Rather than systematically using those expressions as the basis for a Fourier-domain approximation, this paper adopts a physically-motivated \emph{ansatz} to \emph{extrapolate} the form of $\tilde{h}(f)$ from known behavior in two limits: the high order quasi-circular PN approximation and the leading-order PC approximation. Though lacking some eccentricity-dependent modifications beyond leading PN order, this ansatz is particularly easy to construct, it captures the leading-order effects of eccentricity, and (as we show later) it is effective at reproducing the results of time-domain calculations like the $x$-model at suitable PN orders.  

\subsection{Requirements and Construction}

We construct the EPC model with the following requirements:
\begin{itemize}
\item To zeroth order in the eccentricity, the model must recover the {\texttt{TaylorF2}} PN waveform at 3.5 PN order. 
\item To zeroth PN order, the model must recover the PC expansion of~\cite{Yunes:2009}, including
eccentricity corrections up to order \({\cal{O}}(e^8)\).    
\end{itemize}
\vspace{2mm}

There are an infinite number of ways in which one can modify the PC expansion of~\cite{Yunes:2009} to satisfy these two requirements. We choose to employ the same functional form for the Fourier phase as that used in the {\texttt{TaylorF2}} model at 3.5 PN order, Eq.~\eqref{phasef2}. However, we will use a modified velocity function $v(f) \to v_{\rm ecc}(f;e_{0})$ calculated by equating Eq.~\eqref{phasepcexp} at $\ell =2$ with Eq.~\eqref{phasef2} at $i=0$,
which leads to:
 \begin{widetext}
 \begin{eqnarray}
 \label{v_ecc}
v_{ecc}(f;e_{0}) &=& \Bigg\{1+ \frac{471 e_0^2}{1462 }\chi^{-19/9} +  e_0^4\left(\frac{521711}{444448} \chi^{-19/9}-\frac{391963333705}{533796714784}\chi^{-38/9}\right) \\\nonumber &+& e_0^6\left(\frac{265296245}{101334144}\chi^{-19/9} - \frac{1302494157901715}{243411301941504}\chi^{-38/9} +\frac{6142097676388541753}{2135203940630873088}\chi^{-19/3}\right) \\\nonumber&+&   e_0^8 \Bigg\{\frac{1301043440515}{277250217984}\chi^{-19/9}-\frac{9626858181465026345}{443982214741303296}\chi^{-38/9}+\frac{20410190578639124245219}{649101997951785418752}\chi^{-19/3}\\\nonumber&-&\frac{2140356054716884783056259067777}{152020619381675205883794309120} \chi^{-76/9}\Bigg\}  \Bigg\}  \left(\pi\,M\,f\right)^{1/3}\,.
\end{eqnarray}
\end{widetext}

\noindent With this at hand, the EPC model is defined as the PC model for $\tilde{h}(f)$ of Eq.~\eqref{PC-model} but with $\Psi_{\ell} \to \bar{\Psi}_{\ell}$, where 
\begin{equation}
\bar{\Psi}_{\ell} \equiv {\rm{R}}_{e_{0}\ll 1} \left[2 \pi f t_{c} - \ell \phi_{c} - \frac{\pi}{4} + \left(\frac{\ell}{2}\right)^{8/3} \Psi_{\rm{PN}}(v_{ecc})\right]\,,
\end{equation} 

\noindent where ${\rm{R}}\left[\cdot\right]$ stands for re-expanding in $e_{0}$ up to ${\cal{O}}(e_{0}^{8})$. The different PN expressions for $\bar{\Psi}_{2}$ are presented in Appendix~\ref{Apen_PN}. 

\begin{figure}[ht]
\includegraphics[height=0.35\textwidth,  clip]{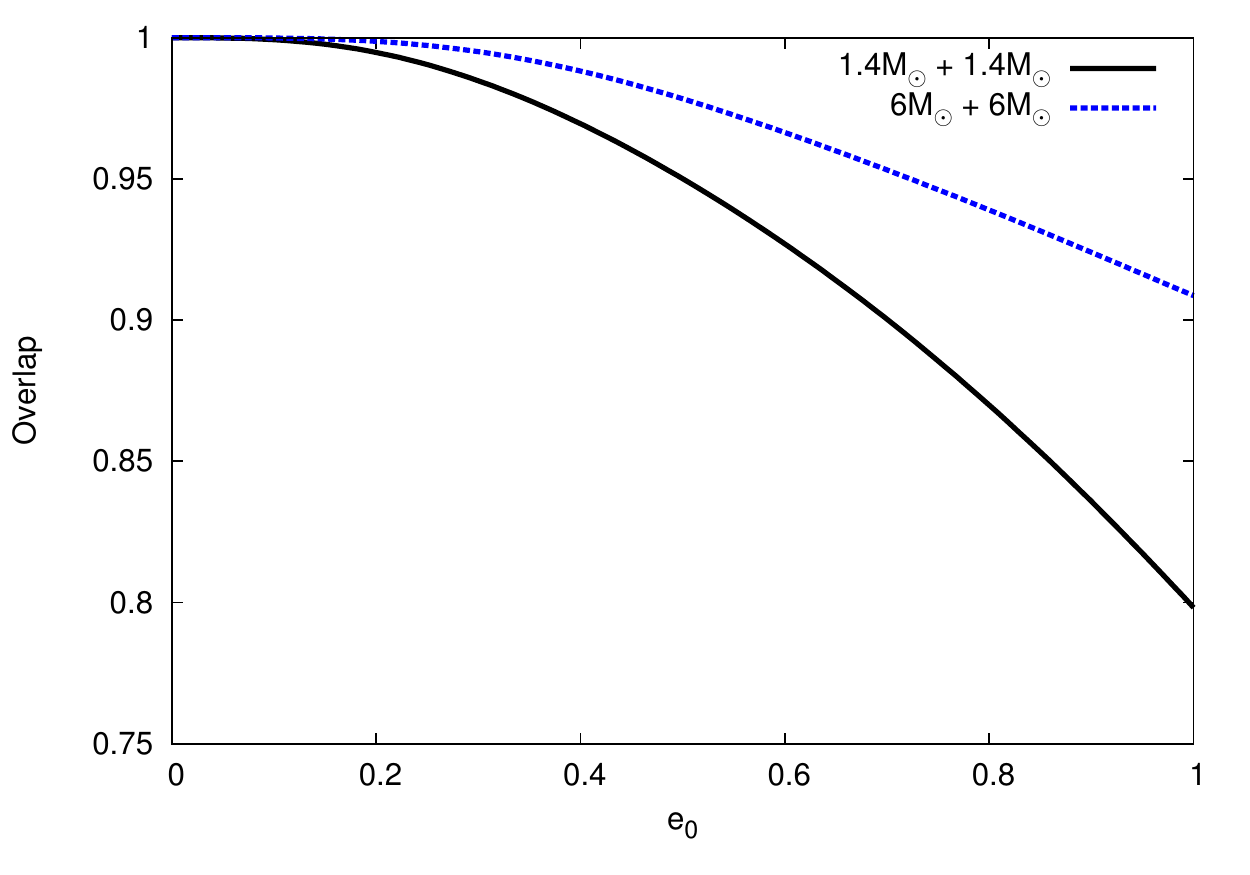}
\caption{(Color Online) We explore the accuracy of the phase prescription of the EPC model by deriving two different phase expressions which include eccentricity corrections up to order \(e^8\) and \(e^6\), respectively. Integrating from a fiducial GW frequency of 10 Hz, the y-axis shows the overlap between both phase approximations. Notice that the overlap between both phase prescriptions is reliable, i.e., the overlap \( \gtrsim0.97\) for  \(e_{0}\lesssim 0.6\) for a (\(6\Msun,\, 6\Msun\)) system,  and for  \(e_{0}\lesssim 0.4\) for a (\(1.4\Msun,\, 1.4\Msun\)) system.}
\label{acc_phase}
\end{figure}

In summary, the EPC model has some appealing features of the two waveform families taken as reference points (the $x$-model at 2 PN order and the {\texttt{TaylorF2}} model at 3.5 PN order). First, as shown in Figure~\ref{acc_phase}, the phase prescription used to construct the EPC model is reliable for  \(e_{0}\lesssim 0.6\) for a (\(6\Msun,\, 6\Msun\)) system,  and for  \(e_{0}\lesssim 0.4\) for a (\(1.4\Msun,\, 1.4\Msun\)) system.  That is, for initial eccentricities at a GW frequency of 10 Hz below these eccentricity values, the error induced by neglecting the ${\cal{O}}(e_{0}^{10})$ terms in the EPC model lead to a loss of overlap of less than $3\%$. Second, like the {\texttt{TaylorF2}} model, the EPC family has the advantage that it is already written analytically in the frequency domain. Therefore, it can be readily used as an efficient and accurate waveform family for future searches of eccentric systems, or to explore the efficiency of current matched-filter algorithms to clearly distinguish instrumental glitches from eccentric signals. Regarding efficiency, we have found that, averaging over 100 iterations, the code we have developed to generate EPC waveforms is two times faster than the \(x\)-model used in~\cite{Huerta:2013a} --- a numerical code that was enhanced by implementing adaptive time-stepping.

The EPC model also has the advantage of encoding high PN order corrections that faithfully describe the dynamical evolution of quasi-circular binaries and high-order eccentricity corrections that reproduce the dynamics of compact sources with low to moderate values of eccentricity. Of course, this model is not perfect, as we will see next, but it can be systematically improved by correcting the eccentricity-dependent, and higher PN order terms in the phase and amplitude modulations, using the results in~\cite{Blanchet:1989M,Gopakumar:2002,Arun:2008,GopakumarandK:2006,Tessmer:2010sh,Tessmer:2010ii}.

\subsection{Comparison to Other Models}
\begin{figure*}[ht]
\includegraphics[height=0.38\textwidth,  clip]{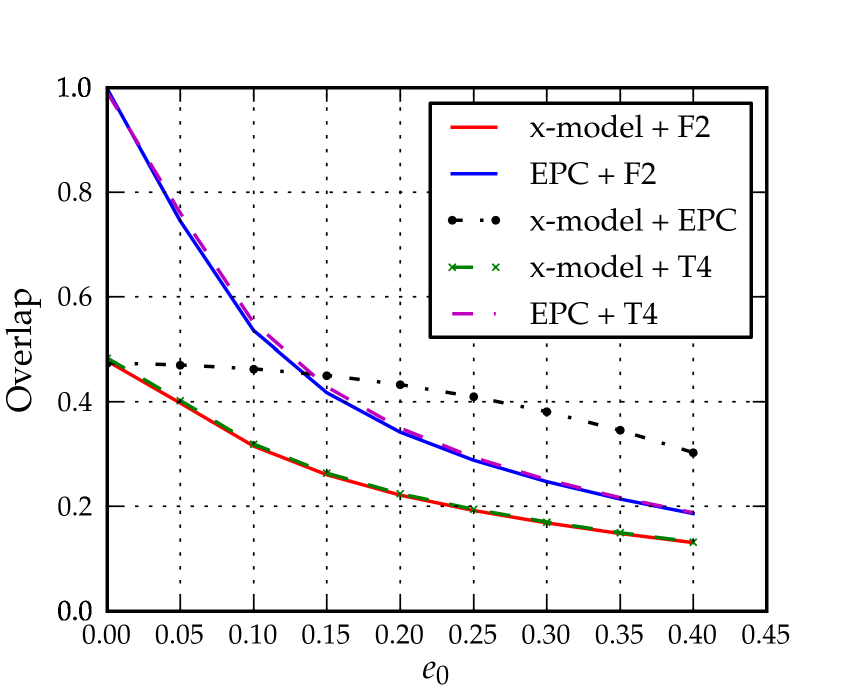}
\includegraphics[height=0.38\textwidth,  clip]{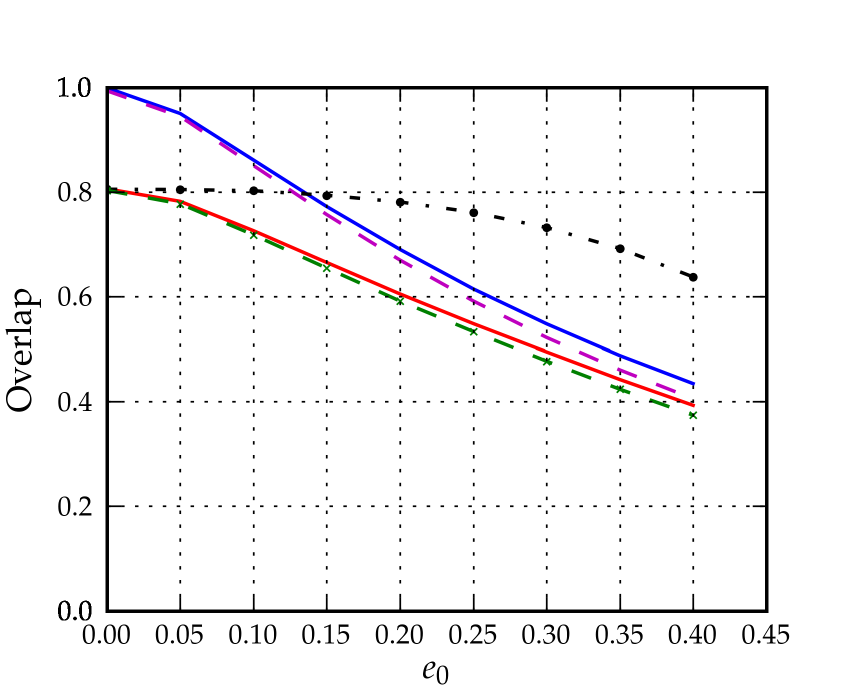}
\\
\includegraphics[height=0.38\textwidth,  clip]{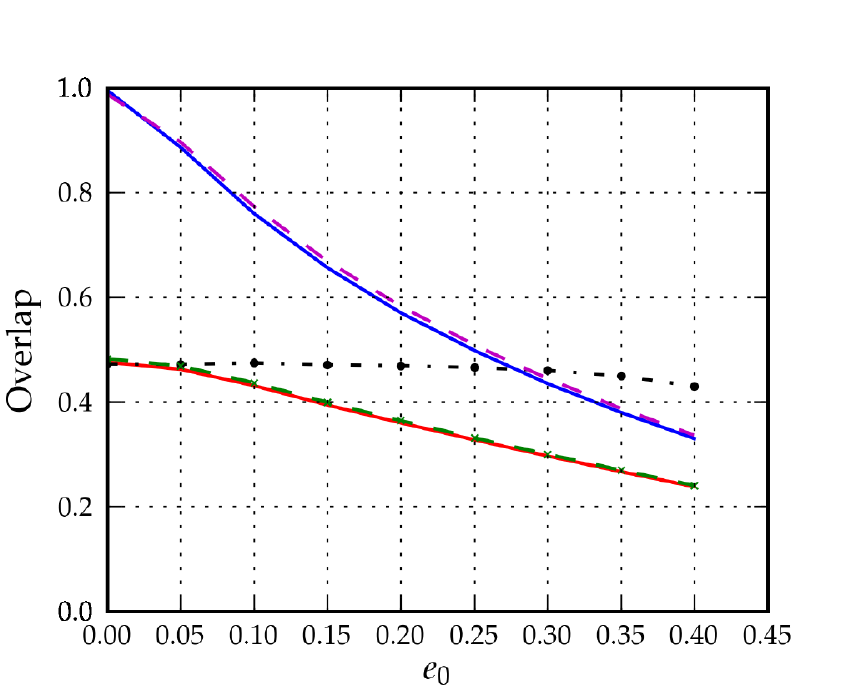}
\includegraphics[height=0.38\textwidth,  clip]{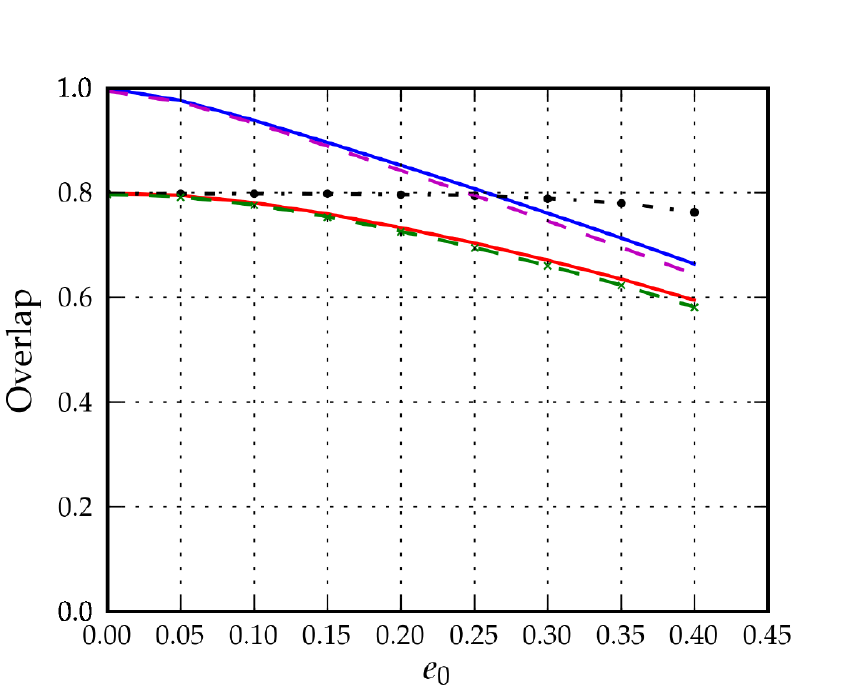}
\caption{(Color Online) Overlap between the EPC waveform and three waveform families: {\texttt{TaylorF2}},  {\texttt{TaylorT4}} and the \(x\)-model; and the \(x\)-model with {\texttt{TaylorF2}} and {\texttt{TaylorT4}}. Note that both  {\texttt{TaylorF2}} and {\texttt{TaylorT4}} include waveform phase corrections at 3.5 PN order (see legend of top-left panel for reference). The different panels show the overlap for binaries with component masses \((1.35\Msun, \,1.35\Msun)]\) (left) and \([(6\Msun, 6\Msun)]\) (right). The top panels assume an aLIGO detector with the ZDHP sensitivity configuration and a lower frequency cut-off of 10 Hz, while the bottom panels use an ET-like detector with the ETB configuration and a lower frequency cut-off of 5Hz. }
\label{overBNSLIGO}
\end{figure*}

Figure~\ref{overBNSLIGO} shows the overlap between different waveform families as a function of GW frequency. We have used 10 Hz and 5 Hz as the initial filtering frequency for aLIGO and ET, respectively. Observe that the EPC model does reduce to the {\texttt{TaylorF2}} 3.5 PN model in the \(e_{0}\rightarrow 0\) limit.  Specifically, the blue curves ({\texttt{TaylorF2}} vs EPC) go to $1$ as $e \to 0$. Observe also that the overlap between the \(x\)-model and either the EPC model or the {\texttt{TaylorF2}} 3.5 PN model is roughly the same as $e \to 0$. Additionally, the $x$-model is systematically unfaithful to the EPC model by a nearly constant amount, which is nearly independent of eccentricity at each mass. These differences arise due to systematic differences between time- and frequency-domain waveforms, as well as by slight differences in PN order used to construct these models.

In Figure~\ref{overBNSLIGO}  we also compare the performance of EPC and the \(x\)-model against the time domain PN-based TaylorT4 3.5 PN approximant, which has been shown to accurately reproduce the features of numerical relativity simulations for the very last few cycles of non-spinning equal mass binary inspirals~\cite{pnbuo,boyle}. Notice that the overlap between EPC and TaylorT4 3.5 PN shows that EPC captures faithfully the features of quasi-circular binaries. In contrast, the low-order PN expansion used to construct the \(x\)-model leads to substantial drops in overlap in the quasi-circular limit.  

Figure~\ref{overBNSLIGO} also demonstrates the significant impact of eccentricity in data analysis.  As also seen in previous studies, the overlap between quasi-circular and eccentric waveforms decreases rapidly as the eccentricity is increased even by a small amount.  In addition, the longer a binary spends in band (i.e., the lower the mass or the initial frequency of the system), the greater is the effect of eccentricity on data analysis, since more cycles accumulate.

In principle, higher order eccentricity corrections in the PN expansion that we have neglected could contaminate our ability to identify the unique impact of eccentricity. In fact, the PN approximation for quasi-circular inspirals is known to converge slowly or even diverge, particularly late in the inspiral~\cite{Yunes:2008tw,Zhang:2011vha,Kocsis:2012,Favata:2014}. In practice, however, eccentric effects are weakest at the end of the inspiral and strongest early on, so we expect that eccentric effects are less susceptible to large systematic errors from unknown higher-order PN terms.

\section{Astrophysics and Cosmology with Eccentric Waveforms}
\label{sec:astro-cons}
 
Having constructed a waveform model that captures the main features of eccentric binaries, in this section we compute
the signal-to-noise ratio (SNR) distribution of a variety of compact binary sources on eccentric orbits whose
mass-ratios represent typical binary BH and NSBH systems. As is well known for
circular~\cite{Prayush:2013a,Colin:2013,Pekowski:2013} and eccentric \cite{Yunes:2009} binaries, the presence of
multiple harmonics provides additional signal power, particularly at frequencies where the dominant harmonic may be
inaccessible.  For this reason, eccentric binaries can potentially be detected with very large masses that are
inaccessible for quasi-circular inspirals.   Due to cosmological redshift, massive eccentric binaries can conceivably be
detected to much greater distances than their quasicircular counterparts.  

The SNR can be computed via
\begin{equation}
\mathrm{SNR} = \sqrt{(h,h)}\,,
\label{effe_SNR}
\end{equation}
where as before, \(f_{\rm{low}} = 1 \rm{Hz}$ and $10 \rm{Hz}\) for ET and aLIGO respectively. Since the PN approximation breaks down as the system approaches merger, we truncate the signal at an orbital frequency corresponding to the innermost stable circular orbit (ISCO) of a test particle in Schwarzschild spacetime; eccentricity-corrections to the ISCO are negligible for SNR calculations.

 As shown in~\cite{Arun:2007}, different GW harmonics will contribute signal over different GW frequency ranges if we terminate the integration of Eq.~\eqref{effe_SNR} at the ISCO orbital frequency. Hence, in order to ensure that the harmonics contribute to the SNR within their region of validity, we will truncate the waveforms using step functions, (\(H(x)=1, {\rm{if}}\, x\geq0\), and \(H(x)=0\) otherwise), as follows~\cite{Yunes:2009}:
\begin{equation}
\tilde{h}={\cal{A}} f^{-7/6}\sum_{\ell=1}^{\ell=10} \left(\frac{\ell}{2}\right)^{2/3} \xi_{\ell} \; e^{i\Psi_{\ell}}\,H(\ell F_{\rm{ISCO}} - 2 F)\,,
\label{trunc_wave}
\end{equation} 
where $F$ is the Keplerian mean orbital frequency and $F_{\rm ISCO}$ is that frequency at ISCO. 
  
\subsection{Increase in Reach}  

\begin{figure*}[htp]
\includegraphics[height=0.33\textwidth,  clip]{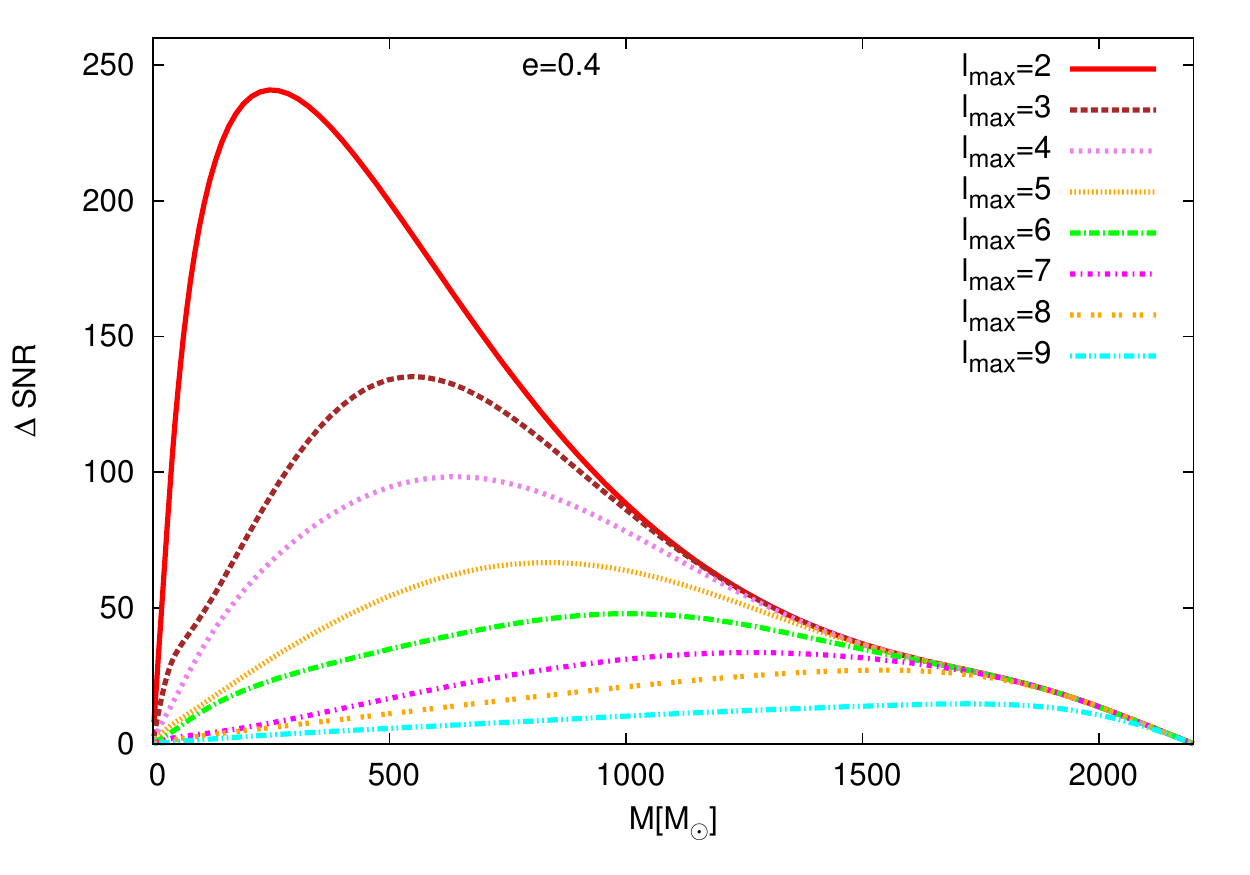}
\includegraphics[height=0.33\textwidth,  clip]{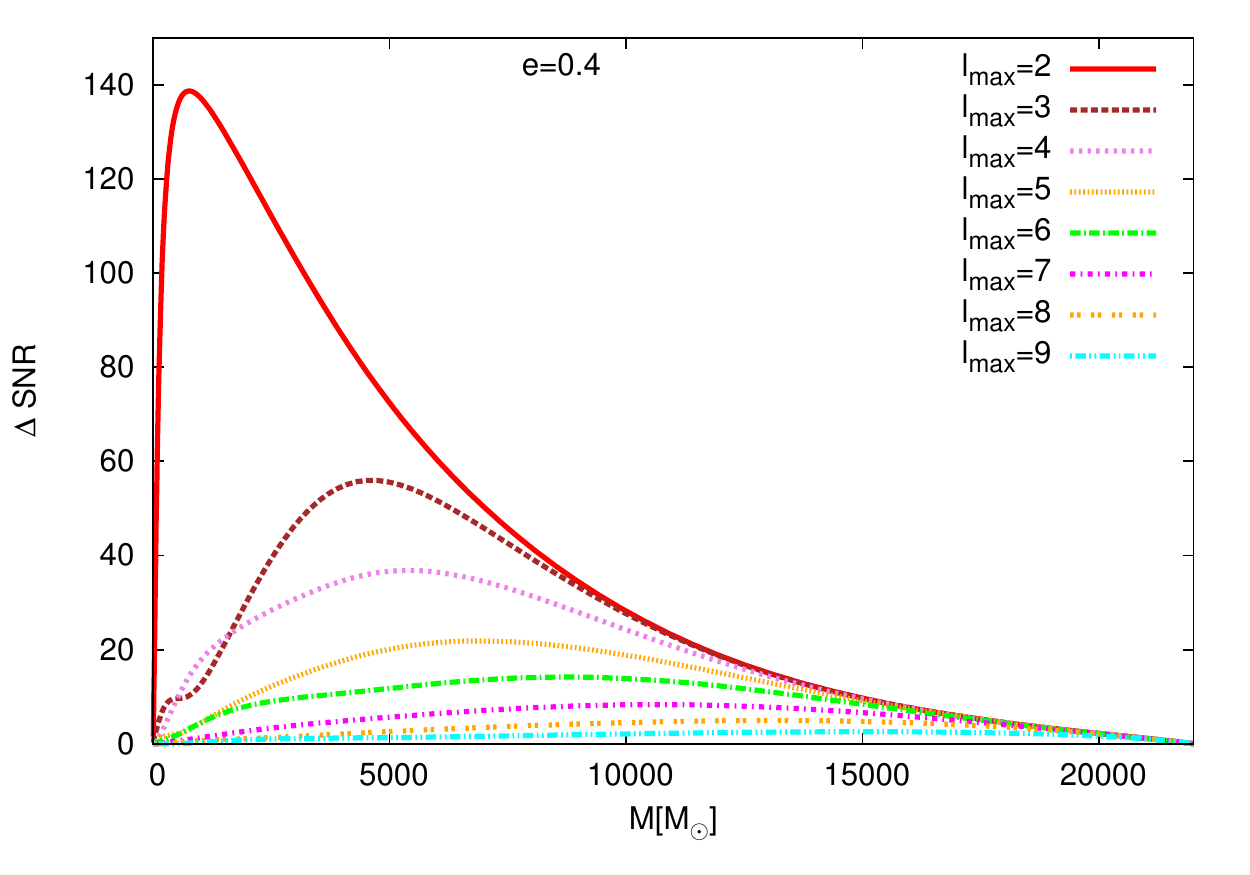}
\caption{(Color Online) Absolute contribution from each harmonic to the total SNR of an optimally oriented
  equal-mass binary directly overhead a single detector (see Eq.~\eqref{partialsnr} for further reference). The left
  Figure assumes a LIGO-type detector and a fixed distance of \(D_L=100\) Mpc. The right Figure assumes an ET-type
  detector and a fixed distance of $D_L = 1$ Gpc. We have assumed that the binaries have an initial eccentricity \(e_0
  =0.4\) at a GW frequency of 1 Hz for ET and 10 Hz for aLIGO. For comparison, Fig.~\ref{one_IFO} shows the total
SNR versus mass.}
\label{elesandsnr}
\end{figure*}

An immediate consequence of the inclusion of eccentricity in the waveform model is the ability to detect systems at a fixed mass farther out, and to detect heavier systems at a fixed distance. The former case was already made in~\cite{Yunes:2009} for space-borne GW detectors, so let us discuss the latter. Figure~\ref{elesandsnr} shows the SNR contribution from each harmonic versus total mass for an optimally oriented equal-mass binary directly overhead for a single aLIGO (left) or ET detector (right). The SNR contribution is obtained from
\begin{equation}
\Delta\rm{SNR} = \sqrt{ |\rm{SNR}^2_{\ell_{\rm{max}}}-\rm{SNR}^2_{\ell_{\rm{max}}=10}|}\,,\,\, \ell_{\rm{max}}={2,\,3\,,...,9}\,, 
\label{partialsnr}
\end{equation}
\noindent where the subscript \(\ell_{\rm{max}}\) indicates the number of harmonics included in the calculation of SNR. 
Observe that for the eccentric inspirals investigated here, the role of higher harmonics is important. 

Based on Fig.~\ref{elesandsnr}, we notice that in the context of aLIGO, the SNR difference peaks at approximately \((200\Msun,\, 500\Msun,\,600\Msun,\, 700\Msun)\) for \(\ell_{\rm{max}}=2,\,3,\,4\) and 5, respectively. In the context of ET, the SNR difference reaches a maximum at approximately \((600\Msun,\, 4500\Msun,\,5500\Msun,\, 7000\Msun)\) for \(\ell_{\rm{max}}=2,\,3,\,4\) and 5, respectively.  We have also found that, both for aLIGO and ET, in order to ensure that the SNR difference (see Eq.~\eqref{partialsnr}) is less than 10, then we need to include up to the ninth harmonic throughout the whole parameter space. 

\begin{figure*}[htp]
\centerline{
\includegraphics[height=0.35\textwidth,  clip]{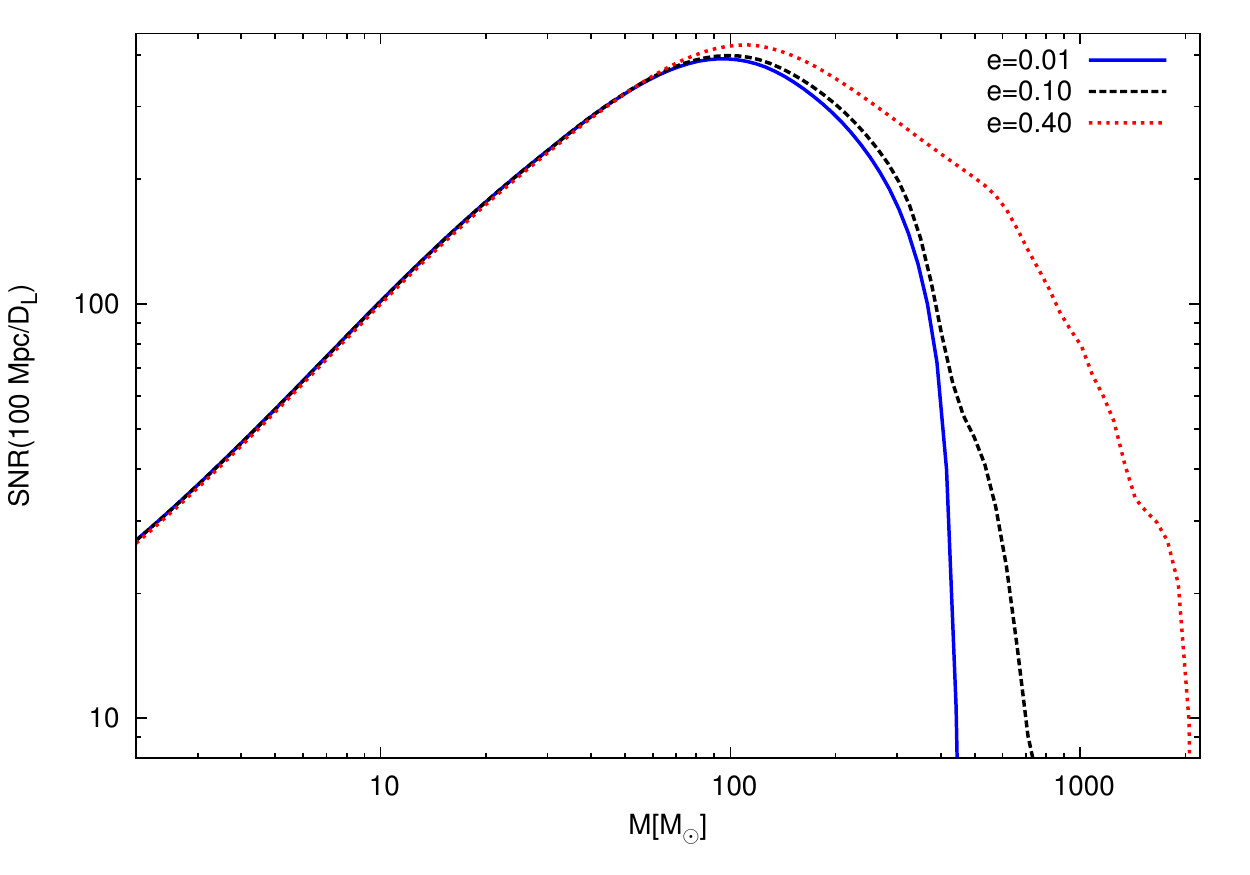}
\includegraphics[height=0.35\textwidth,  clip]{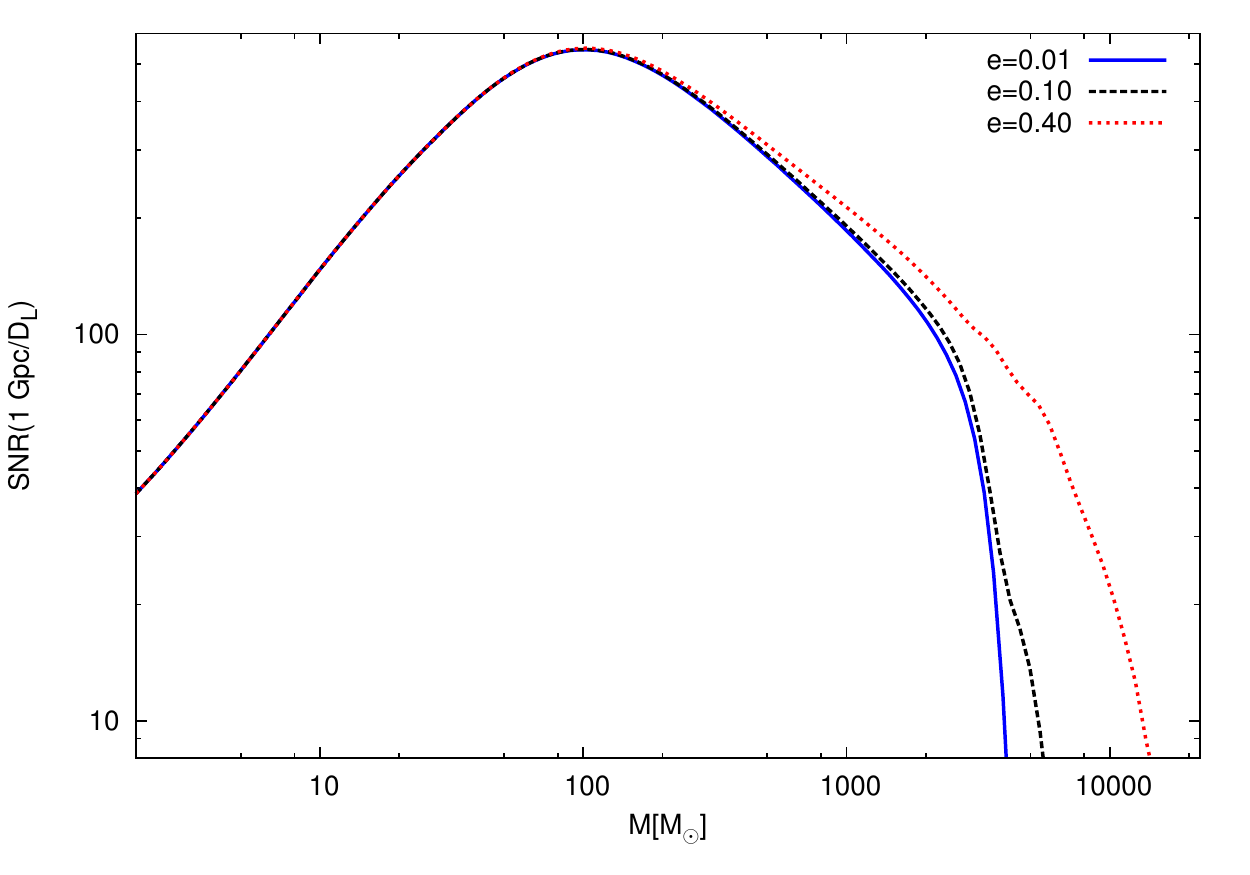}
}
\caption{(Color Online) SNR versus eccentricity for an optimally-oriented, equal-mass binary directly overhead a single detector. Observe that eccentricity can significantly increase the reach of instruments, if all power involved can be captured. For the LIGO-type detector, the SNR is normalized at a distance \(D_L= 100\) Mpc, while for the ET detector we use the ETB configuration and normalize the SNR at \(D_L= 1\) Gpc. }
\label{one_IFO}
\end{figure*}

Figure~\ref{one_IFO} shows the SNR versus total mass and eccentricity for an optimally-oriented, equal-mass binary directly overhead for a single detector. We have used two different detectors --- aLIGO and ET --- to illustrate the increase in reach as a function of sensitivity and eccentricity. We should emphasize, however, that for total mass \(M\gtrsim 300\Msun\), Figure~\ref{elesandsnr} provides a conservative estimate of the SNR that may be expected from these type of events, based on well understood physics only. All of these results are consistent with those presented in~\cite{Yunes:2009}.
 
\subsection{Eccentric Neutron Star-Black Hole Binaries} 
\label{nsbh_sec}
Observational and theoretical evidence suggests that SGRBs may be
associated with NSNS and NSBH mergers~\cite{Paczynski:1986,Eichler:1989,Edo:2013}. 
Our previous calculations show that eccentric mergers can be detected with a larger range than quasicircular
mergers.  

To assess the astrophysical impact of including eccentricity in event  rate calculations, we estimate the
detection rate for a simply-parameterized astrophysical toy model, using EPC waveforms.
Our astrophysical toy model assumes that one to a few eccentric, merging NSBH binaries form per young massive star cluster, over its lifetime \cite{Clausen:2013}.  Following~\cite{Oleary:2007}, we then extrapolate this optimistic formation rate to the entire universe, assuming that the star formation rate (SFR) per unit volume in the Universe, and the cluster formation rate (CFR) are given by
\begin{eqnarray}
\label{cons1}
{\rm{SFR}} &=& 1 \Msun\, \times10^{-2}\, {\rm{galaxies}}\,{\rm{Mpc}}^{-3}\,{\rm{yr}}^{-1}\,,\\
\label{cons2}
{\rm{CFR}} &=& {\rm{SFR}}/10^{6}\Msun\,.
\end{eqnarray}
Assuming that a significant fraction of star formation occurs in clusters, \(g_{\rm{cl}}\),  and that a substantial fraction of clusters undergo evaporation and segregation,  \(g_{\rm{evap}}\), then present-merger rates of compact binaries can be approximated as \(n_{\rm{rate}}=\mathrm{CFR}\,\times\) number per cluster:
\begin{equation}
n_{\rm{rate}} \approx 10^{-2}\,g_{\rm{evap}}\,g_{\rm{cl}} \,{\rm{Mpc}}^{-3}\,{\rm{Myr}}^{-1}\,.
\label{rate_rich}
\end{equation}

Given the lack of a universally accepted model for star formation history, we will present results for merger rates normalized by the factor \(\Gamma  = g_{\rm{evap}}\,g_{\rm{cl}} \), i.e.,
\begin{equation}
n_{\rm{rate}} = 10\,\Gamma\, {\rm{Gpc}}^{-3}\,{\rm{yr}}^{-1}\,.
\label{rate_table}
\end{equation}
Using Eq.~\eqref{rate_table}, we notice that we can readily estimate the number of present-day event rates using the relation
\begin{equation}
n_{\rm{events}} = 10\,\Gamma\,\left(\frac{V_c}{{\rm{Gpc}}^{3}}\right)\, {\rm{yr}}^{-1}\,.
\label{numberofevents}
\end{equation}
If one assumes that \(10\%\) of all clusters survive disruption due to photoionization and supernova gas-driven ejection during  their first \(\sim 10\) Myr of existence, then \(\Gamma\sim 5\times 10^{-2}\). Under more optimistic assumptions, \(\Gamma\) can plausibly be as high as  \(\sim1\).

For a given network of identical GW detectors,  a straightforward calculation provides the detection volume $V_c$ for a given source  \cite{Schutz:2011, Dominik:2014}, including cosmology.   Specifically, for each
source, sky location, and distance $D_L$, we evaluate the SNR in each detector and hence the network SNR $\rho$, carefully accounting for the topology and geographical location of its components~\cite{Schutz:2011}.  For each distance and sky position, a fraction of sources have $\rho> \rho_{\rm threshold}\equiv 10$. Using concordance cosmology, we translate the SNR of sources with  $\rho> \rho_{\rm threshold}$ to a detection range, and finally into a redshift estimate, which in turn determines the detection volume $V_c$.

\begin{table}[htp]
\centering
\begin{tabular}{| >{\centering\arraybackslash}m{1.5cm} | >{\centering\arraybackslash}m{1.5cm} | >{\centering\arraybackslash}m{1.5cm} | >{\centering\arraybackslash}m{1.5cm} | >{\centering\arraybackslash}m{1.5cm} |}
\hline
Network&LIGO Livingston&LIGO Hanford&Virgo &INDIGO\\\cline{1-5}
\(C_1\) & \checkmark &   &  &  \\\cline{1-5}
\(C_2\) &\checkmark  & \checkmark  &  &   \\\cline{1-5}
\(C_3\) &\checkmark  & \checkmark  & \checkmark &   \\\cline{1-5}
\(C_4\) &\checkmark  & \checkmark  & \checkmark & \checkmark   \\\cline{1-5}
\end{tabular}
\centering
\caption{Configurations, \(C_i\), used to compute the SNR distributions of compact binaries that will be targeted by second and third generation ground-based detectors.}
\label{configurations}
\end{table}

We evaluate the detection volume for several combinations of detectors, as described in
Table~\ref{configurations}.
The first detector network is comprised of four LIGO-type L-shaped detectors that operate with the target ZDHP laser configuration, from a low frequency cut-off of 10Hz. The second detector network consists of up to four ET type detectors, including one triangular-shaped detector at the geographic location of Virgo, and up to three L-shaped detectors at the location of LIGO Livingston, LIGO Hanford, and the proposed location for LIGO India (see Table II). We explore two different design sensitivities for these networks, ETB and ETD, and we assume that both configurations operate from a low frequency cut-off of 1Hz. Given that the results obtained by using ETB and ETD are quantitatively similar, we will only quote results derived using ETB in the following. Table~\ref{cosmo_gene}  provides the comoving volume for several compact binaries that may be detected during the advanced detector era and beyond using a four detector network, assuming a uniform distribution of inclination angles and adopting an eccentricity $e_0=0.4$ at $f_{\rm orb}=0.5\, {\rm Hz}$ and $5\,{\rm Hz}$ for ET and aLIGO, respectively.  For context, these tables also provide the median luminosity distance and  associated redshift.  
Based on Table~\ref{cosmo_gene}, we notice that in a conservative or in an optimistic scenario (\(\Gamma \sim
5\times10^{-2}$ or $1\)), an array of four advanced GW detectors could detect from 0.1 to 10 events per year. In Figure~\ref{redshift_genericSNR}  we present the redshift distribution up to which these events may be detected depending on the number of operating GW detectors.  Notice that some of these events could be detected up to cosmological redshifts \(z\sim 0.2\).

\begin{table}[htp]
\centering
\begin{tabular}{| >{\centering\arraybackslash}m{1.5cm} | >{\centering\arraybackslash}m{1.5cm} | >{\centering\arraybackslash}m{1.5cm} | >{\centering\arraybackslash}m{1.5cm} | >{\centering\arraybackslash}m{1.5cm} |}
\hline
\multicolumn{5}{|c|}{aLIGO} \\
\hline
$D_L\, [\rm{Gpc}]$ & $z $ & $ m\,[\Msun] $ & $ M\,[\Msun] $ & $ V_c\,[\rm{Gpc}^3]  $ \\\cline{1-5}
0.425&	0.090&	1.38&	5.50&	0.255 \\\cline{1-5}
0.612&	0.126&	1.33&	10.66&	0.682 \\\cline{1-5}
0.661&	0.135&	1.41&	14.98&	0.834 \\\cline{1-5}
0.726&	0.147&	1.39&	20.05&	1.066 \\\cline{1-5}
\end{tabular}

\begin{tabular}{| >{\centering\arraybackslash}m{1.5cm} | >{\centering\arraybackslash}m{1.5cm} | >{\centering\arraybackslash}m{1.5cm} | >{\centering\arraybackslash}m{1.5cm} | >{\centering\arraybackslash}m{1.5cm} |}
\hline
\multicolumn{5}{|c|}{ET} \\
\hline
$D_L\, [\rm{Gpc}]$ & $z $ & $ m\,[\Msun] $ & $ M\,[\Msun] $ & $ V_c\,[\rm{Gpc}^3]  $ \\\cline{1-5} 15.168&1.917&1.71&5.14   &\(589.596\)  \\\cline{1-5}
17.587&2.164&1.26&9.48   &\(719.221\)  \\\cline{1-5}
18.894&2.296&1.51&14.56  &\(789.238\)  \\\cline{1-5}
16.593&2.063&1.31&19.26   &\(665.925\)  \\\cline{1-5}
\end{tabular}

\centering
\caption{Average range (\(D_L\)), redshift (\(z\)), source-frame masses \([M,\, m]\),  and co-moving volume (\(V_c\)), for several combinations of compact binary systems uniformly distributed in the sky with initial eccentricity \(e_{0} = 0.4\) at a Keplerian mean orbital frequency of $0.5$  Hz and $5$ Hz for ETB and aLIGO respectively, and using the networks described in Table~\ref{configurations} with aLIGO's ZDHP (top) and the ETB (bottom) noise configurations. These results are obtained by fixing the masses and initial eccentricity of the sources, and then running Monte Carlo simulations over random choices of the extrinsic parameters: inclination angles, location of the source in the sky, polarization angle, etc.}
\label{cosmo_gene}
\end{table}

We have found that the SNR distribution of eccentric inspirals is quantitatively similar to that obtained for quasi-circular binaries, even though the former stay in the sensitive frequency band of GW detectors only half of the time the latter do.  
In other words, assuming optimal filtering, the event rate does not increase or decrease significantly due to the
  presence of eccentricity.  Moreover, prior investigations have showed that existing searches will only be slightly
  selection biased against eccentric binaries with  $e_0 \lesssim  0.4$.  Hence, for existing searches, the  relative detection rate  for eccentric and quasi-circular binaries is effectively identical.  By contrast, our work shows that eccentricity produces a significant, measurable effect on radiated GWs,  unless \(e_0\lesssim 10^{-2}\).   Measurement of eccentricity thereby provides a mechanism to help distinguish between different  astrophysical formation scenarios for merging binaries \cite{Samsing:2014,East:2013,Wen:2003,Tai:2014}. 

\subsection{Third generation detectors}
In the context of third generation detectors, Table~\ref{cosmo_gene} shows that an array of four ET-like detectors will enable the detection of NSBH mergers up to redshifts \(z\sim 2.3\).  Therefore, detection of gravitational radiation emitted by high redshift NSBH mergers could be used in conjunction with optical observations to provide further insight on the astrophysical mechanisms that lead to the formation of high-redshift SGRBs.

\begin{figure*}[htp!]
\centerline{
\includegraphics[height=0.35\textwidth,  clip]{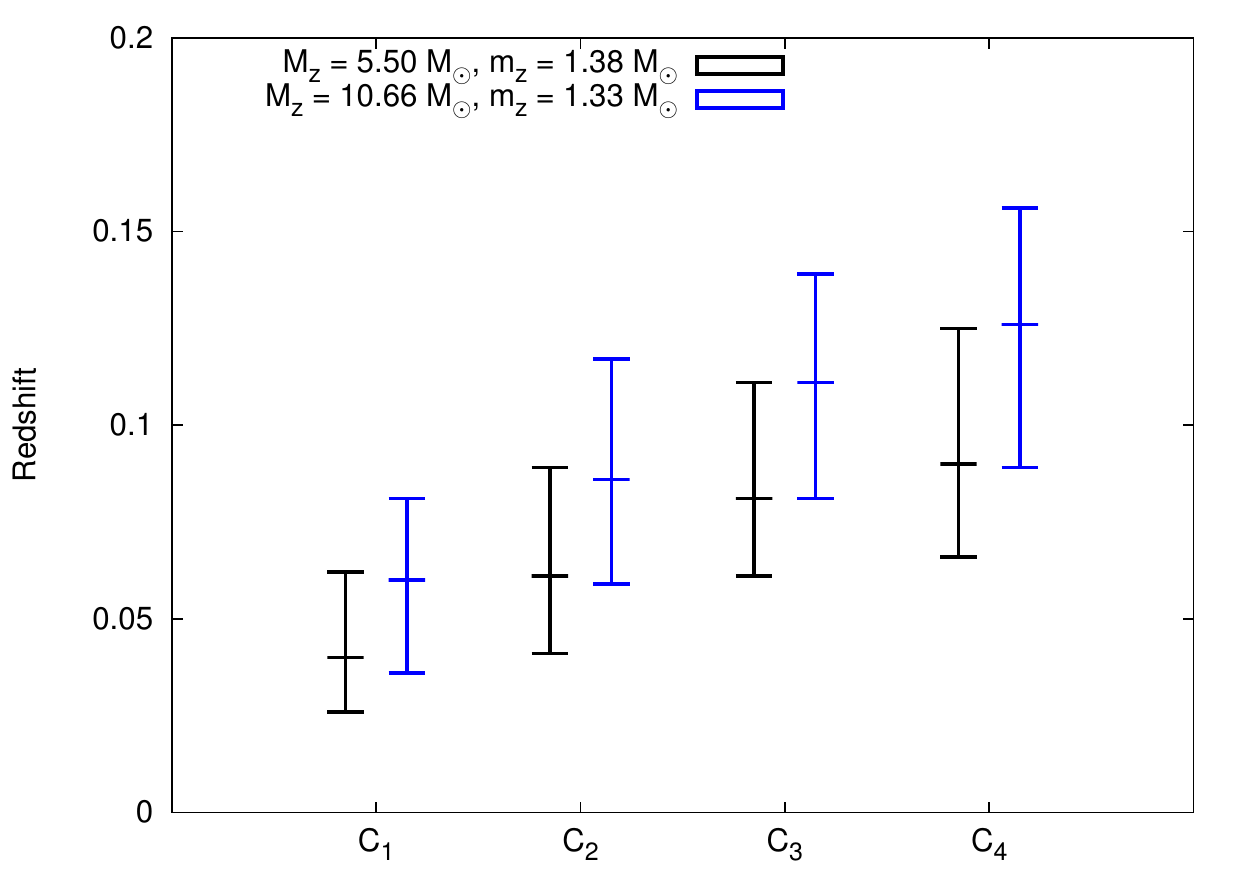}
\includegraphics[height=0.35\textwidth,  clip]{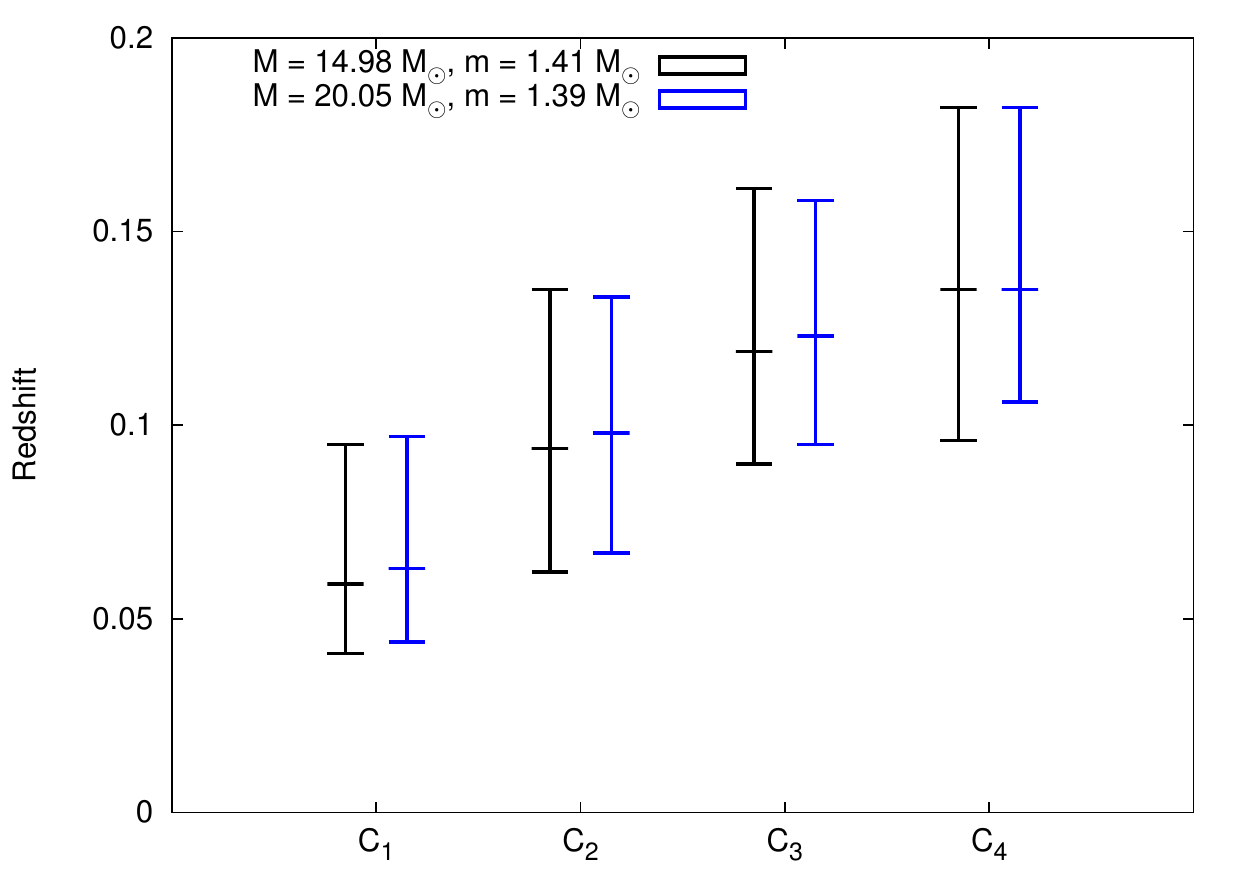}
}
\centerline{
\includegraphics[height=0.35\textwidth,  clip]{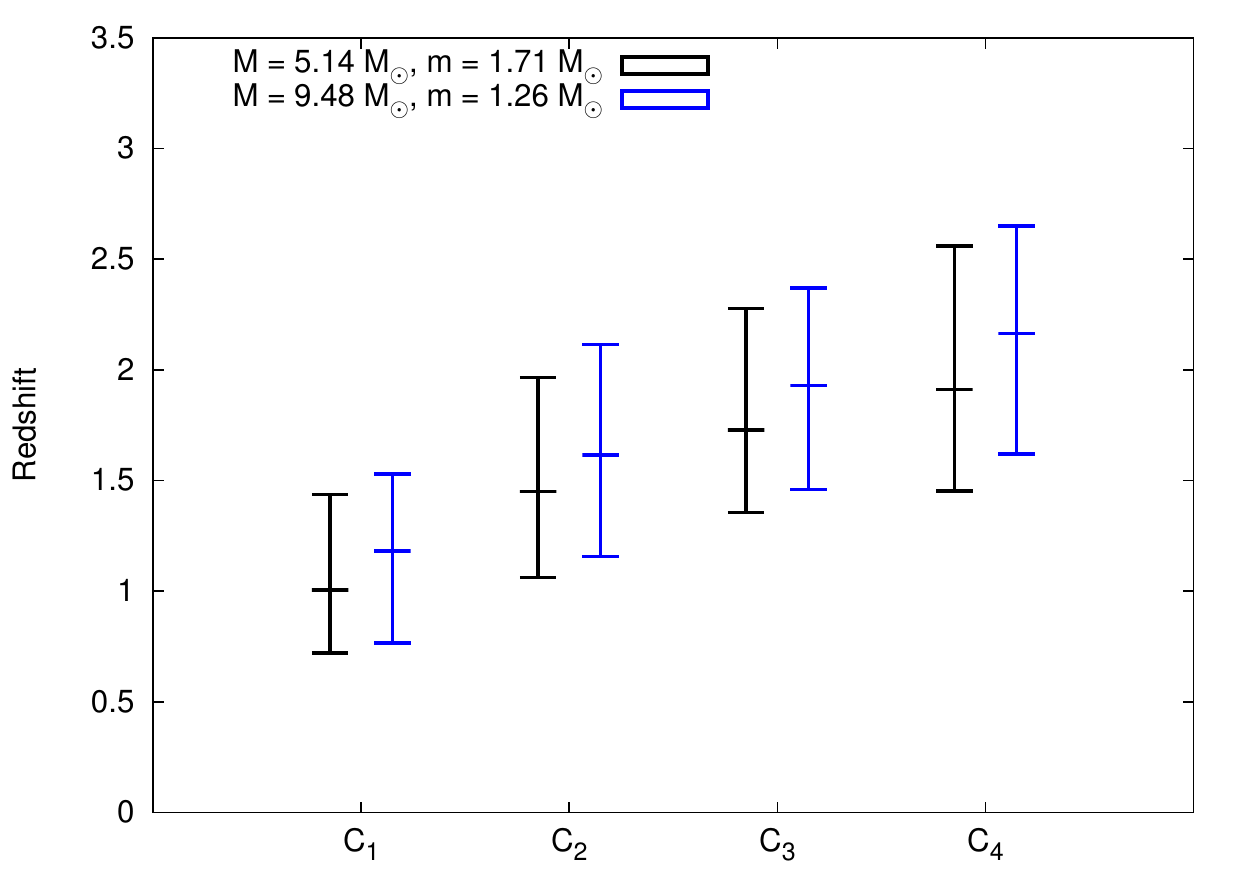}
\includegraphics[height=0.35\textwidth,  clip]{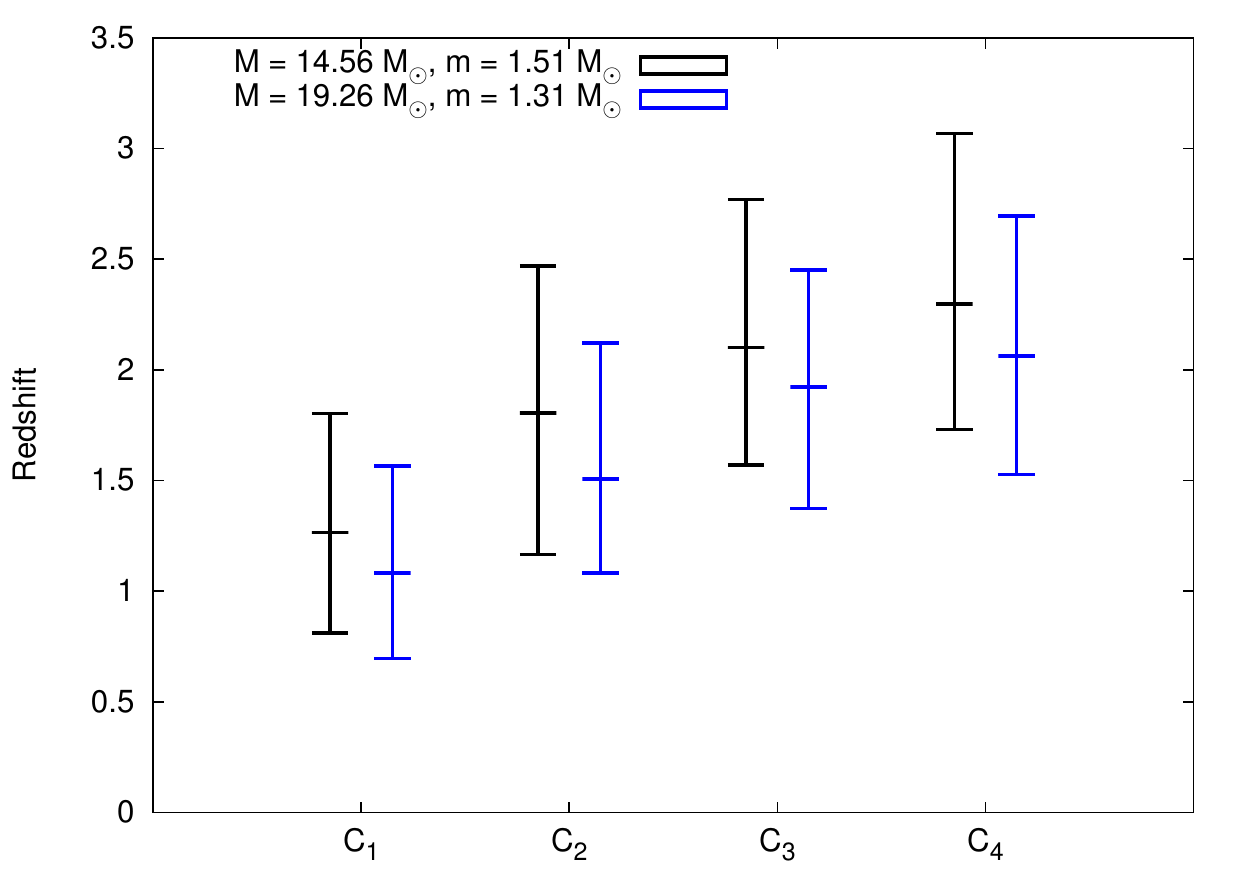}
}
\caption{(Color Online) We run Monte Carlo simulations to estimate the SNR distributions of typical NSBH mergers assuming uniform distributions in the extrinsic parameters of the sources: inclination angles, location of the source in the sky, polarization angle, etc. Using concordance cosmology, we translated the median of the SNR distribution to a detection range, and finally into a redshift estimate. The error bars represent the upper and lower quartiles of the redshift distributions. Each panel shows the redshift distribution for an array of aLIGO detectors --- top panels--- and ET detectors --- bottom panels --- with the sensitivity of the target [ETB, ZDHP] configuration operating from a low frequency cut-off of [1 Hz, 10 Hz], respectively. The \(x\) axis indicates the number of detectors in the array, as defined in Table~\ref{configurations}.}
\label{redshift_genericSNR}
\end{figure*}

If we repeat our earlier exercise, now using Table~\ref{cosmo_gene} in Eq.~\eqref{numberofevents} along with the conservative or the optimistic scenario (\(\Gamma \sim 5\times10^{-2}$ or $1\)), we find that an array of four third generation GW detectors could observe from 60 to 7900 NSBH mergers per year up to redshift \(z\sim 2.3\), again depending on the components masses under consideration (see Figure~\ref{redshift_genericSNR}).

The results presented in this final Section suggest that a network of aLIGO/VIRGO type detectors operating at design sensitivity will be capable of detecting NSBH mergers occurring at cosmological distances, and could enable a joint search of GW sources and their electromagnetic counterparts. Looking ahead, our results also suggest that third generation GW detectors may enable us to discover inspiraling eccentric NSBH binaries at redshifts comparable to the most distant known SGRBs.

\section{Conclusions}
\label{sec:conclusions}

In this paper we have developed and studied a fast and accurate ready-to-use waveform model (the EPC model) to search for compact binary mergers with significant eccentricity ($e < 0.4$) just prior to merger~\cite{East:2013,antonini,Samsing:2014,Thompson:2011}. We have shown that the EPC model includes desirable features of higher-order PN expansions for quasi-circular waveforms, and also includes the same dynamical contributions as time-domain PN-based eccentric waveforms that have been successfully compared with results from numerical relativity.  

Anticipating that matched filtering in the advanced detection era and beyond may be carried out in the frequency domain, the EPC model has been developed in the frequency domain using the SPA. This model is a natural extension of the frequency domain quasicircular PN approximant {\texttt{TaylorF2}} 3.5 PN to now include eccentricity. The EPC model could be used to develop an optimal search that targets moderately eccentric events, which may be important sources for advanced and third generation GW detectors~\cite{aLIGO,virgo,kagra,Freise:2009}, and complementary missions such as NANOGrav~\cite{NANOMaura:2013,Sesana:2013CQG} and space-based missions such as eLISA~\cite{TGU:2013,GairL:2013}. 

We have explored the astrophysics that could be studied if eccentric mergers occur in Nature by computing the improvement in detection range of second and third generation GW detectors. We have also carried out Monte Carlo simulations to explore the SNR distribution of a population of eccentric sources uniformly distributed in the sky, assuming the existence of two types of networks: (i) an array of up to four second generation detectors at the geographical locations of existing and planned LIGO-type detectors, which operate at the target sensitivity of the ZDHP configuration from a low frequency cut-off of 10 Hz, and (ii) an array of up to four ET-type detectors at the geographical locations of existing and planned GW detectors, which operate at the target sensitivity of the ETB and ETD configurations from a low frequency cut-off of 1 Hz. Using the median of the SNR distribution of compact binaries whose source-frame masses represent typical NSBH systems, we have found that GW observations will enable us to observe the merger of eccentric NSBH systems up to redshifts \(z\sim 0.2\) and \(z \sim2.3\), in the context of second and third generation detectors, respectively.  These results suggest that a detector network in the advanced detector era may be capable of testing whether the compact object merger model is the correct description for the generation of SGRBs at cosmological distances~\cite{Inter:2014,SGRBaLIGO:2014}. 

The estimates presented here may be refined once we take into account the SNR that is generated during the merger and ringdown phases of NSBH systems. In order to have a complete picture of the dynamical evolution of these events, we may need to develop a complete inspiral-merger-ringdown model in a similar manner as has been done for quasi-circular binaries with comparable and intermediate mass-ratios~\cite{Huerta:2014a, Huerta:2012,Huerta:2011a, Huerta:2011b,smallbody}. Moreover, the EPC model itself may be improved by consistently accounting for PN corrections to eccentric waveforms in the PC approximation. Such improvements would not only enter in the Fourier phase, but also in relaxing the restricted PN approximation to account for PN amplitude corrections.  

\section*{Acknowledgments}
This material is based on work supported by the National Science Foundation under Award No. PHY-0970074,
PHY-1307429. STM acknowledges the hospitality of the Aspen Center for Physics and NSF grant PHY-1066293. ROS acknowledges the hospitality of the Aspen Center for Physics, support from the UWM Research Growth Initiative, and NSF grant PHY-0970074 and PHY-1307429. NY acknowledges support from NSF grant PHY-1114374 and the NSF CAREER Award PHY-1250636, as well as support provided by the National Aeronautics and Space Administration from grant NNX11AI49G, under sub-award 00001944. PK acknowledges support from the NSF award PHY-0847611. Some calculations were performed on the Syracuse University Gravitation and Relativity cluster, which is supported by NSF awards PHY-1040231 and PHY-1104371 and Syracuse University ITS.

\clearpage
\appendix 

\section{Finite mass-ratio corrections in the EPC model}
\label{Apen_PN}

\allowdisplaybreaks[4]

In this Appendix, we provide the PN corrections introduced to the EPC model. Using the convention

\begin{equation}
\bar{\Psi}_{2}(f) = 2\pi f t_c - \phi_c - \frac{\pi}{4} + \frac{3}{128\,\eta\, (\pi M f)^{5/3}}\sum_{n=0}^{7} \alpha_{n} (\pi M f)^{n/3}\,,
\end{equation}

\noindent and \(x=\left(\pi\,M\,f\right)^{1/3}\),  the \(\alpha_n\) coefficients are given by:

\begin{widetext}
\begin{eqnarray}
\alpha_0&=&1+ \frac{6505217202575  e_0^8}{277250217984 \chi
   ^{19/9}}-\frac{34527299926564117885  e_0^8}{443982214741303296 \chi
   ^{38/9}}+\frac{70854027727533517312175  e_0^8}{649101997951785418752
   \chi ^{19/3}}\\\nonumber&-&\frac{1480771978371648379295884378817
    e_0^8}{30404123876335041176758861824 \chi ^{76/9}}+\frac{1326481225
    e_0^6}{101334144 \chi ^{19/9}}-\frac{4671472830986095
    e_0^6}{243411301941504 \chi ^{38/9}}+\frac{21322307471421461725
    e_0^6}{2135203940630873088 \chi ^{19/3}}\\\nonumber&+&\frac{2608555
    e_0^4}{444448 \chi ^{19/9}}-\frac{1405799828765
    e_0^4}{533796714784 \chi ^{38/9}}+\frac{2355  e_0^2}{1462 \chi
   ^{19/9}}\,,\\
   \alpha_2&=&\frac{71557389228325  e_0^8 \eta }{831750653952 \chi
   ^{19/9}}-\frac{454638749586390872855  e_0^8 \eta }{1331946644223909888
   \chi ^{38/9}}+\frac{942278545508138530341925  e_0^8 \eta
   }{1947305993855356256256 \chi ^{19/3}}\\\nonumber&-&\frac{19670327207880872553538752792427
    e_0^8 \eta }{91212371629005123530276585472 \chi
   ^{76/9}}+\frac{690482340216175  e_0^8}{9981007847424 \chi
   ^{19/9}}-\frac{30708780994789856230115  e_0^8}{111883518114808430592
   \chi ^{38/9}}\\\nonumber&+&\frac{63646632664776993458550025
    e_0^8}{163573703483849925525504 \chi
   ^{19/3}}-\frac{1328641192314135300661753938615751
    e_0^8}{7661839216836430376543233179648 \chi ^{76/9}}+\frac{14591293475
    e_0^6 \eta }{304002432 \chi ^{19/9}}\\\nonumber&-&\frac{61511689912720685
    e_0^6 \eta }{730233905824512 \chi ^{38/9}}+\frac{283562607736424474975
    e_0^6 \eta }{6405611821892619264 \chi ^{19/3}}+\frac{985575550175
    e_0^6}{25536204288 \chi ^{19/9}}-\frac{4154835055013769905
    e_0^6}{61339648089259008 \chi ^{38/9}}\\\nonumber&+&\frac{19153365231651216809675
    e_0^6}{538071393038980018176 \chi ^{19/3}}+\frac{28694105  e_0^4
   \eta }{1333344 \chi ^{19/9}}-\frac{18510890735095  e_0^4 \eta
   }{1601390144352 \chi ^{38/9}}+\frac{1938156365  e_0^4}{112000896 \chi
   ^{19/9}}-\frac{1250326528743235  e_0^4}{134516772125568 \chi
   ^{38/9}}\\\nonumber&+&\frac{8635  e_0^2 \eta }{1462 \chi ^{19/9}}+\frac{583255
    e_0^2}{122808 \chi ^{19/9}}+\frac{55 \eta }{9}+\frac{3715}{756}\,,
\\  
 \alpha_3&=&-\frac{1301043440515 \pi   e_0^8}{8664069312 \chi
   ^{19/9}}+\frac{8946508632426975653 \pi   e_0^8}{13874444210665728 \chi
   ^{38/9}}-\frac{18731723667094131800179 \pi   e_0^8}{20284437435993294336
   \chi ^{19/3}}\\\nonumber&+&\frac{115205353044886385251076880001 \pi 
    e_0^8}{279449667981020599051092480 \chi ^{76/9}}-\frac{265296245 \pi 
    e_0^6}{3166692 \chi ^{19/9}}+\frac{1210444259975591 \pi 
    e_0^6}{7606603185672 \chi ^{38/9}}-\frac{5636991774629591273 \pi 
    e_0^6}{66725123144714784 \chi ^{19/3}}\\\nonumber&-&\frac{521711 \pi 
    e_0^4}{13889 \chi ^{19/9}}+\frac{364262491717 \pi 
    e_0^4}{16681147337 \chi ^{38/9}}-\frac{7536 \pi   e_0^2}{731 \chi
   ^{19/9}}-16 \pi\,, \\
   \alpha_4& =& \frac{4013719013988775  e_0^8 \eta ^2}{19962015694848 \chi
   ^{19/9}}-\frac{29698857489819606274325  e_0^8 \eta
   ^2}{31966719461373837312 \chi ^{38/9}}+\frac{62965437935101698296500615
    e_0^8 \eta ^2}{46735343852528550150144 \chi
   ^{19/3}}\\\nonumber&-&\frac{1320599685760317911145711844818409  e_0^8 \eta
   ^2}{2189096919096122964726638051328 \chi ^{76/9}}+\frac{5045260598968525
    e_0^8 \eta }{19962015694848 \chi
   ^{19/9}}-\frac{261321065335868140135025  e_0^8 \eta
   }{223767036229616861184 \chi ^{38/9}}\\\nonumber&+&\frac{554034623257159027636469755
    e_0^8 \eta }{327147406967699851051008 \chi
   ^{19/3}}-\frac{11619993021057967487212430478961333  e_0^8 \eta
   }{15323678433672860753086466359296 \chi ^{76/9}}+\frac{2842476030950240425
    e_0^8}{20121711820406784 \chi
   ^{19/9}}\\\nonumber&-&\frac{147227055972380882628700925
    e_0^8}{225557172519453796073472 \chi
   ^{38/9}}+\frac{312140494238689330362483671935
    e_0^8}{329764586223441449859416064 \chi
   ^{19/3}}\\\nonumber&-&\frac{6546649274949058130045037091614679921
    e_0^8}{15446267861142243639111158090170368 \chi
   ^{76/9}}+\frac{818438915825  e_0^6 \eta ^2}{7296058368 \chi
   ^{19/9}}-\frac{4018194477126790775  e_0^6 \eta ^2}{17525613739788288
   \chi ^{38/9}}\\\nonumber&+&\frac{18948371331658651308005  e_0^6 \eta
   ^2}{153734683725422862336 \chi ^{19/3}}+\frac{7201466570525  e_0^6 \eta
   }{51072408576 \chi ^{19/9}}-\frac{35356203916242053675  e_0^6 \eta
   }{122679296178518016 \chi ^{38/9}}+\frac{166727241425566965885185  e_0^6
   \eta }{1076142786077960036352 \chi ^{19/3}}\\\nonumber&+&\frac{4057272307914425
    e_0^6}{51480987844608 \chi ^{19/9}}-\frac{19919518567158561620975
    e_0^6}{123660730547946160128 \chi
   ^{38/9}}+\frac{93933341630661850846368845
    e_0^6}{1084751928366583716642816 \chi ^{19/3}}+\frac{1609478435
    e_0^4 \eta ^2}{32000256 \chi ^{19/9}}\\\nonumber&-&\frac{1209206884479925
    e_0^4 \eta ^2}{38433363464448 \chi ^{38/9}}+\frac{14161845095
    e_0^4 \eta }{224001792 \chi ^{19/9}}-\frac{10639844693422225
    e_0^4 \eta }{269033544251136 \chi ^{38/9}}+\frac{7978716747515
    e_0^4}{225793806336 \chi ^{19/9}}\\\nonumber&-&\frac{5994438328967367325
    e_0^4}{271185812605145088 \chi ^{38/9}}+\frac{484345  e_0^2 \eta
   ^2}{35088 \chi ^{19/9}}+\frac{4261765  e_0^2 \eta }{245616 \chi
   ^{19/9}}+\frac{2401058305  e_0^2}{247580928 \chi ^{19/9}}+\frac{3085
   \eta ^2}{72}+\frac{27145 \eta }{504}+\frac{15293365}{508032}\\
   \alpha_5&=&-\frac{84567823633475 \pi   e_0^8 \eta }{831750653952 \chi
   ^{19/9}}+\frac{669968502482700007405 \pi   e_0^8 \eta
   }{1331946644223909888 \chi ^{38/9}}-\frac{1440902965169982699005075 \pi 
    e_0^8 \eta }{1947305993855356256256 \chi
   ^{19/3}}\\\nonumber&+&\frac{2334216112662079584736091244017 \pi   e_0^8 \eta
   }{7016336279154240271559737344 \chi ^{76/9}}+\frac{7182689108386025 \pi 
    e_0^8}{9981007847424 \chi ^{19/9}}-\frac{56903148963613058870695 \pi 
    e_0^8}{15983359730686918656 \chi
   ^{38/9}}\\\nonumber&+&\frac{122381747448338420666046425 \pi 
    e_0^8}{23367671926264275075072 \chi
   ^{19/3}}-\frac{18041156334765213110425249225007393 \pi 
    e_0^8}{7661839216836430376543233179648 \chi ^{76/9}}-\frac{17244255925
   \pi   e_0^6 \eta }{304002432 \chi ^{19/9}}\\\nonumber&+&\frac{90645363628809535 \pi 
    e_0^6 \eta }{730233905824512 \chi ^{38/9}}-\frac{433615096349678814025
   \pi   e_0^6 \eta }{6405611821892619264 \chi
   ^{19/3}}+\frac{10252373388025 \pi   e_0^6}{25536204288 \chi
   ^{19/9}}-\frac{7698879291066691165 \pi   e_0^6}{8762806869894144 \chi
   ^{38/9}}\\\nonumber&+&\frac{36828693183369973116475 \pi   e_0^6}{76867341862711431168
   \chi ^{19/3}}-\frac{33911215 \pi   e_0^4 \eta }{1333344 \chi
   ^{19/9}}+\frac{27278171420045 \pi   e_0^4 \eta }{1601390144352 \chi
   ^{38/9}}+\frac{20161521595 \pi   e_0^4}{112000896 \chi
   ^{19/9}}-\frac{2316846009950855 \pi   e_0^4}{19216681732224 \chi
   ^{38/9}}\\\nonumber&-&\frac{10205 \pi   e_0^2 \eta }{1462 \chi ^{19/9}}+\frac{6067265
   \pi   e_0^2}{122808 \chi ^{19/9}}-\frac{65 \pi  \eta }{9}-\frac{65}{3}
   \pi  \eta  \log \left(\sqrt{6} x\right)+\frac{38645}{252} \pi  \log
   \left(\sqrt{6} x\right)+\frac{38645 \pi }{756}\,,\\
   \alpha_6&=&\frac{19887378305015 \log (4 x)  e_0^8}{12996103968 \chi
   ^{19/9}}-\frac{1175668628910900667003 \log (4 x)
    e_0^8}{145681664211990144 \chi ^{38/9}}+\frac{2568467130783058548836609
   \log (4 x)  e_0^8}{212986593077929590528 \chi
   ^{19/3}}\\\nonumber&-&\frac{271831420459973064883705332651019 \log (4 x)
    e_0^8}{49881765734612176930620007680 \chi
   ^{76/9}}+\frac{166305877783829875 \eta ^3  e_0^8}{359316282507264 \chi
   ^{19/9}}-\frac{98950858868368325 \eta ^2  e_0^8}{479088376676352 \chi
   ^{19/9}}\\\nonumber&+&\frac{19887378305015 \gamma   e_0^8}{12996103968 \chi
   ^{19/9}}-\frac{2933852958361325 \pi ^2 \eta   e_0^8}{3327002615808 \chi
   ^{19/9}}+\frac{2925073821111304814575 \eta   e_0^8}{120730270922440704
   \chi ^{19/9}}+\frac{6505217202575 \pi ^2  e_0^8}{6498051984 \chi
   ^{19/9}}\\\nonumber&-&\frac{487482116356903781533459  e_0^8}{37184923444111736832
   \chi ^{19/9}}-\frac{1404484509257344651959425 \eta ^3
    e_0^8}{575400950304729071616 \chi
   ^{38/9}}+\frac{15767144696141518291115 \eta ^2
    e_0^8}{14475495605150416896 \chi ^{38/9}}\\\nonumber&-&\frac{1175668628910900667003
   \gamma   e_0^8}{145681664211990144 \chi
   ^{38/9}}+\frac{24776941665365243028895 \pi ^2 \eta 
    e_0^8}{5327786576895639552 \chi
   ^{38/9}}-\frac{172919601366556448542601792915 \eta 
    e_0^8}{1353343035116722776440832 \chi
   ^{38/9}}\\\nonumber&-&\frac{54937786397705638645 \pi ^2  e_0^8}{10405833157999296
   \chi ^{38/9}}+\frac{145132197850314788879078394300739
    e_0^8}{2084148274079753075718881280 \chi
   ^{38/9}}+\frac{3068358046657424850514388275 \eta ^3
    e_0^8}{841236189345513902702592 \chi
   ^{19/3}}\\\nonumber&-&\frac{34446264791342888014771345 \eta ^2
    e_0^8}{21163174574729909501952 \chi
   ^{19/3}}+\frac{2568467130783058548836609 \gamma 
    e_0^8}{212986593077929590528 \chi
   ^{19/3}}-\frac{54129844672110252594640685 \pi ^2 \eta 
    e_0^8}{7789223975421425025024 \chi
   ^{19/3}}\\\nonumber&+&\frac{377775081733314667944614526636745 \eta 
    e_0^8}{1978587517340648699156496384 \chi
   ^{19/3}}+\frac{120021828541264418169935 \pi ^2
    e_0^8}{15213328076994970752 \chi
   ^{19/3}}
   \\\nonumber&-&\frac{317665040175671089647133596190236377
    e_0^8}{3047024776704598996701004431360 \chi
   ^{19/3}}-\frac{64947385645413190689270344198348605 \eta ^3
    e_0^8}{39403744543730213365079484923904 \chi
   ^{76/9}}\\\nonumber&+&\frac{729117922168338968426387200662079 \eta ^2
    e_0^8}{991289170911451908555458740224 \chi
   ^{76/9}}-\frac{271831420459973064883705332651019 \gamma 
    e_0^8}{49881765734612176930620007680 \chi
   ^{76/9}}\\\nonumber&+&\frac{1145756734835961236098608458183267 \pi ^2 \eta 
    e_0^8}{364849486516020494121106341888 \chi
   ^{76/9}}-\frac{7996297546594766343088603842357850964359 \eta 
    e_0^8}{92677607166853461834666948541022208 \chi
   ^{76/9}}\\\nonumber&-&\frac{2540480565046477241903788155617 \pi ^2
    e_0^8}{712596653351602527580285824 \chi
   ^{76/9}}+\frac{33641288006290009112960754500321337034087507
    e_0^8}{713617575184771656126935503765871001600 \chi
   ^{76/9}}\\\nonumber&+&\frac{28386698215 \log (4 x)  e_0^6}{33250266 \chi
   ^{19/9}}-\frac{159065553051674041 \log (4 x)  e_0^6}{79869333449556 \chi
   ^{38/9}}+\frac{772936241583827429683 \log (4 x)
    e_0^6}{700613793019505232 \chi ^{19/3}}+\frac{33911492517125 \eta ^3
    e_0^6}{131329050624 \chi ^{19/9}}\\\nonumber&-&\frac{20177105913475 \eta ^2
    e_0^6}{175105400832 \chi ^{19/9}}+\frac{28386698215 \gamma 
    e_0^6}{33250266 \chi ^{19/9}}-\frac{598243032475 \pi ^2 \eta 
    e_0^6}{1216009728 \chi ^{19/9}}+\frac{4175170127655540575 \eta 
    e_0^6}{308885927067648 \chi ^{19/9}}+\frac{1326481225 \pi ^2
    e_0^6}{2375019 \chi ^{19/9}}\\\nonumber&-&\frac{695818599616190365379
    e_0^6}{95136865536835584 \chi ^{19/9}}-\frac{190023872138600320475 \eta
   ^3  e_0^6}{315461047316189184 \chi ^{38/9}}+\frac{2133262323636936905
   \eta ^2  e_0^6}{7936126976507904 \chi ^{38/9}}-\frac{159065553051674041
   \gamma   e_0^6}{79869333449556 \chi ^{38/9}}\\\nonumber&+&\frac{3352269365715186565
   \pi ^2 \eta   e_0^6}{2920935623298048 \chi
   ^{38/9}}-\frac{23395667238587898146461505 \eta 
    e_0^6}{741964383287676960768 \chi ^{38/9}}-\frac{7432969768769815 \pi
   ^2  e_0^6}{5704952389254 \chi
   ^{38/9}}\\\nonumber&+&\frac{19636088561834767348202868433
    e_0^6}{1142625150263022519582720 \chi
   ^{38/9}}+\frac{923369860564978889712425 \eta ^3
    e_0^6}{2767224307057611522048 \chi
   ^{19/3}}-\frac{10366014081054134220515 \eta ^2
    e_0^6}{69615705837927333888 \chi ^{19/3}}\\\nonumber&+&\frac{772936241583827429683
   \gamma   e_0^6}{700613793019505232 \chi
   ^{19/3}}-\frac{16289450698799353775095 \pi ^2 \eta 
    e_0^6}{25622447287570477056 \chi
   ^{19/3}}+\frac{113684947858355301819023330315 \eta 
    e_0^6}{6508511570199502299856896 \chi
   ^{19/3}}\\\nonumber&+&\frac{36118515961861094845 \pi ^2  e_0^6}{50043842358536088
   \chi ^{19/3}}-\frac{95595859216271769379215647363899
    e_0^6}{10023107818107233541779619840 \chi ^{19/3}}+\frac{111646154 \log
   (4 x)  e_0^4}{291669 \chi ^{19/9}}\\\nonumber&-&\frac{95736113783734 \log (4 x)
    e_0^4}{350304094077 \chi ^{38/9}}+\frac{66687708575 \eta ^3
    e_0^4}{576004608 \chi ^{19/9}}-\frac{39678730105 \eta ^2
    e_0^4}{768006144 \chi ^{19/9}}+\frac{111646154 \gamma 
    e_0^4}{291669 \chi ^{19/9}}-\frac{1176458305 \pi ^2 \eta 
    e_0^4}{5333376 \chi ^{19/9}}\\\nonumber&+&\frac{8210565447201485 \eta 
    e_0^4}{1354762838016 \chi ^{19/9}}+\frac{10434220 \pi ^2
    e_0^4}{41667 \chi ^{19/9}}-\frac{6841714201879530781
    e_0^4}{2086334770544640 \chi ^{19/9}}-\frac{57184433385073825 \eta ^3
    e_0^4}{691800542360064 \chi ^{38/9}}\\\nonumber&+&\frac{641968800372235 \eta ^2
    e_0^4}{17403787229184 \chi ^{38/9}}-\frac{95736113783734 \gamma 
    e_0^4}{350304094077 \chi ^{38/9}}+\frac{1008808114870655 \pi ^2 \eta 
    e_0^4}{6405560577408 \chi ^{38/9}}-\frac{7040525801561209192435 \eta 
    e_0^4}{1627114875630870528 \chi ^{38/9}}\\\nonumber&-&\frac{8947300353620 \pi ^2
    e_0^4}{50043442011 \chi ^{38/9}}+\frac{5909144917795596553777571
    e_0^4}{2505756908471540613120 \chi ^{38/9}}+\frac{537568 \log (4 x)
    e_0^2}{5117 \chi ^{19/9}}+\frac{20068525 \eta ^3  e_0^2}{631584
   \chi ^{19/9}}-\frac{11940635 \eta ^2 e_0^2}{842112 \chi
   ^{19/9}}\\\nonumber&+&\frac{537568 \gamma   e_0^2}{5117 \chi ^{19/9}}-\frac{354035
   \pi ^2 \eta   e_0^2}{5848 \chi ^{19/9}}+\frac{2470829204695 \eta 
    e_0^2}{1485485568 \chi ^{19/9}}+\frac{50240 \pi ^2  e_0^2}{731
   \chi ^{19/9}}-\frac{2058896840770247  e_0^2}{2287647774720 \chi
   ^{19/9}}-\frac{127825 \eta ^3}{1296}+\frac{76055 \eta ^2}{1728}\\\nonumber&-&\frac{6848
   \gamma }{21}+\frac{2255 \pi ^2 \eta }{12}-\frac{15737765635 \eta
   }{3048192}-\frac{6848}{21} \log (4 x)-\frac{640 \pi
   ^2}{3}+\frac{11583231236531}{4694215680}\,,\\
   \alpha_7&=& \frac{13762251650419025 \pi   e_0^8 \eta ^2}{14971511771136 \chi
   ^{19/9}}-\frac{863950161122145266182945 \pi   e_0^8 \eta
   ^2}{167825277172212645888 \chi ^{38/9}}+\frac{1919251873874486617026907175
   \pi   e_0^8 \eta ^2}{245360555225774888288256 \chi
   ^{19/3}}\\\nonumber&-&\frac{40866306288716811342278348158188713 \pi   e_0^8 \eta
   ^2}{11492758825254645564814849769472 \chi ^{76/9}}-\frac{70352065412362175
   \pi   e_0^8 \eta }{29943023542272 \chi
   ^{19/9}}+\frac{4416477753219647720024815 \pi   e_0^8 \eta
   }{335650554344425291776 \chi ^{38/9}}\\\nonumber&-&\frac{9811136782221639568423793225 \pi 
    e_0^8 \eta }{490721110451549776576512 \chi
   ^{19/3}}+\frac{208906879936169138297285285341303271 \pi   e_0^8 \eta
   }{22985517650509291129629699538944 \chi ^{76/9}}-\frac{14329446184895255375
   \pi   e_0^8}{5030427955101696 \chi
   ^{19/9}}\\\nonumber&+&\frac{899556820693249630490850175 \pi 
    e_0^8}{56389293129863449018368 \chi
   ^{38/9}}-\frac{1998351515473594240053655597625 \pi 
    e_0^8}{82441146555860362464854016 \chi
   ^{19/3}}\\\nonumber&+&\frac{42550561609719173085177224221604751095 \pi 
    e_0^8}{3861566965285560909777789522542592 \chi
   ^{76/9}}+\frac{19643860461025 \pi   e_0^6 \eta ^2}{38304306432 \chi
   ^{19/9}}-\frac{116890683997652041915 \pi   e_0^6 \eta
   ^2}{92009472133888512 \chi ^{38/9}}\\\nonumber&+&\frac{577566016814470844726725 \pi 
    e_0^6 \eta ^2}{807107089558470027264 \chi
   ^{19/3}}-\frac{100418608176175 \pi   e_0^6 \eta }{76608612864 \chi
   ^{19/9}}+\frac{597540377518688130805 \pi   e_0^6 \eta
   }{184018944267777024 \chi ^{38/9}}\\\nonumber&-&\frac{2952493765339042903528075 \pi 
    e_0^6 \eta }{1614214179116940054528 \chi
   ^{19/3}}-\frac{20453458379485375 \pi   e_0^6}{12870246961152 \chi
   ^{19/9}}+\frac{121708191973727871410725 \pi 
    e_0^6}{30915182636986540032 \chi
   ^{38/9}}\\\nonumber&-&\frac{601369700714292579011030875 \pi 
    e_0^6}{271187982091645929160704 \chi ^{19/3}}+\frac{38630090995 \pi 
    e_0^4 \eta ^2}{168001344 \chi ^{19/9}}-\frac{35176251579191105 \pi 
    e_0^4 \eta ^2}{201775158188352 \chi ^{38/9}}-\frac{197475439165 \pi 
    e_0^4 \eta }{336002688 \chi ^{19/9}}
    \\\nonumber&+&\frac{179819553872611535 \pi 
    e_0^4 \eta }{403550316376704 \chi ^{38/9}}-\frac{40222183410925 \pi 
    e_0^4}{56448451584 \chi ^{19/9}}+\frac{36625998186496500575 \pi 
    e_0^4}{67796453151286272 \chi ^{38/9}}+\frac{11625065 \pi   e_0^2
   \eta ^2}{184212 \chi ^{19/9}}
   \\\nonumber&-&\frac{59426855 \pi   e_0^2 \eta }{368424
   \chi ^{19/9}}-\frac{12104177975 \pi   e_0^2}{61895232 \chi
   ^{19/9}}-\frac{74045 \pi  \eta ^2}{756}+\frac{378515 \pi  \eta
   }{1512}+\frac{77096675 \pi }{254016}\,,
   \end{eqnarray}
\end{widetext}
and recall $\chi = f/f_{0}$ with $f_{0}$ the GW frequency of the $\ell=2$ harmonic at which the
eccentricity equals $e_{0}$.

\bibliography{references}

\begin{thebibliography}{114}%
\makeatletter
\providecommand \@ifxundefined [1]{%
 \@ifx{#1\undefined}
}%
\providecommand \@ifnum [1]{%
 \ifnum #1\expandafter \@firstoftwo
 \else \expandafter \@secondoftwo
 \fi
}%
\providecommand \@ifx [1]{%
 \ifx #1\expandafter \@firstoftwo
 \else \expandafter \@secondoftwo
 \fi
}%
\providecommand \natexlab [1]{#1}%
\providecommand \enquote  [1]{``#1''}%
\providecommand \bibnamefont  [1]{#1}%
\providecommand \bibfnamefont [1]{#1}%
\providecommand \citenamefont [1]{#1}%
\providecommand \href@noop [0]{\@secondoftwo}%
\providecommand \href [0]{\begingroup \@sanitize@url \@href}%
\providecommand \@href[1]{\@@startlink{#1}\@@href}%
\providecommand \@@href[1]{\endgroup#1\@@endlink}%
\providecommand \@sanitize@url [0]{\catcode `\\12\catcode `\$12\catcode
  `\&12\catcode `\#12\catcode `\^12\catcode `\_12\catcode `\%12\relax}%
\providecommand \@@startlink[1]{}%
\providecommand \@@endlink[0]{}%
\providecommand \url  [0]{\begingroup\@sanitize@url \@url }%
\providecommand \@url [1]{\endgroup\@href {#1}{\urlprefix }}%
\providecommand \urlprefix  [0]{URL }%
\providecommand \Eprint [0]{\href }%
\providecommand \doibase [0]{http://dx.doi.org/}%
\providecommand \selectlanguage [0]{\@gobble}%
\providecommand \bibinfo  [0]{\@secondoftwo}%
\providecommand \bibfield  [0]{\@secondoftwo}%
\providecommand \translation [1]{[#1]}%
\providecommand \BibitemOpen [0]{}%
\providecommand \bibitemStop [0]{}%
\providecommand \bibitemNoStop [0]{.\EOS\space}%
\providecommand \EOS [0]{\spacefactor3000\relax}%
\providecommand \BibitemShut  [1]{\csname bibitem#1\endcsname}%
\let\auto@bib@innerbib\@empty
\bibitem [{\citenamefont {{Pols}}\ \emph {et~al.}(1998)\citenamefont {{Pols}},
  \citenamefont {{Schr{\"o}der}}, \citenamefont {{Hurley}}, \citenamefont
  {{Tout}},\ and\ \citenamefont {{Eggleton}}}]{Pols:1998}%
  \BibitemOpen
  \bibfield  {author} {\bibinfo {author} {\bibfnamefont {O.~R.}\ \bibnamefont
  {{Pols}}}, \bibinfo {author} {\bibfnamefont {K.-P.}\ \bibnamefont
  {{Schr{\"o}der}}}, \bibinfo {author} {\bibfnamefont {J.~R.}\ \bibnamefont
  {{Hurley}}}, \bibinfo {author} {\bibfnamefont {C.~A.}\ \bibnamefont
  {{Tout}}}, \ and\ \bibinfo {author} {\bibfnamefont {P.~P.}\ \bibnamefont
  {{Eggleton}}},\ }\href {\doibase 10.1046/j.1365-8711.1998.01658.x} {\bibfield
   {journal} {\bibinfo  {journal} {\mnras}\ }\textbf {\bibinfo {volume}
  {298}},\ \bibinfo {pages} {525} (\bibinfo {year} {1998})}\BibitemShut
  {NoStop}%
\bibitem [{\citenamefont {{Harry}}\ and\ \citenamefont {{LIGO Scientific
  Collaboration}}(2010)}]{aLIGO}%
  \BibitemOpen
  \bibfield  {author} {\bibinfo {author} {\bibfnamefont {G.~M.}\ \bibnamefont
  {{Harry}}}\ and\ \bibinfo {author} {\bibnamefont {{LIGO Scientific
  Collaboration}}},\ }\href {\doibase 10.1088/0264-9381/27/8/084006} {\bibfield
   {journal} {\bibinfo  {journal} {\CQG}\ }\textbf {\bibinfo {volume} {27}},\
  \bibinfo {pages} {084006} (\bibinfo {year} {2010})}\BibitemShut {NoStop}%
\bibitem [{\citenamefont {{Acernese}}\ \emph {et~al.}(2008)\citenamefont
  {{Acernese}} \emph {et~al.}}]{virgo}%
  \BibitemOpen
  \bibfield  {author} {\bibinfo {author} {\bibfnamefont {F.}~\bibnamefont
  {{Acernese}}} \emph {et~al.},\ }\href {\doibase
  10.1088/0264-9381/25/18/184001} {\bibfield  {journal} {\bibinfo  {journal}
  {\CQG}\ }\textbf {\bibinfo {volume} {25}},\ \bibinfo {pages} {184001}
  (\bibinfo {year} {2008})}\BibitemShut {NoStop}%
\bibitem [{\citenamefont {{Somiya}}\ and\ \citenamefont {{the KAGRA
  Collaboration}}(2012)}]{kagra}%
  \BibitemOpen
  \bibfield  {author} {\bibinfo {author} {\bibfnamefont {K.}~\bibnamefont
  {{Somiya}}}\ and\ \bibinfo {author} {\bibnamefont {{the KAGRA
  Collaboration}}},\ }\href@noop {} {\bibfield  {journal} {\bibinfo  {journal}
  {\CQG}\ }\textbf {\bibinfo {volume} {29}},\ \bibinfo {pages} {124007}
  (\bibinfo {year} {2012})}\BibitemShut {NoStop}%
\bibitem [{\citenamefont {{Freise}}\ \emph {et~al.}(2009)\citenamefont
  {{Freise}}, \citenamefont {{Chelkowski}}, \citenamefont {{Hild}},
  \citenamefont {{Del Pozzo}}, \citenamefont {{Perreca}},\ and\ \citenamefont
  {{Vecchio}}}]{Freise:2009}%
  \BibitemOpen
  \bibfield  {author} {\bibinfo {author} {\bibfnamefont {A.}~\bibnamefont
  {{Freise}}}, \bibinfo {author} {\bibfnamefont {S.}~\bibnamefont
  {{Chelkowski}}}, \bibinfo {author} {\bibfnamefont {S.}~\bibnamefont
  {{Hild}}}, \bibinfo {author} {\bibfnamefont {W.}~\bibnamefont {{Del Pozzo}}},
  \bibinfo {author} {\bibfnamefont {A.}~\bibnamefont {{Perreca}}}, \ and\
  \bibinfo {author} {\bibfnamefont {A.}~\bibnamefont {{Vecchio}}},\ }\href
  {\doibase 10.1088/0264-9381/26/8/085012} {\bibfield  {journal} {\bibinfo
  {journal} {\CQG}\ }\textbf {\bibinfo {volume} {26}},\ \bibinfo {pages}
  {085012} (\bibinfo {year} {2009})},\ \Eprint {http://arxiv.org/abs/0804.1036}
  {arXiv:0804.1036} \BibitemShut {NoStop}%
\bibitem [{\citenamefont {Peters}(1964)}]{Peters:1964}%
  \BibitemOpen
  \bibfield  {author} {\bibinfo {author} {\bibfnamefont {P.~C.}\ \bibnamefont
  {Peters}},\ }\href {\doibase 10.1103/PhysRev.136.B1224} {\bibfield  {journal}
  {\bibinfo  {journal} {Phys. Rev.}\ }\textbf {\bibinfo {volume} {136}},\
  \bibinfo {pages} {B1224} (\bibinfo {year} {1964})}\BibitemShut {NoStop}%
\bibitem [{\citenamefont {{Antonini}}\ and\ \citenamefont
  {{Perets}}(2012)}]{antonini}%
  \BibitemOpen
  \bibfield  {author} {\bibinfo {author} {\bibfnamefont {F.}~\bibnamefont
  {{Antonini}}}\ and\ \bibinfo {author} {\bibfnamefont {H.~B.}\ \bibnamefont
  {{Perets}}},\ }\href {\doibase 10.1088/0004-637X/757/1/27} {\bibfield
  {journal} {\bibinfo  {journal} {\apj}\ }\textbf {\bibinfo {volume} {757}},\
  \bibinfo {eid} {27} (\bibinfo {year} {2012})}\BibitemShut {NoStop}%
\bibitem [{\citenamefont {{Samsing}}\ \emph {et~al.}(2014)\citenamefont
  {{Samsing}}, \citenamefont {{MacLeod}},\ and\ \citenamefont
  {{Ramirez-Ruiz}}}]{Samsing:2014}%
  \BibitemOpen
  \bibfield  {author} {\bibinfo {author} {\bibfnamefont {J.}~\bibnamefont
  {{Samsing}}}, \bibinfo {author} {\bibfnamefont {M.}~\bibnamefont
  {{MacLeod}}}, \ and\ \bibinfo {author} {\bibfnamefont {E.}~\bibnamefont
  {{Ramirez-Ruiz}}},\ }\href {\doibase 10.1088/0004-637X/784/1/71} {\bibfield
  {journal} {\bibinfo  {journal} {\apj}\ }\textbf {\bibinfo {volume} {784}},\
  \bibinfo {eid} {71} (\bibinfo {year} {2014})},\ \Eprint
  {http://arxiv.org/abs/1308.2964} {arXiv:1308.2964 [astro-ph.HE]} \BibitemShut
  {NoStop}%
\bibitem [{\citenamefont {{Thompson}}(2011)}]{Thompson:2011}%
  \BibitemOpen
  \bibfield  {author} {\bibinfo {author} {\bibfnamefont {T.~A.}\ \bibnamefont
  {{Thompson}}},\ }\href {\doibase 10.1088/0004-637X/741/2/82} {\bibfield
  {journal} {\bibinfo  {journal} {\apj}\ }\textbf {\bibinfo {volume} {741}},\
  \bibinfo {eid} {82} (\bibinfo {year} {2011})},\ \Eprint
  {http://arxiv.org/abs/1011.4322} {arXiv:1011.4322 [astro-ph.HE]} \BibitemShut
  {NoStop}%
\bibitem [{\citenamefont {{East}}\ \emph {et~al.}(2013)\citenamefont {{East}},
  \citenamefont {{McWilliams}}, \citenamefont {{Levin}},\ and\ \citenamefont
  {{Pretorius}}}]{East:2013}%
  \BibitemOpen
  \bibfield  {author} {\bibinfo {author} {\bibfnamefont {W.~E.}\ \bibnamefont
  {{East}}}, \bibinfo {author} {\bibfnamefont {S.~T.}\ \bibnamefont
  {{McWilliams}}}, \bibinfo {author} {\bibfnamefont {J.}~\bibnamefont
  {{Levin}}}, \ and\ \bibinfo {author} {\bibfnamefont {F.}~\bibnamefont
  {{Pretorius}}},\ }\href {\doibase 10.1103/PhysRevD.87.043004} {\bibfield
  {journal} {\bibinfo  {journal} {\prd}\ }\textbf {\bibinfo {volume} {87}},\
  \bibinfo {eid} {043004} (\bibinfo {year} {2013})},\ \Eprint
  {http://arxiv.org/abs/1212.0837} {arXiv:1212.0837 [gr-qc]} \BibitemShut
  {NoStop}%
\bibitem [{\citenamefont {{O'Leary}}\ \emph {et~al.}(2009)\citenamefont
  {{O'Leary}}, \citenamefont {{Kocsis}},\ and\ \citenamefont
  {{Loeb}}}]{Leary:2009}%
  \BibitemOpen
  \bibfield  {author} {\bibinfo {author} {\bibfnamefont {R.~M.}\ \bibnamefont
  {{O'Leary}}}, \bibinfo {author} {\bibfnamefont {B.}~\bibnamefont {{Kocsis}}},
  \ and\ \bibinfo {author} {\bibfnamefont {A.}~\bibnamefont {{Loeb}}},\ }\href
  {\doibase 10.1111/j.1365-2966.2009.14653.x} {\bibfield  {journal} {\bibinfo
  {journal} {\mnras}\ }\textbf {\bibinfo {volume} {395}},\ \bibinfo {pages}
  {2127} (\bibinfo {year} {2009})},\ \Eprint {http://arxiv.org/abs/0807.2638}
  {arXiv:0807.2638} \BibitemShut {NoStop}%
\bibitem [{\citenamefont {{Kocsis}}\ and\ \citenamefont
  {{Levin}}(2012)}]{Kocsis:2012}%
  \BibitemOpen
  \bibfield  {author} {\bibinfo {author} {\bibfnamefont {B.}~\bibnamefont
  {{Kocsis}}}\ and\ \bibinfo {author} {\bibfnamefont {J.}~\bibnamefont
  {{Levin}}},\ }\href {\doibase 10.1103/PhysRevD.85.123005} {\bibfield
  {journal} {\bibinfo  {journal} {\prd}\ }\textbf {\bibinfo {volume} {85}},\
  \bibinfo {eid} {123005} (\bibinfo {year} {2012})},\ \Eprint
  {http://arxiv.org/abs/1109.4170} {arXiv:1109.4170 [astro-ph.CO]} \BibitemShut
  {NoStop}%
\bibitem [{\citenamefont {{Tsang}}(2013)}]{Tsang:2013}%
  \BibitemOpen
  \bibfield  {author} {\bibinfo {author} {\bibfnamefont {D.}~\bibnamefont
  {{Tsang}}},\ }\href {\doibase 10.1088/0004-637X/777/2/103} {\bibfield
  {journal} {\bibinfo  {journal} {\apj}\ }\textbf {\bibinfo {volume} {777}},\
  \bibinfo {eid} {103} (\bibinfo {year} {2013})},\ \Eprint
  {http://arxiv.org/abs/1307.3554} {arXiv:1307.3554 [astro-ph.HE]} \BibitemShut
  {NoStop}%
\bibitem [{\citenamefont {{Wen}}(2003)}]{Wen:2003}%
  \BibitemOpen
  \bibfield  {author} {\bibinfo {author} {\bibfnamefont {L.}~\bibnamefont
  {{Wen}}},\ }\href {\doibase 10.1086/378794} {\bibfield  {journal} {\bibinfo
  {journal} {\apj}\ }\textbf {\bibinfo {volume} {598}},\ \bibinfo {pages} {419}
  (\bibinfo {year} {2003})},\ \Eprint
  {http://arxiv.org/abs/arXiv:astro-ph/0211492} {arXiv:astro-ph/0211492}
  \BibitemShut {NoStop}%
\bibitem [{\citenamefont {{East}}\ \emph {et~al.}(2012)\citenamefont {{East}},
  \citenamefont {{Pretorius}},\ and\ \citenamefont {{Stephens}}}]{east:2012a}%
  \BibitemOpen
  \bibfield  {author} {\bibinfo {author} {\bibfnamefont {W.~E.}\ \bibnamefont
  {{East}}}, \bibinfo {author} {\bibfnamefont {F.}~\bibnamefont {{Pretorius}}},
  \ and\ \bibinfo {author} {\bibfnamefont {B.~C.}\ \bibnamefont {{Stephens}}},\
  }\href {\doibase 10.1103/PhysRevD.85.124009} {\bibfield  {journal} {\bibinfo
  {journal} {\prd}\ }\textbf {\bibinfo {volume} {85}},\ \bibinfo {eid} {124009}
  (\bibinfo {year} {2012})},\ \Eprint {http://arxiv.org/abs/1111.3055}
  {arXiv:1111.3055 [astro-ph.HE]} \BibitemShut {NoStop}%
\bibitem [{\citenamefont {{Pooley}}\ \emph {et~al.}(2003)\citenamefont
  {{Pooley}}, \citenamefont {{Lewin}}, \citenamefont {{Anderson}},
  \citenamefont {{Baumgardt}}, \citenamefont {{Filippenko}}, \citenamefont
  {{Gaensler}}, \citenamefont {{Homer}}, \citenamefont {{Hut}}, \citenamefont
  {{Kaspi}}, \citenamefont {{Makino}}, \citenamefont {{Margon}}, \citenamefont
  {{McMillan}}, \citenamefont {{Portegies Zwart}}, \citenamefont {{van der
  Klis}},\ and\ \citenamefont {{Verbunt}}}]{Pooley:2003}%
  \BibitemOpen
  \bibfield  {author} {\bibinfo {author} {\bibfnamefont {D.}~\bibnamefont
  {{Pooley}}}, \bibinfo {author} {\bibfnamefont {W.~H.~G.}\ \bibnamefont
  {{Lewin}}}, \bibinfo {author} {\bibfnamefont {S.~F.}\ \bibnamefont
  {{Anderson}}}, \bibinfo {author} {\bibfnamefont {H.}~\bibnamefont
  {{Baumgardt}}}, \bibinfo {author} {\bibfnamefont {A.~V.}\ \bibnamefont
  {{Filippenko}}}, \bibinfo {author} {\bibfnamefont {B.~M.}\ \bibnamefont
  {{Gaensler}}}, \bibinfo {author} {\bibfnamefont {L.}~\bibnamefont {{Homer}}},
  \bibinfo {author} {\bibfnamefont {P.}~\bibnamefont {{Hut}}}, \bibinfo
  {author} {\bibfnamefont {V.~M.}\ \bibnamefont {{Kaspi}}}, \bibinfo {author}
  {\bibfnamefont {J.}~\bibnamefont {{Makino}}}, \bibinfo {author}
  {\bibfnamefont {B.}~\bibnamefont {{Margon}}}, \bibinfo {author}
  {\bibfnamefont {S.}~\bibnamefont {{McMillan}}}, \bibinfo {author}
  {\bibfnamefont {S.}~\bibnamefont {{Portegies Zwart}}}, \bibinfo {author}
  {\bibfnamefont {M.}~\bibnamefont {{van der Klis}}}, \ and\ \bibinfo {author}
  {\bibfnamefont {F.}~\bibnamefont {{Verbunt}}},\ }\href {\doibase
  10.1086/377074} {\bibfield  {journal} {\bibinfo  {journal} {\apjl}\ }\textbf
  {\bibinfo {volume} {591}},\ \bibinfo {pages} {L131} (\bibinfo {year}
  {2003})},\ \Eprint {http://arxiv.org/abs/astro-ph/0305003} {astro-ph/0305003}
  \BibitemShut {NoStop}%
\bibitem [{\citenamefont {{Freitag}}\ \emph {et~al.}(2006)\citenamefont
  {{Freitag}}, \citenamefont {{Amaro-Seoane}},\ and\ \citenamefont
  {{Kalogera}}}]{Freitag:2006A}%
  \BibitemOpen
  \bibfield  {author} {\bibinfo {author} {\bibfnamefont {M.}~\bibnamefont
  {{Freitag}}}, \bibinfo {author} {\bibfnamefont {P.}~\bibnamefont
  {{Amaro-Seoane}}}, \ and\ \bibinfo {author} {\bibfnamefont {V.}~\bibnamefont
  {{Kalogera}}},\ }\href {\doibase 10.1086/506193} {\bibfield  {journal}
  {\bibinfo  {journal} {\apj}\ }\textbf {\bibinfo {volume} {649}},\ \bibinfo
  {pages} {91} (\bibinfo {year} {2006})},\ \Eprint
  {http://arxiv.org/abs/astro-ph/0603280} {astro-ph/0603280} \BibitemShut
  {NoStop}%
\bibitem [{\citenamefont {{Antonini}}\ and\ \citenamefont
  {{Merritt}}(2012)}]{Anton:2012}%
  \BibitemOpen
  \bibfield  {author} {\bibinfo {author} {\bibfnamefont {F.}~\bibnamefont
  {{Antonini}}}\ and\ \bibinfo {author} {\bibfnamefont {D.}~\bibnamefont
  {{Merritt}}},\ }\href {\doibase 10.1088/0004-637X/745/1/83} {\bibfield
  {journal} {\bibinfo  {journal} {\apj}\ }\textbf {\bibinfo {volume} {745}},\
  \bibinfo {eid} {83} (\bibinfo {year} {2012})},\ \Eprint
  {http://arxiv.org/abs/1108.1163} {arXiv:1108.1163 [astro-ph.GA]} \BibitemShut
  {NoStop}%
\bibitem [{\citenamefont {{Antonini}}(2014)}]{Antonini:2014}%
  \BibitemOpen
  \bibfield  {author} {\bibinfo {author} {\bibfnamefont {F.}~\bibnamefont
  {{Antonini}}},\ }\href@noop {} {\bibfield  {journal} {\bibinfo  {journal}
  {ArXiv e-prints}\ } (\bibinfo {year} {2014})},\ \Eprint
  {http://arxiv.org/abs/1402.4865} {arXiv:1402.4865 [astro-ph.GA]} \BibitemShut
  {NoStop}%
\bibitem [{\citenamefont {{Hopman}}\ and\ \citenamefont
  {{Alexander}}(2006)}]{Hopman:2006}%
  \BibitemOpen
  \bibfield  {author} {\bibinfo {author} {\bibfnamefont {C.}~\bibnamefont
  {{Hopman}}}\ and\ \bibinfo {author} {\bibfnamefont {T.}~\bibnamefont
  {{Alexander}}},\ }\href {\doibase 10.1086/506273} {\bibfield  {journal}
  {\bibinfo  {journal} {\apjl}\ }\textbf {\bibinfo {volume} {645}},\ \bibinfo
  {pages} {L133} (\bibinfo {year} {2006})},\ \Eprint
  {http://arxiv.org/abs/astro-ph/0603324} {astro-ph/0603324} \BibitemShut
  {NoStop}%
\bibitem [{\citenamefont {{Kozai}}(1962)}]{Kozai:1962}%
  \BibitemOpen
  \bibfield  {author} {\bibinfo {author} {\bibfnamefont {Y.}~\bibnamefont
  {{Kozai}}},\ }\href {\doibase 10.1086/108790} {\bibfield  {journal} {\bibinfo
   {journal} {Astronomical Journal}\ }\textbf {\bibinfo {volume} {67}},\
  \bibinfo {pages} {591} (\bibinfo {year} {1962})}\BibitemShut {NoStop}%
\bibitem [{\citenamefont {{Lidov}}(1962)}]{Lidov:1962}%
  \BibitemOpen
  \bibfield  {author} {\bibinfo {author} {\bibfnamefont {M.~L.}\ \bibnamefont
  {{Lidov}}},\ }\href {\doibase 10.1016/0032-0633(62)90129-0} {\bibfield
  {journal} {\bibinfo  {journal} {Planetary and Space Science}\ }\textbf
  {\bibinfo {volume} {9}},\ \bibinfo {pages} {719} (\bibinfo {year}
  {1962})}\BibitemShut {NoStop}%
\bibitem [{\citenamefont {{Antonini}}\ \emph {et~al.}(2014)\citenamefont
  {{Antonini}}, \citenamefont {{Murray}},\ and\ \citenamefont
  {{Mikkola}}}]{Anton:2014}%
  \BibitemOpen
  \bibfield  {author} {\bibinfo {author} {\bibfnamefont {F.}~\bibnamefont
  {{Antonini}}}, \bibinfo {author} {\bibfnamefont {N.}~\bibnamefont
  {{Murray}}}, \ and\ \bibinfo {author} {\bibfnamefont {S.}~\bibnamefont
  {{Mikkola}}},\ }\href {\doibase 10.1088/0004-637X/781/1/45} {\bibfield
  {journal} {\bibinfo  {journal} {\apj}\ }\textbf {\bibinfo {volume} {781}},\
  \bibinfo {eid} {45} (\bibinfo {year} {2014})},\ \Eprint
  {http://arxiv.org/abs/1308.3674} {arXiv:1308.3674 [astro-ph.HE]} \BibitemShut
  {NoStop}%
\bibitem [{\citenamefont {{Aarseth}}(2012)}]{Aarseth:2012}%
  \BibitemOpen
  \bibfield  {author} {\bibinfo {author} {\bibfnamefont {S.~J.}\ \bibnamefont
  {{Aarseth}}},\ }\href {\doibase 10.1111/j.1365-2966.2012.20666.x} {\bibfield
  {journal} {\bibinfo  {journal} {\mnras}\ }\textbf {\bibinfo {volume} {422}},\
  \bibinfo {pages} {841} (\bibinfo {year} {2012})},\ \Eprint
  {http://arxiv.org/abs/1202.4688} {arXiv:1202.4688 [astro-ph.SR]} \BibitemShut
  {NoStop}%
\bibitem [{\citenamefont {{Miller}}\ and\ \citenamefont
  {{Hamilton}}(2002)}]{Miller:2002}%
  \BibitemOpen
  \bibfield  {author} {\bibinfo {author} {\bibfnamefont {M.~C.}\ \bibnamefont
  {{Miller}}}\ and\ \bibinfo {author} {\bibfnamefont {D.~P.}\ \bibnamefont
  {{Hamilton}}},\ }\href {\doibase 10.1086/341788} {\bibfield  {journal}
  {\bibinfo  {journal} {\apj}\ }\textbf {\bibinfo {volume} {576}},\ \bibinfo
  {pages} {894} (\bibinfo {year} {2002})},\ \Eprint
  {http://arxiv.org/abs/arXiv:astro-ph/0202298} {arXiv:astro-ph/0202298}
  \BibitemShut {NoStop}%
\bibitem [{\citenamefont {{Maccarone}}\ \emph {et~al.}(2007)\citenamefont
  {{Maccarone}}, \citenamefont {{Kundu}}, \citenamefont {{Zepf}},\ and\
  \citenamefont {{Rhode}}}]{Maccarone:2007}%
  \BibitemOpen
  \bibfield  {author} {\bibinfo {author} {\bibfnamefont {T.~J.}\ \bibnamefont
  {{Maccarone}}}, \bibinfo {author} {\bibfnamefont {A.}~\bibnamefont
  {{Kundu}}}, \bibinfo {author} {\bibfnamefont {S.~E.}\ \bibnamefont {{Zepf}}},
  \ and\ \bibinfo {author} {\bibfnamefont {K.~L.}\ \bibnamefont {{Rhode}}},\
  }\href {\doibase 10.1038/nature05434} {\bibfield  {journal} {\bibinfo
  {journal} {\nat}\ }\textbf {\bibinfo {volume} {445}},\ \bibinfo {pages} {183}
  (\bibinfo {year} {2007})},\ \Eprint {http://arxiv.org/abs/astro-ph/0701310}
  {astro-ph/0701310} \BibitemShut {NoStop}%
\bibitem [{\citenamefont {{Irwin}}\ \emph {et~al.}(2010)\citenamefont
  {{Irwin}}, \citenamefont {{Brink}}, \citenamefont {{Bregman}},\ and\
  \citenamefont {{Roberts}}}]{Irwing:2010}%
  \BibitemOpen
  \bibfield  {author} {\bibinfo {author} {\bibfnamefont {J.~A.}\ \bibnamefont
  {{Irwin}}}, \bibinfo {author} {\bibfnamefont {T.~G.}\ \bibnamefont
  {{Brink}}}, \bibinfo {author} {\bibfnamefont {J.~N.}\ \bibnamefont
  {{Bregman}}}, \ and\ \bibinfo {author} {\bibfnamefont {T.~P.}\ \bibnamefont
  {{Roberts}}},\ }\href {\doibase 10.1088/2041-8205/712/1/L1} {\bibfield
  {journal} {\bibinfo  {journal} {\apjl}\ }\textbf {\bibinfo {volume} {712}},\
  \bibinfo {pages} {L1} (\bibinfo {year} {2010})},\ \Eprint
  {http://arxiv.org/abs/0908.1115} {arXiv:0908.1115 [astro-ph.HE]} \BibitemShut
  {NoStop}%
\bibitem [{\citenamefont {{Strader}}\ \emph {et~al.}(2012)\citenamefont
  {{Strader}}, \citenamefont {{Chomiuk}}, \citenamefont {{Maccarone}},
  \citenamefont {{Miller-Jones}},\ and\ \citenamefont {{Seth}}}]{Strader:2012}%
  \BibitemOpen
  \bibfield  {author} {\bibinfo {author} {\bibfnamefont {J.}~\bibnamefont
  {{Strader}}}, \bibinfo {author} {\bibfnamefont {L.}~\bibnamefont
  {{Chomiuk}}}, \bibinfo {author} {\bibfnamefont {T.~J.}\ \bibnamefont
  {{Maccarone}}}, \bibinfo {author} {\bibfnamefont {J.~C.~A.}\ \bibnamefont
  {{Miller-Jones}}}, \ and\ \bibinfo {author} {\bibfnamefont {A.~C.}\
  \bibnamefont {{Seth}}},\ }\href {\doibase 10.1038/nature11490} {\bibfield
  {journal} {\bibinfo  {journal} {\nat}\ }\textbf {\bibinfo {volume} {490}},\
  \bibinfo {pages} {71} (\bibinfo {year} {2012})},\ \Eprint
  {http://arxiv.org/abs/1210.0901} {arXiv:1210.0901 [astro-ph.HE]} \BibitemShut
  {NoStop}%
\bibitem [{\citenamefont {{Sippel}}\ and\ \citenamefont
  {{Hurley}}(2013)}]{Sippel:2013}%
  \BibitemOpen
  \bibfield  {author} {\bibinfo {author} {\bibfnamefont {A.~C.}\ \bibnamefont
  {{Sippel}}}\ and\ \bibinfo {author} {\bibfnamefont {J.~R.}\ \bibnamefont
  {{Hurley}}},\ }\href {\doibase 10.1093/mnrasl/sls044} {\bibfield  {journal}
  {\bibinfo  {journal} {\mnras}\ }\textbf {\bibinfo {volume} {430}},\ \bibinfo
  {pages} {L30} (\bibinfo {year} {2013})},\ \Eprint
  {http://arxiv.org/abs/1211.6608} {arXiv:1211.6608 [astro-ph.GA]} \BibitemShut
  {NoStop}%
\bibitem [{\citenamefont {{Morscher}}\ \emph {et~al.}(2013)\citenamefont
  {{Morscher}}, \citenamefont {{Umbreit}}, \citenamefont {{Farr}},\ and\
  \citenamefont {{Rasio}}}]{Morscher:2013}%
  \BibitemOpen
  \bibfield  {author} {\bibinfo {author} {\bibfnamefont {M.}~\bibnamefont
  {{Morscher}}}, \bibinfo {author} {\bibfnamefont {S.}~\bibnamefont
  {{Umbreit}}}, \bibinfo {author} {\bibfnamefont {W.~M.}\ \bibnamefont
  {{Farr}}}, \ and\ \bibinfo {author} {\bibfnamefont {F.~A.}\ \bibnamefont
  {{Rasio}}},\ }\href {\doibase 10.1088/2041-8205/763/1/L15} {\bibfield
  {journal} {\bibinfo  {journal} {\apjl}\ }\textbf {\bibinfo {volume} {763}},\
  \bibinfo {eid} {L15} (\bibinfo {year} {2013})},\ \Eprint
  {http://arxiv.org/abs/1211.3372} {arXiv:1211.3372 [astro-ph.GA]} \BibitemShut
  {NoStop}%
\bibitem [{\citenamefont {{Lee}}\ \emph {et~al.}(2010)\citenamefont {{Lee}},
  \citenamefont {{Ramirez-Ruiz}},\ and\ \citenamefont {{van de
  Ven}}}]{Lee:2010}%
  \BibitemOpen
  \bibfield  {author} {\bibinfo {author} {\bibfnamefont {W.~H.}\ \bibnamefont
  {{Lee}}}, \bibinfo {author} {\bibfnamefont {E.}~\bibnamefont
  {{Ramirez-Ruiz}}}, \ and\ \bibinfo {author} {\bibfnamefont {G.}~\bibnamefont
  {{van de Ven}}},\ }\href@noop {} {\bibfield  {journal} {\bibinfo  {journal}
  {\apj}\ }\textbf {\bibinfo {volume} {720}},\ \bibinfo {pages} {953} (\bibinfo
  {year} {2010})}\BibitemShut {NoStop}%
\bibitem [{\citenamefont {{Murphy}}\ \emph {et~al.}(2011)\citenamefont
  {{Murphy}}, \citenamefont {{Cohn}},\ and\ \citenamefont
  {{Lugger}}}]{Murphy:2011}%
  \BibitemOpen
  \bibfield  {author} {\bibinfo {author} {\bibfnamefont {B.~W.}\ \bibnamefont
  {{Murphy}}}, \bibinfo {author} {\bibfnamefont {H.~N.}\ \bibnamefont
  {{Cohn}}}, \ and\ \bibinfo {author} {\bibfnamefont {P.~M.}\ \bibnamefont
  {{Lugger}}},\ }\href {\doibase 10.1088/0004-637X/732/2/67} {\bibfield
  {journal} {\bibinfo  {journal} {\apj}\ }\textbf {\bibinfo {volume} {732}},\
  \bibinfo {eid} {67} (\bibinfo {year} {2011})},\ \Eprint
  {http://arxiv.org/abs/1205.1049} {arXiv:1205.1049 [astro-ph.GA]} \BibitemShut
  {NoStop}%
\bibitem [{\citenamefont {{Pfahl}}\ \emph {et~al.}(2002)\citenamefont
  {{Pfahl}}, \citenamefont {{Rappaport}},\ and\ \citenamefont
  {{Podsiadlowski}}}]{Pfahl:2002}%
  \BibitemOpen
  \bibfield  {author} {\bibinfo {author} {\bibfnamefont {E.}~\bibnamefont
  {{Pfahl}}}, \bibinfo {author} {\bibfnamefont {S.}~\bibnamefont
  {{Rappaport}}}, \ and\ \bibinfo {author} {\bibfnamefont {P.}~\bibnamefont
  {{Podsiadlowski}}},\ }\href {\doibase 10.1086/340494} {\bibfield  {journal}
  {\bibinfo  {journal} {\apj}\ }\textbf {\bibinfo {volume} {573}},\ \bibinfo
  {pages} {283} (\bibinfo {year} {2002})},\ \Eprint
  {http://arxiv.org/abs/astro-ph/0106141} {astro-ph/0106141} \BibitemShut
  {NoStop}%
\bibitem [{\citenamefont {{O'Leary}}\ \emph {et~al.}(2006)\citenamefont
  {{O'Leary}}, \citenamefont {{Rasio}}, \citenamefont {{Fregeau}},
  \citenamefont {{Ivanova}},\ and\ \citenamefont
  {{O'Shaughnessy}}}]{Oleary:2006}%
  \BibitemOpen
  \bibfield  {author} {\bibinfo {author} {\bibfnamefont {R.~M.}\ \bibnamefont
  {{O'Leary}}}, \bibinfo {author} {\bibfnamefont {F.~A.}\ \bibnamefont
  {{Rasio}}}, \bibinfo {author} {\bibfnamefont {J.~M.}\ \bibnamefont
  {{Fregeau}}}, \bibinfo {author} {\bibfnamefont {N.}~\bibnamefont
  {{Ivanova}}}, \ and\ \bibinfo {author} {\bibfnamefont {R.}~\bibnamefont
  {{O'Shaughnessy}}},\ }\href {\doibase 10.1086/498446} {\bibfield  {journal}
  {\bibinfo  {journal} {\apj}\ }\textbf {\bibinfo {volume} {637}},\ \bibinfo
  {pages} {937} (\bibinfo {year} {2006})},\ \Eprint
  {http://arxiv.org/abs/arXiv:astro-ph/0508224} {arXiv:astro-ph/0508224}
  \BibitemShut {NoStop}%
\bibitem [{\citenamefont {{Kulkarni}}\ \emph {et~al.}(1993)\citenamefont
  {{Kulkarni}}, \citenamefont {{Hut}},\ and\ \citenamefont
  {{McMillan}}}]{Kulkarni:1993}%
  \BibitemOpen
  \bibfield  {author} {\bibinfo {author} {\bibfnamefont {S.~R.}\ \bibnamefont
  {{Kulkarni}}}, \bibinfo {author} {\bibfnamefont {P.}~\bibnamefont {{Hut}}}, \
  and\ \bibinfo {author} {\bibfnamefont {S.}~\bibnamefont {{McMillan}}},\
  }\href {\doibase 10.1038/364421a0} {\bibfield  {journal} {\bibinfo  {journal}
  {\nat}\ }\textbf {\bibinfo {volume} {364}},\ \bibinfo {pages} {421} (\bibinfo
  {year} {1993})}\BibitemShut {NoStop}%
\bibitem [{\citenamefont {{Sigurdsson}}\ and\ \citenamefont
  {{Hernquist}}(1993)}]{nat}%
  \BibitemOpen
  \bibfield  {author} {\bibinfo {author} {\bibfnamefont {S.}~\bibnamefont
  {{Sigurdsson}}}\ and\ \bibinfo {author} {\bibfnamefont {L.}~\bibnamefont
  {{Hernquist}}},\ }\href {\doibase 10.1038/364423a0} {\bibfield  {journal}
  {\bibinfo  {journal} {\nat}\ }\textbf {\bibinfo {volume} {364}},\ \bibinfo
  {pages} {423} (\bibinfo {year} {1993})}\BibitemShut {NoStop}%
\bibitem [{\citenamefont {{Portegies Zwart}}\ and\ \citenamefont
  {{McMillan}}(2000)}]{Portegies:2000}%
  \BibitemOpen
  \bibfield  {author} {\bibinfo {author} {\bibfnamefont {S.~F.}\ \bibnamefont
  {{Portegies Zwart}}}\ and\ \bibinfo {author} {\bibfnamefont {S.~L.~W.}\
  \bibnamefont {{McMillan}}},\ }\href {\doibase 10.1086/312422} {\bibfield
  {journal} {\bibinfo  {journal} {\apjl}\ }\textbf {\bibinfo {volume} {528}},\
  \bibinfo {pages} {L17} (\bibinfo {year} {2000})},\ \Eprint
  {http://arxiv.org/abs/astro-ph/9910061} {astro-ph/9910061} \BibitemShut
  {NoStop}%
\bibitem [{\citenamefont {{Martel}}\ and\ \citenamefont
  {{Poisson}}(1999)}]{Martel:1999}%
  \BibitemOpen
  \bibfield  {author} {\bibinfo {author} {\bibfnamefont {K.}~\bibnamefont
  {{Martel}}}\ and\ \bibinfo {author} {\bibfnamefont {E.}~\bibnamefont
  {{Poisson}}},\ }\href {\doibase 10.1103/PhysRevD.60.124008} {\bibfield
  {journal} {\bibinfo  {journal} {\prd}\ }\textbf {\bibinfo {volume} {60}},\
  \bibinfo {eid} {124008} (\bibinfo {year} {1999})},\ \Eprint
  {http://arxiv.org/abs/gr-qc/9907006} {gr-qc/9907006} \BibitemShut {NoStop}%
\bibitem [{\citenamefont {{Cokelaer}}\ and\ \citenamefont
  {{Pathak}}(2009)}]{Cokelaer:2009}%
  \BibitemOpen
  \bibfield  {author} {\bibinfo {author} {\bibfnamefont {T.}~\bibnamefont
  {{Cokelaer}}}\ and\ \bibinfo {author} {\bibfnamefont {D.}~\bibnamefont
  {{Pathak}}},\ }\href {\doibase 10.1088/0264-9381/26/4/045013} {\bibfield
  {journal} {\bibinfo  {journal} {Classical and Quantum Gravity}\ }\textbf
  {\bibinfo {volume} {26}},\ \bibinfo {eid} {045013} (\bibinfo {year}
  {2009})},\ \Eprint {http://arxiv.org/abs/0903.4791} {arXiv:0903.4791 [gr-qc]}
  \BibitemShut {NoStop}%
\bibitem [{\citenamefont {{Brown}}\ and\ \citenamefont
  {{Zimmerman}}(2010)}]{Brown:2010}%
  \BibitemOpen
  \bibfield  {author} {\bibinfo {author} {\bibfnamefont {D.~A.}\ \bibnamefont
  {{Brown}}}\ and\ \bibinfo {author} {\bibfnamefont {P.~J.}\ \bibnamefont
  {{Zimmerman}}},\ }\href {\doibase 10.1103/PhysRevD.81.024007} {\bibfield
  {journal} {\bibinfo  {journal} {\prd}\ }\textbf {\bibinfo {volume} {81}},\
  \bibinfo {eid} {024007} (\bibinfo {year} {2010})}\BibitemShut {NoStop}%
\bibitem [{\citenamefont {{Huerta}}\ and\ \citenamefont
  {{Brown}}(2013)}]{Huerta:2013a}%
  \BibitemOpen
  \bibfield  {author} {\bibinfo {author} {\bibfnamefont {E.~A.}\ \bibnamefont
  {{Huerta}}}\ and\ \bibinfo {author} {\bibfnamefont {D.~A.}\ \bibnamefont
  {{Brown}}},\ }\href {\doibase 10.1103/PhysRevD.87.127501} {\bibfield
  {journal} {\bibinfo  {journal} {\prd}\ }\textbf {\bibinfo {volume} {87}},\
  \bibinfo {eid} {127501} (\bibinfo {year} {2013})},\ \Eprint
  {http://arxiv.org/abs/1301.1895} {arXiv:1301.1895 [gr-qc]} \BibitemShut
  {NoStop}%
\bibitem [{\citenamefont {{Babak}}\ \emph {et~al.}(2007)\citenamefont
  {{Babak}}, \citenamefont {{Fang}}, \citenamefont {{Gair}}, \citenamefont
  {{Glampedakis}},\ and\ \citenamefont {{Hughes}}}]{kludge}%
  \BibitemOpen
  \bibfield  {author} {\bibinfo {author} {\bibfnamefont {S.}~\bibnamefont
  {{Babak}}}, \bibinfo {author} {\bibfnamefont {H.}~\bibnamefont {{Fang}}},
  \bibinfo {author} {\bibfnamefont {J.~R.}\ \bibnamefont {{Gair}}}, \bibinfo
  {author} {\bibfnamefont {K.}~\bibnamefont {{Glampedakis}}}, \ and\ \bibinfo
  {author} {\bibfnamefont {S.~A.}\ \bibnamefont {{Hughes}}},\ }\href {\doibase
  10.1103/PhysRevD.75.024005} {\bibfield  {journal} {\bibinfo  {journal}
  {\PRD}\ }\textbf {\bibinfo {volume} {75}},\ \bibinfo {pages} {024005}
  (\bibinfo {year} {2007})},\ \Eprint
  {http://arxiv.org/abs/arXiv:gr-qc/0607007} {arXiv:gr-qc/0607007} \BibitemShut
  {NoStop}%
\bibitem [{\citenamefont {{Gair}}\ and\ \citenamefont
  {{Glampedakis}}(2006)}]{improved}%
  \BibitemOpen
  \bibfield  {author} {\bibinfo {author} {\bibfnamefont {J.~R.}\ \bibnamefont
  {{Gair}}}\ and\ \bibinfo {author} {\bibfnamefont {K.}~\bibnamefont
  {{Glampedakis}}},\ }\href {\doibase 10.1103/PhysRevD.73.064037} {\bibfield
  {journal} {\bibinfo  {journal} {\PRD}\ }\textbf {\bibinfo {volume} {73}},\
  \bibinfo {pages} {064037} (\bibinfo {year} {2006})},\ \Eprint
  {http://arxiv.org/abs/arXiv:gr-qc/0510129} {arXiv:gr-qc/0510129} \BibitemShut
  {NoStop}%
\bibitem [{\citenamefont {Glampedakis}\ and\ \citenamefont
  {Kennefick}(2002)}]{gla}%
  \BibitemOpen
  \bibfield  {author} {\bibinfo {author} {\bibfnamefont {K.}~\bibnamefont
  {Glampedakis}}\ and\ \bibinfo {author} {\bibfnamefont {D.}~\bibnamefont
  {Kennefick}},\ }\href {\doibase 10.1103/PhysRevD.66.044002} {\bibfield
  {journal} {\bibinfo  {journal} {\PRD}\ }\textbf {\bibinfo {volume} {66}},\
  \bibinfo {pages} {044002} (\bibinfo {year} {2002})}\BibitemShut {NoStop}%
\bibitem [{\citenamefont {{Sundararajan}}\ \emph {et~al.}(2008)\citenamefont
  {{Sundararajan}}, \citenamefont {{Khanna}}, \citenamefont {{Hughes}},\ and\
  \citenamefont {{Drasco}}}]{Sunda:2008}%
  \BibitemOpen
  \bibfield  {author} {\bibinfo {author} {\bibfnamefont {P.~A.}\ \bibnamefont
  {{Sundararajan}}}, \bibinfo {author} {\bibfnamefont {G.}~\bibnamefont
  {{Khanna}}}, \bibinfo {author} {\bibfnamefont {S.~A.}\ \bibnamefont
  {{Hughes}}}, \ and\ \bibinfo {author} {\bibfnamefont {S.}~\bibnamefont
  {{Drasco}}},\ }\href {\doibase 10.1103/PhysRevD.78.024022} {\bibfield
  {journal} {\bibinfo  {journal} {\prd}\ }\textbf {\bibinfo {volume} {78}},\
  \bibinfo {eid} {024022} (\bibinfo {year} {2008})},\ \Eprint
  {http://arxiv.org/abs/0803.0317} {arXiv:0803.0317 [gr-qc]} \BibitemShut
  {NoStop}%
\bibitem [{\citenamefont {{Ganz}}\ \emph {et~al.}(2007)\citenamefont {{Ganz}},
  \citenamefont {{Hikida}}, \citenamefont {{Nakano}}, \citenamefont {{Sago}},\
  and\ \citenamefont {{Tanaka}}}]{Ganz:2007}%
  \BibitemOpen
  \bibfield  {author} {\bibinfo {author} {\bibfnamefont {K.}~\bibnamefont
  {{Ganz}}}, \bibinfo {author} {\bibfnamefont {W.}~\bibnamefont {{Hikida}}},
  \bibinfo {author} {\bibfnamefont {H.}~\bibnamefont {{Nakano}}}, \bibinfo
  {author} {\bibfnamefont {N.}~\bibnamefont {{Sago}}}, \ and\ \bibinfo {author}
  {\bibfnamefont {T.}~\bibnamefont {{Tanaka}}},\ }\href {\doibase
  10.1143/PTP.117.1041} {\bibfield  {journal} {\bibinfo  {journal} {Progress of
  Theoretical Physics}\ }\textbf {\bibinfo {volume} {117}},\ \bibinfo {pages}
  {1041} (\bibinfo {year} {2007})},\ \Eprint
  {http://arxiv.org/abs/gr-qc/0702054} {gr-qc/0702054} \BibitemShut {NoStop}%
\bibitem [{\citenamefont {{Mino}}\ \emph
  {et~al.}(1997{\natexlab{a}})\citenamefont {{Mino}}, \citenamefont
  {{Sasaki}},\ and\ \citenamefont {{Tanaka}}}]{MMT}%
  \BibitemOpen
  \bibfield  {author} {\bibinfo {author} {\bibfnamefont {Y.}~\bibnamefont
  {{Mino}}}, \bibinfo {author} {\bibfnamefont {M.}~\bibnamefont {{Sasaki}}}, \
  and\ \bibinfo {author} {\bibfnamefont {T.}~\bibnamefont {{Tanaka}}},\ }\href
  {\doibase 10.1143/PTPS.128.373} {\bibfield  {journal} {\bibinfo  {journal}
  {Progress of Theoretical Physics Supplement}\ }\textbf {\bibinfo {volume}
  {128}},\ \bibinfo {pages} {373} (\bibinfo {year} {1997}{\natexlab{a}})},\
  \Eprint {http://arxiv.org/abs/arXiv:gr-qc/9712056} {arXiv:gr-qc/9712056}
  \BibitemShut {NoStop}%
\bibitem [{\citenamefont {{Mino}}\ \emph
  {et~al.}(1997{\natexlab{b}})\citenamefont {{Mino}}, \citenamefont
  {{Sasaki}},\ and\ \citenamefont {{Tanaka}}}]{mino}%
  \BibitemOpen
  \bibfield  {author} {\bibinfo {author} {\bibfnamefont {Y.}~\bibnamefont
  {{Mino}}}, \bibinfo {author} {\bibfnamefont {M.}~\bibnamefont {{Sasaki}}}, \
  and\ \bibinfo {author} {\bibfnamefont {T.}~\bibnamefont {{Tanaka}}},\
  }\href@noop {} {\bibfield  {journal} {\bibinfo  {journal} {\PRD}\ }\textbf
  {\bibinfo {volume} {55}},\ \bibinfo {pages} {3457} (\bibinfo {year}
  {1997}{\natexlab{b}})},\ \Eprint {http://arxiv.org/abs/arXiv:gr-qc/9606018}
  {arXiv:gr-qc/9606018} \BibitemShut {NoStop}%
\bibitem [{\citenamefont {{Mino}}\ \emph
  {et~al.}(1997{\natexlab{c}})\citenamefont {{Mino}}, \citenamefont {{Sasaki}},
  \citenamefont {{Shibata}}, \citenamefont {{Tagoshi}},\ and\ \citenamefont
  {{Tanaka}}}]{Mino:1997}%
  \BibitemOpen
  \bibfield  {author} {\bibinfo {author} {\bibfnamefont {Y.}~\bibnamefont
  {{Mino}}}, \bibinfo {author} {\bibfnamefont {M.}~\bibnamefont {{Sasaki}}},
  \bibinfo {author} {\bibfnamefont {M.}~\bibnamefont {{Shibata}}}, \bibinfo
  {author} {\bibfnamefont {H.}~\bibnamefont {{Tagoshi}}}, \ and\ \bibinfo
  {author} {\bibfnamefont {T.}~\bibnamefont {{Tanaka}}},\ }\href {\doibase
  10.1143/PTPS.128.1} {\bibfield  {journal} {\bibinfo  {journal} {Progress of
  Theoretical Physics Supplement}\ }\textbf {\bibinfo {volume} {128}},\
  \bibinfo {pages} {1} (\bibinfo {year} {1997}{\natexlab{c}})},\ \Eprint
  {http://arxiv.org/abs/arXiv:gr-qc/9712057} {arXiv:gr-qc/9712057} \BibitemShut
  {NoStop}%
\bibitem [{\citenamefont {{Berti}}(2006)}]{Berti:2006C}%
  \BibitemOpen
  \bibfield  {author} {\bibinfo {author} {\bibfnamefont {E.}~\bibnamefont
  {{Berti}}},\ }\href {\doibase 10.1088/0264-9381/23/19/S17} {\bibfield
  {journal} {\bibinfo  {journal} {Classical and Quantum Gravity}\ }\textbf
  {\bibinfo {volume} {23}},\ \bibinfo {pages} {785} (\bibinfo {year} {2006})},\
  \Eprint {http://arxiv.org/abs/astro-ph/0602470} {astro-ph/0602470}
  \BibitemShut {NoStop}%
\bibitem [{\citenamefont {{Campanelli}}\ \emph {et~al.}(2009)\citenamefont
  {{Campanelli}}, \citenamefont {{Lousto}}, \citenamefont {{Nakano}},\ and\
  \citenamefont {{Zlochower}}}]{CampLou:2009}%
  \BibitemOpen
  \bibfield  {author} {\bibinfo {author} {\bibfnamefont {M.}~\bibnamefont
  {{Campanelli}}}, \bibinfo {author} {\bibfnamefont {C.~O.}\ \bibnamefont
  {{Lousto}}}, \bibinfo {author} {\bibfnamefont {H.}~\bibnamefont {{Nakano}}},
  \ and\ \bibinfo {author} {\bibfnamefont {Y.}~\bibnamefont {{Zlochower}}},\
  }\href {\doibase 10.1103/PhysRevD.79.084010} {\bibfield  {journal} {\bibinfo
  {journal} {\prd}\ }\textbf {\bibinfo {volume} {79}},\ \bibinfo {eid} {084010}
  (\bibinfo {year} {2009})},\ \Eprint {http://arxiv.org/abs/0808.0713}
  {arXiv:0808.0713 [gr-qc]} \BibitemShut {NoStop}%
\bibitem [{\citenamefont {{Barack}}\ and\ \citenamefont
  {{Cutler}}(2004)}]{cutler}%
  \BibitemOpen
  \bibfield  {author} {\bibinfo {author} {\bibfnamefont {L.}~\bibnamefont
  {{Barack}}}\ and\ \bibinfo {author} {\bibfnamefont {C.}~\bibnamefont
  {{Cutler}}},\ }\href {\doibase 10.1103/PhysRevD.69.082005} {\bibfield
  {journal} {\bibinfo  {journal} {\PRD}\ }\textbf {\bibinfo {volume} {69}},\
  \bibinfo {pages} {082005} (\bibinfo {year} {2004})},\ \Eprint
  {http://arxiv.org/abs/arXiv:gr-qc/0310125} {arXiv:gr-qc/0310125} \BibitemShut
  {NoStop}%
\bibitem [{\citenamefont {{Cornish}}\ and\ \citenamefont
  {{Key}}(2010)}]{NCornish:2010}%
  \BibitemOpen
  \bibfield  {author} {\bibinfo {author} {\bibfnamefont {N.~J.}\ \bibnamefont
  {{Cornish}}}\ and\ \bibinfo {author} {\bibfnamefont {J.~S.}\ \bibnamefont
  {{Key}}},\ }\href {\doibase 10.1103/PhysRevD.82.044028} {\bibfield  {journal}
  {\bibinfo  {journal} {\prd}\ }\textbf {\bibinfo {volume} {82}},\ \bibinfo
  {eid} {044028} (\bibinfo {year} {2010})},\ \Eprint
  {http://arxiv.org/abs/1004.5322} {arXiv:1004.5322 [gr-qc]} \BibitemShut
  {NoStop}%
\bibitem [{\citenamefont {{Key}}\ and\ \citenamefont
  {{Cornish}}(2011)}]{Key:2011J}%
  \BibitemOpen
  \bibfield  {author} {\bibinfo {author} {\bibfnamefont {J.~S.}\ \bibnamefont
  {{Key}}}\ and\ \bibinfo {author} {\bibfnamefont {N.~J.}\ \bibnamefont
  {{Cornish}}},\ }\href {\doibase 10.1103/PhysRevD.83.083001} {\bibfield
  {journal} {\bibinfo  {journal} {\prd}\ }\textbf {\bibinfo {volume} {83}},\
  \bibinfo {eid} {083001} (\bibinfo {year} {2011})},\ \Eprint
  {http://arxiv.org/abs/1006.3759} {arXiv:1006.3759 [gr-qc]} \BibitemShut
  {NoStop}%
\bibitem [{\citenamefont {{Levin}}\ \emph {et~al.}(2011)\citenamefont
  {{Levin}}, \citenamefont {{McWilliams}},\ and\ \citenamefont
  {{Contreras}}}]{Levin:2011C}%
  \BibitemOpen
  \bibfield  {author} {\bibinfo {author} {\bibfnamefont {J.}~\bibnamefont
  {{Levin}}}, \bibinfo {author} {\bibfnamefont {S.~T.}\ \bibnamefont
  {{McWilliams}}}, \ and\ \bibinfo {author} {\bibfnamefont {H.}~\bibnamefont
  {{Contreras}}},\ }\href {\doibase 10.1088/0264-9381/28/17/175001} {\bibfield
  {journal} {\bibinfo  {journal} {Classical and Quantum Gravity}\ }\textbf
  {\bibinfo {volume} {28}},\ \bibinfo {eid} {175001} (\bibinfo {year}
  {2011})},\ \Eprint {http://arxiv.org/abs/1009.2533} {arXiv:1009.2533 [gr-qc]}
  \BibitemShut {NoStop}%
\bibitem [{\citenamefont {{Mik{\'o}czi}}\ \emph {et~al.}(2012)\citenamefont
  {{Mik{\'o}czi}}, \citenamefont {{Kocsis}}, \citenamefont {{Forg{\'a}cs}},\
  and\ \citenamefont {{Vas{\'u}th}}}]{MikKoc:2012}%
  \BibitemOpen
  \bibfield  {author} {\bibinfo {author} {\bibfnamefont {B.}~\bibnamefont
  {{Mik{\'o}czi}}}, \bibinfo {author} {\bibfnamefont {B.}~\bibnamefont
  {{Kocsis}}}, \bibinfo {author} {\bibfnamefont {P.}~\bibnamefont
  {{Forg{\'a}cs}}}, \ and\ \bibinfo {author} {\bibfnamefont {M.}~\bibnamefont
  {{Vas{\'u}th}}},\ }\href {\doibase 10.1103/PhysRevD.86.104027} {\bibfield
  {journal} {\bibinfo  {journal} {\prd}\ }\textbf {\bibinfo {volume} {86}},\
  \bibinfo {eid} {104027} (\bibinfo {year} {2012})},\ \Eprint
  {http://arxiv.org/abs/1206.5786} {arXiv:1206.5786 [gr-qc]} \BibitemShut
  {NoStop}%
\bibitem [{\citenamefont {{Arun}}\ \emph
  {et~al.}(2008{\natexlab{a}})\citenamefont {{Arun}}, \citenamefont
  {{Blanchet}}, \citenamefont {{Iyer}},\ and\ \citenamefont
  {{Qusailah}}}]{ArunBlanchet:2008}%
  \BibitemOpen
  \bibfield  {author} {\bibinfo {author} {\bibfnamefont {K.~G.}\ \bibnamefont
  {{Arun}}}, \bibinfo {author} {\bibfnamefont {L.}~\bibnamefont {{Blanchet}}},
  \bibinfo {author} {\bibfnamefont {B.~R.}\ \bibnamefont {{Iyer}}}, \ and\
  \bibinfo {author} {\bibfnamefont {M.~S.~S.}\ \bibnamefont {{Qusailah}}},\
  }\href {\doibase 10.1103/PhysRevD.77.064034} {\bibfield  {journal} {\bibinfo
  {journal} {\prd}\ }\textbf {\bibinfo {volume} {77}},\ \bibinfo {eid} {064034}
  (\bibinfo {year} {2008}{\natexlab{a}})},\ \Eprint
  {http://arxiv.org/abs/0711.0250} {arXiv:0711.0250 [gr-qc]} \BibitemShut
  {NoStop}%
\bibitem [{\citenamefont {{Gold}}\ and\ \citenamefont
  {{Br{\"u}gmann}}(2013)}]{Gold:2013}%
  \BibitemOpen
  \bibfield  {author} {\bibinfo {author} {\bibfnamefont {R.}~\bibnamefont
  {{Gold}}}\ and\ \bibinfo {author} {\bibfnamefont {B.}~\bibnamefont
  {{Br{\"u}gmann}}},\ }\href {\doibase 10.1103/PhysRevD.88.064051} {\bibfield
  {journal} {\bibinfo  {journal} {\prd}\ }\textbf {\bibinfo {volume} {88}},\
  \bibinfo {eid} {064051} (\bibinfo {year} {2013})},\ \Eprint
  {http://arxiv.org/abs/1209.4085} {arXiv:1209.4085 [gr-qc]} \BibitemShut
  {NoStop}%
\bibitem [{\citenamefont {{Hinder}}\ \emph {et~al.}(2010)\citenamefont
  {{Hinder}}, \citenamefont {{Herrmann}}, \citenamefont {{Laguna}},\ and\
  \citenamefont {{Shoemaker}}}]{Hinder:2010}%
  \BibitemOpen
  \bibfield  {author} {\bibinfo {author} {\bibfnamefont {I.}~\bibnamefont
  {{Hinder}}}, \bibinfo {author} {\bibfnamefont {F.}~\bibnamefont
  {{Herrmann}}}, \bibinfo {author} {\bibfnamefont {P.}~\bibnamefont
  {{Laguna}}}, \ and\ \bibinfo {author} {\bibfnamefont {D.}~\bibnamefont
  {{Shoemaker}}},\ }\href {\doibase 10.1103/PhysRevD.82.024033} {\bibfield
  {journal} {\bibinfo  {journal} {\prd}\ }\textbf {\bibinfo {volume} {82}},\
  \bibinfo {eid} {024033} (\bibinfo {year} {2010})}\BibitemShut {NoStop}%
\bibitem [{\citenamefont {{Hinderer}}\ \emph {et~al.}(2013)\citenamefont
  {{Hinderer}}, \citenamefont {{Buonanno}}, \citenamefont {{Mrou{\'e}}},
  \citenamefont {{Hemberger}}, \citenamefont {{Lovelace}}, \citenamefont
  {{Pfeiffer}}, \citenamefont {{Kidder}}, \citenamefont {{Scheel}},
  \citenamefont {{Szilagyi}}, \citenamefont {{Taylor}},\ and\ \citenamefont
  {{Teukolsky}}}]{Tania:2013}%
  \BibitemOpen
  \bibfield  {author} {\bibinfo {author} {\bibfnamefont {T.}~\bibnamefont
  {{Hinderer}}}, \bibinfo {author} {\bibfnamefont {A.}~\bibnamefont
  {{Buonanno}}}, \bibinfo {author} {\bibfnamefont {A.~H.}\ \bibnamefont
  {{Mrou{\'e}}}}, \bibinfo {author} {\bibfnamefont {D.~A.}\ \bibnamefont
  {{Hemberger}}}, \bibinfo {author} {\bibfnamefont {G.}~\bibnamefont
  {{Lovelace}}}, \bibinfo {author} {\bibfnamefont {H.~P.}\ \bibnamefont
  {{Pfeiffer}}}, \bibinfo {author} {\bibfnamefont {L.~E.}\ \bibnamefont
  {{Kidder}}}, \bibinfo {author} {\bibfnamefont {M.~A.}\ \bibnamefont
  {{Scheel}}}, \bibinfo {author} {\bibfnamefont {B.}~\bibnamefont
  {{Szilagyi}}}, \bibinfo {author} {\bibfnamefont {N.~W.}\ \bibnamefont
  {{Taylor}}}, \ and\ \bibinfo {author} {\bibfnamefont {S.~A.}\ \bibnamefont
  {{Teukolsky}}},\ }\href {\doibase 10.1103/PhysRevD.88.084005} {\bibfield
  {journal} {\bibinfo  {journal} {\prd}\ }\textbf {\bibinfo {volume} {88}},\
  \bibinfo {eid} {084005} (\bibinfo {year} {2013})},\ \Eprint
  {http://arxiv.org/abs/1309.0544} {arXiv:1309.0544 [gr-qc]} \BibitemShut
  {NoStop}%
\bibitem [{\citenamefont {{Yunes}}\ \emph {et~al.}(2009)\citenamefont
  {{Yunes}}, \citenamefont {{Arun}}, \citenamefont {{Berti}},\ and\
  \citenamefont {{Will}}}]{Yunes:2009}%
  \BibitemOpen
  \bibfield  {author} {\bibinfo {author} {\bibfnamefont {N.}~\bibnamefont
  {{Yunes}}}, \bibinfo {author} {\bibfnamefont {K.~G.}\ \bibnamefont {{Arun}}},
  \bibinfo {author} {\bibfnamefont {E.}~\bibnamefont {{Berti}}}, \ and\
  \bibinfo {author} {\bibfnamefont {C.~M.}\ \bibnamefont {{Will}}},\ }\href
  {\doibase 10.1103/PhysRevD.80.084001} {\bibfield  {journal} {\bibinfo
  {journal} {\prd}\ }\textbf {\bibinfo {volume} {80}},\ \bibinfo {eid} {084001}
  (\bibinfo {year} {2009})},\ \Eprint {http://arxiv.org/abs/0906.0313}
  {arXiv:0906.0313 [gr-qc]} \BibitemShut {NoStop}%
\bibitem [{\citenamefont {{Blanchet}}(2006)}]{Blanchet:2006}%
  \BibitemOpen
  \bibfield  {author} {\bibinfo {author} {\bibfnamefont {L.}~\bibnamefont
  {{Blanchet}}},\ }\href {\doibase 10.12942/lrr-2006-4} {\bibfield  {journal}
  {\bibinfo  {journal} {Living Reviews in Relativity}\ }\textbf {\bibinfo
  {volume} {9}},\ \bibinfo {pages} {4} (\bibinfo {year} {2006})}\BibitemShut
  {NoStop}%
\bibitem [{\citenamefont {{Boyle}}\ \emph {et~al.}(2007)\citenamefont
  {{Boyle}}, \citenamefont {{Brown}}, \citenamefont {{Kidder}}, \citenamefont
  {{Mrou{\'e}}}, \citenamefont {{Pfeiffer}}, \citenamefont {{Scheel}},
  \citenamefont {{Cook}},\ and\ \citenamefont {{Teukolsky}}}]{boyle}%
  \BibitemOpen
  \bibfield  {author} {\bibinfo {author} {\bibfnamefont {M.}~\bibnamefont
  {{Boyle}}}, \bibinfo {author} {\bibfnamefont {D.~A.}\ \bibnamefont
  {{Brown}}}, \bibinfo {author} {\bibfnamefont {L.~E.}\ \bibnamefont
  {{Kidder}}}, \bibinfo {author} {\bibfnamefont {A.~H.}\ \bibnamefont
  {{Mrou{\'e}}}}, \bibinfo {author} {\bibfnamefont {H.~P.}\ \bibnamefont
  {{Pfeiffer}}}, \bibinfo {author} {\bibfnamefont {M.~A.}\ \bibnamefont
  {{Scheel}}}, \bibinfo {author} {\bibfnamefont {G.~B.}\ \bibnamefont
  {{Cook}}}, \ and\ \bibinfo {author} {\bibfnamefont {S.~A.}\ \bibnamefont
  {{Teukolsky}}},\ }\href {\doibase 10.1103/PhysRevD.76.124038} {\bibfield
  {journal} {\bibinfo  {journal} {\prd}\ }\textbf {\bibinfo {volume} {76}},\
  \bibinfo {eid} {124038} (\bibinfo {year} {2007})}\BibitemShut {NoStop}%
\bibitem [{\citenamefont {{Bender}}\ and\ \citenamefont
  {{Orszag}}(1999)}]{Bender:1999}%
  \BibitemOpen
  \bibfield  {author} {\bibinfo {author} {\bibfnamefont {C.~M.}\ \bibnamefont
  {{Bender}}}\ and\ \bibinfo {author} {\bibfnamefont {S.~A.}\ \bibnamefont
  {{Orszag}}},\ }\href@noop {} {\emph {\bibinfo {title} {Advanced mathematical
  methods for scientists and engineers I: asymptotic methods and perturbation
  theory.}}}\ (\bibinfo  {publisher} {Springer-Verlag},\ \bibinfo {year}
  {1999})\BibitemShut {NoStop}%
\bibitem [{\citenamefont {Tessmer}\ and\ \citenamefont
  {Schaefer}(2010)}]{Tessmer:2010sh}%
  \BibitemOpen
  \bibfield  {author} {\bibinfo {author} {\bibfnamefont {M.}~\bibnamefont
  {Tessmer}}\ and\ \bibinfo {author} {\bibfnamefont {G.}~\bibnamefont
  {Schaefer}},\ }\href {\doibase 10.1103/PhysRevD.82.124064} {\bibfield
  {journal} {\bibinfo  {journal} {Phys.Rev.}\ }\textbf {\bibinfo {volume}
  {D82}},\ \bibinfo {pages} {124064} (\bibinfo {year} {2010})},\ \Eprint
  {http://arxiv.org/abs/1006.3714} {arXiv:1006.3714 [gr-qc]} \BibitemShut
  {NoStop}%
\bibitem [{\citenamefont {Tessmer}\ and\ \citenamefont
  {Schaefer}(2011)}]{Tessmer:2010ii}%
  \BibitemOpen
  \bibfield  {author} {\bibinfo {author} {\bibfnamefont {M.}~\bibnamefont
  {Tessmer}}\ and\ \bibinfo {author} {\bibfnamefont {G.}~\bibnamefont
  {Schaefer}},\ }\href {\doibase 10.1002/andp.201100007} {\bibfield  {journal}
  {\bibinfo  {journal} {Annalen Phys.}\ }\textbf {\bibinfo {volume} {523}},\
  \bibinfo {pages} {813} (\bibinfo {year} {2011})},\ \Eprint
  {http://arxiv.org/abs/1012.3894} {arXiv:1012.3894 [gr-qc]} \BibitemShut
  {NoStop}%
\bibitem [{\citenamefont {{Arun}}\ \emph
  {et~al.}(2008{\natexlab{b}})\citenamefont {{Arun}}, \citenamefont
  {{Blanchet}}, \citenamefont {{Iyer}},\ and\ \citenamefont
  {{Qusailah}}}]{Arun:2008}%
  \BibitemOpen
  \bibfield  {author} {\bibinfo {author} {\bibfnamefont {K.~G.}\ \bibnamefont
  {{Arun}}}, \bibinfo {author} {\bibfnamefont {L.}~\bibnamefont {{Blanchet}}},
  \bibinfo {author} {\bibfnamefont {B.~R.}\ \bibnamefont {{Iyer}}}, \ and\
  \bibinfo {author} {\bibfnamefont {M.~S.~S.}\ \bibnamefont {{Qusailah}}},\
  }\href {\doibase 10.1103/PhysRevD.77.064035} {\bibfield  {journal} {\bibinfo
  {journal} {\prd}\ }\textbf {\bibinfo {volume} {77}},\ \bibinfo {eid} {064035}
  (\bibinfo {year} {2008}{\natexlab{b}})},\ \Eprint
  {http://arxiv.org/abs/0711.0302} {arXiv:0711.0302 [gr-qc]} \BibitemShut
  {NoStop}%
\bibitem [{\citenamefont {{Brown}}\ \emph {et~al.}(2013)\citenamefont
  {{Brown}}, \citenamefont {{Kumar}},\ and\ \citenamefont
  {{Nitz}}}]{Prayush:2013a}%
  \BibitemOpen
  \bibfield  {author} {\bibinfo {author} {\bibfnamefont {D.~A.}\ \bibnamefont
  {{Brown}}}, \bibinfo {author} {\bibfnamefont {P.}~\bibnamefont {{Kumar}}}, \
  and\ \bibinfo {author} {\bibfnamefont {A.~H.}\ \bibnamefont {{Nitz}}},\
  }\href {\doibase 10.1103/PhysRevD.87.082004} {\bibfield  {journal} {\bibinfo
  {journal} {\prd}\ }\textbf {\bibinfo {volume} {87}},\ \bibinfo {eid} {082004}
  (\bibinfo {year} {2013})},\ \Eprint {http://arxiv.org/abs/1211.6184}
  {arXiv:1211.6184 [gr-qc]} \BibitemShut {NoStop}%
\bibitem [{\citenamefont {{Buonanno}}\ \emph {et~al.}(2009)\citenamefont
  {{Buonanno}}, \citenamefont {{Iyer}}, \citenamefont {{Ochsner}},
  \citenamefont {{Pan}},\ and\ \citenamefont {{Sathyaprakash}}}]{pnbuo}%
  \BibitemOpen
  \bibfield  {author} {\bibinfo {author} {\bibfnamefont {A.}~\bibnamefont
  {{Buonanno}}}, \bibinfo {author} {\bibfnamefont {B.~R.}\ \bibnamefont
  {{Iyer}}}, \bibinfo {author} {\bibfnamefont {E.}~\bibnamefont {{Ochsner}}},
  \bibinfo {author} {\bibfnamefont {Y.}~\bibnamefont {{Pan}}}, \ and\ \bibinfo
  {author} {\bibfnamefont {B.~S.}\ \bibnamefont {{Sathyaprakash}}},\
  }\href@noop {} {\bibfield  {journal} {\bibinfo  {journal} {\prd}\ }\textbf
  {\bibinfo {volume} {80}},\ \bibinfo {pages} {084043} (\bibinfo {year}
  {2009})}\BibitemShut {NoStop}%
\bibitem [{\citenamefont {{Van Den Broeck}}\ and\ \citenamefont
  {{Sengupta}}(2007{\natexlab{a}})}]{VanDen:2007}%
  \BibitemOpen
  \bibfield  {author} {\bibinfo {author} {\bibfnamefont {C.}~\bibnamefont {{Van
  Den Broeck}}}\ and\ \bibinfo {author} {\bibfnamefont {A.~S.}\ \bibnamefont
  {{Sengupta}}},\ }\href {\doibase 10.1088/0264-9381/24/1/009} {\bibfield
  {journal} {\bibinfo  {journal} {Classical and Quantum Gravity}\ }\textbf
  {\bibinfo {volume} {24}},\ \bibinfo {pages} {155} (\bibinfo {year}
  {2007}{\natexlab{a}})},\ \Eprint {http://arxiv.org/abs/gr-qc/0607092}
  {gr-qc/0607092} \BibitemShut {NoStop}%
\bibitem [{\citenamefont {{Arun}}\ \emph
  {et~al.}(2007{\natexlab{a}})\citenamefont {{Arun}}, \citenamefont {{Iyer}},
  \citenamefont {{Sathyaprakash}},\ and\ \citenamefont {{Sinha}}}]{Arun:2007}%
  \BibitemOpen
  \bibfield  {author} {\bibinfo {author} {\bibfnamefont {K.~G.}\ \bibnamefont
  {{Arun}}}, \bibinfo {author} {\bibfnamefont {B.~R.}\ \bibnamefont {{Iyer}}},
  \bibinfo {author} {\bibfnamefont {B.~S.}\ \bibnamefont {{Sathyaprakash}}}, \
  and\ \bibinfo {author} {\bibfnamefont {S.}~\bibnamefont {{Sinha}}},\ }\href
  {\doibase 10.1103/PhysRevD.75.124002} {\bibfield  {journal} {\bibinfo
  {journal} {\prd}\ }\textbf {\bibinfo {volume} {75}},\ \bibinfo {eid} {124002}
  (\bibinfo {year} {2007}{\natexlab{a}})},\ \Eprint
  {http://arxiv.org/abs/0704.1086} {arXiv:0704.1086 [gr-qc]} \BibitemShut
  {NoStop}%
\bibitem [{\citenamefont {{Trias}}\ and\ \citenamefont
  {{Sintes}}(2008)}]{Trias:2008}%
  \BibitemOpen
  \bibfield  {author} {\bibinfo {author} {\bibfnamefont {M.}~\bibnamefont
  {{Trias}}}\ and\ \bibinfo {author} {\bibfnamefont {A.~M.}\ \bibnamefont
  {{Sintes}}},\ }\href {\doibase 10.1103/PhysRevD.77.024030} {\bibfield
  {journal} {\bibinfo  {journal} {\prd}\ }\textbf {\bibinfo {volume} {77}},\
  \bibinfo {eid} {024030} (\bibinfo {year} {2008})},\ \Eprint
  {http://arxiv.org/abs/0707.4434} {arXiv:0707.4434 [gr-qc]} \BibitemShut
  {NoStop}%
\bibitem [{\citenamefont {{Sintes}}\ and\ \citenamefont
  {{Vecchio}}(2000{\natexlab{a}})}]{Sintes:2000}%
  \BibitemOpen
  \bibfield  {author} {\bibinfo {author} {\bibfnamefont {A.~M.}\ \bibnamefont
  {{Sintes}}}\ and\ \bibinfo {author} {\bibfnamefont {A.}~\bibnamefont
  {{Vecchio}}},\ }\href@noop {} {\bibfield  {journal} {\bibinfo  {journal}
  {ArXiv General Relativity and Quantum Cosmology e-prints}\ } (\bibinfo {year}
  {2000}{\natexlab{a}})},\ \Eprint {http://arxiv.org/abs/gr-qc/0005058}
  {gr-qc/0005058} \BibitemShut {NoStop}%
\bibitem [{\citenamefont {{Sintes}}\ and\ \citenamefont
  {{Vecchio}}(2000{\natexlab{b}})}]{Sintes:2000L}%
  \BibitemOpen
  \bibfield  {author} {\bibinfo {author} {\bibfnamefont {A.~M.}\ \bibnamefont
  {{Sintes}}}\ and\ \bibinfo {author} {\bibfnamefont {A.}~\bibnamefont
  {{Vecchio}}},\ }in\ \href {\doibase 10.1063/1.1291890} {\emph {\bibinfo
  {booktitle} {American Institute of Physics Conference Series}}},\ \bibinfo
  {series} {American Institute of Physics Conference Series}, Vol.\ \bibinfo
  {volume} {523},\ \bibinfo {editor} {edited by\ \bibinfo {editor}
  {\bibfnamefont {S.}~\bibnamefont {{Meshkov}}}}\ (\bibinfo {year} {2000})\
  pp.\ \bibinfo {pages} {403--404},\ \Eprint
  {http://arxiv.org/abs/gr-qc/0005059} {gr-qc/0005059} \BibitemShut {NoStop}%
\bibitem [{\citenamefont {{Moore}}\ and\ \citenamefont
  {{Hellings}}(2002)}]{Moore:2002}%
  \BibitemOpen
  \bibfield  {author} {\bibinfo {author} {\bibfnamefont {T.~A.}\ \bibnamefont
  {{Moore}}}\ and\ \bibinfo {author} {\bibfnamefont {R.~W.}\ \bibnamefont
  {{Hellings}}},\ }\href {\doibase 10.1103/PhysRevD.65.062001} {\bibfield
  {journal} {\bibinfo  {journal} {\prd}\ }\textbf {\bibinfo {volume} {65}},\
  \bibinfo {eid} {062001} (\bibinfo {year} {2002})},\ \Eprint
  {http://arxiv.org/abs/gr-qc/9910116} {gr-qc/9910116} \BibitemShut {NoStop}%
\bibitem [{\citenamefont {{Hellings}}\ and\ \citenamefont
  {{Moore}}(2003)}]{Hellings:2003}%
  \BibitemOpen
  \bibfield  {author} {\bibinfo {author} {\bibfnamefont {R.~W.}\ \bibnamefont
  {{Hellings}}}\ and\ \bibinfo {author} {\bibfnamefont {T.~A.}\ \bibnamefont
  {{Moore}}},\ }\href {\doibase 10.1088/0264-9381/20/10/321} {\bibfield
  {journal} {\bibinfo  {journal} {Classical and Quantum Gravity}\ }\textbf
  {\bibinfo {volume} {20}},\ \bibinfo {pages} {181} (\bibinfo {year} {2003})},\
  \Eprint {http://arxiv.org/abs/gr-qc/0207102} {gr-qc/0207102} \BibitemShut
  {NoStop}%
\bibitem [{\citenamefont {{Van Den Broeck}}\ and\ \citenamefont
  {{Sengupta}}(2007{\natexlab{b}})}]{VanDen:2007a}%
  \BibitemOpen
  \bibfield  {author} {\bibinfo {author} {\bibfnamefont {C.}~\bibnamefont {{Van
  Den Broeck}}}\ and\ \bibinfo {author} {\bibfnamefont {A.~S.}\ \bibnamefont
  {{Sengupta}}},\ }\href {\doibase 10.1088/0264-9381/24/5/005} {\bibfield
  {journal} {\bibinfo  {journal} {Classical and Quantum Gravity}\ }\textbf
  {\bibinfo {volume} {24}},\ \bibinfo {pages} {1089} (\bibinfo {year}
  {2007}{\natexlab{b}})},\ \Eprint {http://arxiv.org/abs/gr-qc/0610126}
  {gr-qc/0610126} \BibitemShut {NoStop}%
\bibitem [{\citenamefont {{Arun}}\ \emph
  {et~al.}(2007{\natexlab{b}})\citenamefont {{Arun}}, \citenamefont {{Iyer}},
  \citenamefont {{Sathyaprakash}}, \citenamefont {{Sinha}},\ and\ \citenamefont
  {{van den Broeck}}}]{Arun:2007a}%
  \BibitemOpen
  \bibfield  {author} {\bibinfo {author} {\bibfnamefont {K.~G.}\ \bibnamefont
  {{Arun}}}, \bibinfo {author} {\bibfnamefont {B.~R.}\ \bibnamefont {{Iyer}}},
  \bibinfo {author} {\bibfnamefont {B.~S.}\ \bibnamefont {{Sathyaprakash}}},
  \bibinfo {author} {\bibfnamefont {S.}~\bibnamefont {{Sinha}}}, \ and\
  \bibinfo {author} {\bibfnamefont {C.}~\bibnamefont {{van den Broeck}}},\
  }\href {\doibase 10.1103/PhysRevD.76.104016} {\bibfield  {journal} {\bibinfo
  {journal} {\prd}\ }\textbf {\bibinfo {volume} {76}},\ \bibinfo {eid} {104016}
  (\bibinfo {year} {2007}{\natexlab{b}})},\ \Eprint
  {http://arxiv.org/abs/0707.3920} {arXiv:0707.3920} \BibitemShut {NoStop}%
\bibitem [{\citenamefont {{Porter}}\ and\ \citenamefont
  {{Cornish}}(2008)}]{Porter:2008}%
  \BibitemOpen
  \bibfield  {author} {\bibinfo {author} {\bibfnamefont {E.~K.}\ \bibnamefont
  {{Porter}}}\ and\ \bibinfo {author} {\bibfnamefont {N.~J.}\ \bibnamefont
  {{Cornish}}},\ }\href {\doibase 10.1103/PhysRevD.78.064005} {\bibfield
  {journal} {\bibinfo  {journal} {\prd}\ }\textbf {\bibinfo {volume} {78}},\
  \bibinfo {eid} {064005} (\bibinfo {year} {2008})},\ \Eprint
  {http://arxiv.org/abs/0804.0332} {arXiv:0804.0332 [gr-qc]} \BibitemShut
  {NoStop}%
\bibitem [{\citenamefont {{Arun}}\ \emph {et~al.}(2009)\citenamefont {{Arun}},
  \citenamefont {{Buonanno}}, \citenamefont {{Faye}},\ and\ \citenamefont
  {{Ochsner}}}]{Arun:2009}%
  \BibitemOpen
  \bibfield  {author} {\bibinfo {author} {\bibfnamefont {K.~G.}\ \bibnamefont
  {{Arun}}}, \bibinfo {author} {\bibfnamefont {A.}~\bibnamefont {{Buonanno}}},
  \bibinfo {author} {\bibfnamefont {G.}~\bibnamefont {{Faye}}}, \ and\ \bibinfo
  {author} {\bibfnamefont {E.}~\bibnamefont {{Ochsner}}},\ }\href {\doibase
  10.1103/PhysRevD.79.104023} {\bibfield  {journal} {\bibinfo  {journal}
  {\prd}\ }\textbf {\bibinfo {volume} {79}},\ \bibinfo {eid} {104023} (\bibinfo
  {year} {2009})},\ \Eprint {http://arxiv.org/abs/0810.5336} {arXiv:0810.5336
  [gr-qc]} \BibitemShut {NoStop}%
\bibitem [{\citenamefont {{Paczynski}}(1986)}]{Paczynski:1986}%
  \BibitemOpen
  \bibfield  {author} {\bibinfo {author} {\bibfnamefont {B.}~\bibnamefont
  {{Paczynski}}},\ }\href {\doibase 10.1086/184740} {\bibfield  {journal}
  {\bibinfo  {journal} {\apjl}\ }\textbf {\bibinfo {volume} {308}},\ \bibinfo
  {pages} {L43} (\bibinfo {year} {1986})}\BibitemShut {NoStop}%
\bibitem [{\citenamefont {{Eichler}}\ \emph {et~al.}(1989)\citenamefont
  {{Eichler}}, \citenamefont {{Livio}}, \citenamefont {{Piran}},\ and\
  \citenamefont {{Schramm}}}]{Eichler:1989}%
  \BibitemOpen
  \bibfield  {author} {\bibinfo {author} {\bibfnamefont {D.}~\bibnamefont
  {{Eichler}}}, \bibinfo {author} {\bibfnamefont {M.}~\bibnamefont {{Livio}}},
  \bibinfo {author} {\bibfnamefont {T.}~\bibnamefont {{Piran}}}, \ and\
  \bibinfo {author} {\bibfnamefont {D.~N.}\ \bibnamefont {{Schramm}}},\ }\href
  {\doibase 10.1038/340126a0} {\bibfield  {journal} {\bibinfo  {journal}
  {\nat}\ }\textbf {\bibinfo {volume} {340}},\ \bibinfo {pages} {126} (\bibinfo
  {year} {1989})}\BibitemShut {NoStop}%
\bibitem [{\citenamefont {{Barthelmy}}\ \emph {et~al.}(2005)\citenamefont
  {{Barthelmy}} \emph {et~al.}}]{Bart:2005}%
  \BibitemOpen
  \bibfield  {author} {\bibinfo {author} {\bibfnamefont {S.}~\bibnamefont
  {{Barthelmy}}} \emph {et~al.},\ }\href@noop {} {\bibfield  {journal}
  {\bibinfo  {journal} {\nat}\ }\textbf {\bibinfo {volume} {438}},\ \bibinfo
  {pages} {994} (\bibinfo {year} {2005})}\BibitemShut {NoStop}%
\bibitem [{\citenamefont {{Grindlay}}\ \emph {et~al.}(2006)\citenamefont
  {{Grindlay}}, \citenamefont {{Portegies Zwart}},\ and\ \citenamefont
  {{McMillan}}}]{grindlay}%
  \BibitemOpen
  \bibfield  {author} {\bibinfo {author} {\bibfnamefont {J.}~\bibnamefont
  {{Grindlay}}}, \bibinfo {author} {\bibfnamefont {S.}~\bibnamefont {{Portegies
  Zwart}}}, \ and\ \bibinfo {author} {\bibfnamefont {S.}~\bibnamefont
  {{McMillan}}},\ }\href@noop {} {\bibfield  {journal} {\bibinfo  {journal}
  {Nature Physics}\ }\textbf {\bibinfo {volume} {2}},\ \bibinfo {pages} {116}
  (\bibinfo {year} {2006})}\BibitemShut {NoStop}%
\bibitem [{\citenamefont {{Troja}}\ \emph {et~al.}(2010)\citenamefont
  {{Troja}}, \citenamefont {{Rosswog}},\ and\ \citenamefont
  {{Gehrels}}}]{Troja:2010}%
  \BibitemOpen
  \bibfield  {author} {\bibinfo {author} {\bibfnamefont {E.}~\bibnamefont
  {{Troja}}}, \bibinfo {author} {\bibfnamefont {S.}~\bibnamefont {{Rosswog}}},
  \ and\ \bibinfo {author} {\bibfnamefont {N.}~\bibnamefont {{Gehrels}}},\
  }\href {\doibase 10.1088/0004-637X/723/2/1711} {\bibfield  {journal}
  {\bibinfo  {journal} {\apj}\ }\textbf {\bibinfo {volume} {723}},\ \bibinfo
  {pages} {1711} (\bibinfo {year} {2010})},\ \Eprint
  {http://arxiv.org/abs/1009.1385} {arXiv:1009.1385 [astro-ph.HE]} \BibitemShut
  {NoStop}%
\bibitem [{\citenamefont {{O'Leary}}\ \emph {et~al.}(2007)\citenamefont
  {{O'Leary}}, \citenamefont {{O'Shaughnessy}},\ and\ \citenamefont
  {{Rasio}}}]{Oleary:2007}%
  \BibitemOpen
  \bibfield  {author} {\bibinfo {author} {\bibfnamefont {R.~M.}\ \bibnamefont
  {{O'Leary}}}, \bibinfo {author} {\bibfnamefont {R.}~\bibnamefont
  {{O'Shaughnessy}}}, \ and\ \bibinfo {author} {\bibfnamefont {F.~A.}\
  \bibnamefont {{Rasio}}},\ }\href {\doibase 10.1103/PhysRevD.76.061504}
  {\bibfield  {journal} {\bibinfo  {journal} {\prd}\ }\textbf {\bibinfo
  {volume} {76}},\ \bibinfo {eid} {061504} (\bibinfo {year} {2007})},\ \Eprint
  {http://arxiv.org/abs/arXiv:astro-ph/0701887} {arXiv:astro-ph/0701887}
  \BibitemShut {NoStop}%
\bibitem [{\citenamefont {{Misner}}\ \emph {et~al.}(1973)\citenamefont
  {{Misner}}, \citenamefont {{Thorne}},\ and\ \citenamefont {{Wheeler}}}]{mtw}%
  \BibitemOpen
  \bibfield  {author} {\bibinfo {author} {\bibfnamefont {C.~W.}\ \bibnamefont
  {{Misner}}}, \bibinfo {author} {\bibfnamefont {K.~S.}\ \bibnamefont
  {{Thorne}}}, \ and\ \bibinfo {author} {\bibfnamefont {J.~A.}\ \bibnamefont
  {{Wheeler}}},\ }\href@noop {} {\emph {\bibinfo {title}
  {{\textit{Gravitation}}}}}\ (\bibinfo  {publisher} {San Francisco:
  W.H.~Freeman and Co., 1973},\ \bibinfo {year} {1973})\BibitemShut {NoStop}%
\bibitem [{\citenamefont {{K{\"o}nigsd{\"o}rffer}}\ and\ \citenamefont
  {{Gopakumar}}(2005)}]{Gopakumar:2005b}%
  \BibitemOpen
  \bibfield  {author} {\bibinfo {author} {\bibfnamefont {C.}~\bibnamefont
  {{K{\"o}nigsd{\"o}rffer}}}\ and\ \bibinfo {author} {\bibfnamefont
  {A.}~\bibnamefont {{Gopakumar}}},\ }\href {\doibase
  10.1103/PhysRevD.71.024039} {\bibfield  {journal} {\bibinfo  {journal}
  {\prd}\ }\textbf {\bibinfo {volume} {71}},\ \bibinfo {eid} {024039} (\bibinfo
  {year} {2005})},\ \Eprint {http://arxiv.org/abs/gr-qc/0501011}
  {gr-qc/0501011} \BibitemShut {NoStop}%
\bibitem [{\citenamefont {{Will}}\ and\ \citenamefont {{Wiseman}}(1996)}]{w1}%
  \BibitemOpen
  \bibfield  {author} {\bibinfo {author} {\bibfnamefont {C.~M.}\ \bibnamefont
  {{Will}}}\ and\ \bibinfo {author} {\bibfnamefont {A.~G.}\ \bibnamefont
  {{Wiseman}}},\ }\href {\doibase 10.1103/PhysRevD.54.4813} {\bibfield
  {journal} {\bibinfo  {journal} {\PRD}\ }\textbf {\bibinfo {volume} {54}},\
  \bibinfo {pages} {4813} (\bibinfo {year} {1996})},\ \Eprint
  {http://arxiv.org/abs/arXiv:gr-qc/9608012} {arXiv:gr-qc/9608012} \BibitemShut
  {NoStop}%
\bibitem [{\citenamefont {{Cutler}}\ \emph {et~al.}(1994)\citenamefont
  {{Cutler}}, \citenamefont {{Kennefick}},\ and\ \citenamefont
  {{Poisson}}}]{Cutler:1994}%
  \BibitemOpen
  \bibfield  {author} {\bibinfo {author} {\bibfnamefont {C.}~\bibnamefont
  {{Cutler}}}, \bibinfo {author} {\bibfnamefont {D.}~\bibnamefont
  {{Kennefick}}}, \ and\ \bibinfo {author} {\bibfnamefont {E.}~\bibnamefont
  {{Poisson}}},\ }\href {\doibase 10.1103/PhysRevD.50.3816} {\bibfield
  {journal} {\bibinfo  {journal} {\prd}\ }\textbf {\bibinfo {volume} {50}},\
  \bibinfo {pages} {3816} (\bibinfo {year} {1994})}\BibitemShut {NoStop}%
\bibitem [{\citenamefont {{Blanchet}}\ and\ \citenamefont
  {{Schaefer}}(1989)}]{Blanchet:1989M}%
  \BibitemOpen
  \bibfield  {author} {\bibinfo {author} {\bibfnamefont {L.}~\bibnamefont
  {{Blanchet}}}\ and\ \bibinfo {author} {\bibfnamefont {G.}~\bibnamefont
  {{Schaefer}}},\ }\href@noop {} {\bibfield  {journal} {\bibinfo  {journal}
  {\mnras}\ }\textbf {\bibinfo {volume} {239}},\ \bibinfo {pages} {845}
  (\bibinfo {year} {1989})}\BibitemShut {NoStop}%
\bibitem [{\citenamefont {{Gopakumar}}\ and\ \citenamefont
  {{Iyer}}(2002)}]{Gopakumar:2002}%
  \BibitemOpen
  \bibfield  {author} {\bibinfo {author} {\bibfnamefont {A.}~\bibnamefont
  {{Gopakumar}}}\ and\ \bibinfo {author} {\bibfnamefont {B.~R.}\ \bibnamefont
  {{Iyer}}},\ }\href {\doibase 10.1103/PhysRevD.65.084011} {\bibfield
  {journal} {\bibinfo  {journal} {\prd}\ }\textbf {\bibinfo {volume} {65}},\
  \bibinfo {eid} {084011} (\bibinfo {year} {2002})},\ \Eprint
  {http://arxiv.org/abs/gr-qc/0110100} {gr-qc/0110100} \BibitemShut {NoStop}%
\bibitem [{\citenamefont {{K{\"o}nigsd{\"o}rffer}}\ and\ \citenamefont
  {{Gopakumar}}(2006)}]{GopakumarandK:2006}%
  \BibitemOpen
  \bibfield  {author} {\bibinfo {author} {\bibfnamefont {C.}~\bibnamefont
  {{K{\"o}nigsd{\"o}rffer}}}\ and\ \bibinfo {author} {\bibfnamefont
  {A.}~\bibnamefont {{Gopakumar}}},\ }\href {\doibase
  10.1103/PhysRevD.73.124012} {\bibfield  {journal} {\bibinfo  {journal}
  {\prd}\ }\textbf {\bibinfo {volume} {73}},\ \bibinfo {eid} {124012} (\bibinfo
  {year} {2006})},\ \Eprint {http://arxiv.org/abs/gr-qc/0603056}
  {gr-qc/0603056} \BibitemShut {NoStop}%
\bibitem [{\citenamefont {Yunes}\ and\ \citenamefont
  {Berti}(2008)}]{Yunes:2008tw}%
  \BibitemOpen
  \bibfield  {author} {\bibinfo {author} {\bibfnamefont {N.}~\bibnamefont
  {Yunes}}\ and\ \bibinfo {author} {\bibfnamefont {E.}~\bibnamefont {Berti}},\
  }\href {\doibase 10.1103/PhysRevD.77.124006, 10.1103/PhysRevD.83.109901}
  {\bibfield  {journal} {\bibinfo  {journal} {Phys.Rev.}\ }\textbf {\bibinfo
  {volume} {D77}},\ \bibinfo {pages} {124006} (\bibinfo {year} {2008})},\
  \Eprint {http://arxiv.org/abs/0803.1853} {arXiv:0803.1853 [gr-qc]}
  \BibitemShut {NoStop}%
\bibitem [{\citenamefont {Zhang}\ \emph {et~al.}(2011)\citenamefont {Zhang},
  \citenamefont {Yunes},\ and\ \citenamefont {Berti}}]{Zhang:2011vha}%
  \BibitemOpen
  \bibfield  {author} {\bibinfo {author} {\bibfnamefont {Z.}~\bibnamefont
  {Zhang}}, \bibinfo {author} {\bibfnamefont {N.}~\bibnamefont {Yunes}}, \ and\
  \bibinfo {author} {\bibfnamefont {E.}~\bibnamefont {Berti}},\ }\href
  {\doibase 10.1103/PhysRevD.84.024029} {\bibfield  {journal} {\bibinfo
  {journal} {Phys.Rev.}\ }\textbf {\bibinfo {volume} {D84}},\ \bibinfo {pages}
  {024029} (\bibinfo {year} {2011})},\ \Eprint {http://arxiv.org/abs/1103.6041}
  {arXiv:1103.6041 [gr-qc]} \BibitemShut {NoStop}%
\bibitem [{\citenamefont {{Favata}}(2014)}]{Favata:2014}%
  \BibitemOpen
  \bibfield  {author} {\bibinfo {author} {\bibfnamefont {M.}~\bibnamefont
  {{Favata}}},\ }\href {\doibase 10.1103/PhysRevLett.112.101101} {\bibfield
  {journal} {\bibinfo  {journal} {Physical Review Letters}\ }\textbf {\bibinfo
  {volume} {112}},\ \bibinfo {eid} {101101} (\bibinfo {year} {2014})},\ \Eprint
  {http://arxiv.org/abs/1310.8288} {arXiv:1310.8288 [gr-qc]} \BibitemShut
  {NoStop}%
\bibitem [{\citenamefont {{Capano}}\ \emph {et~al.}(2013)\citenamefont
  {{Capano}}, \citenamefont {{Pan}},\ and\ \citenamefont
  {{Buonanno}}}]{Colin:2013}%
  \BibitemOpen
  \bibfield  {author} {\bibinfo {author} {\bibfnamefont {C.}~\bibnamefont
  {{Capano}}}, \bibinfo {author} {\bibfnamefont {Y.}~\bibnamefont {{Pan}}}, \
  and\ \bibinfo {author} {\bibfnamefont {A.}~\bibnamefont {{Buonanno}}},\
  }\href@noop {} {\bibfield  {journal} {\bibinfo  {journal} {ArXiv e-prints}\ }
  (\bibinfo {year} {2013})},\ \Eprint {http://arxiv.org/abs/1311.1286}
  {arXiv:1311.1286 [gr-qc]} \BibitemShut {NoStop}%
\bibitem [{\citenamefont {{Pekowsky}}\ \emph {et~al.}(2013)\citenamefont
  {{Pekowsky}}, \citenamefont {{Healy}}, \citenamefont {{Shoemaker}},\ and\
  \citenamefont {{Laguna}}}]{Pekowski:2013}%
  \BibitemOpen
  \bibfield  {author} {\bibinfo {author} {\bibfnamefont {L.}~\bibnamefont
  {{Pekowsky}}}, \bibinfo {author} {\bibfnamefont {J.}~\bibnamefont {{Healy}}},
  \bibinfo {author} {\bibfnamefont {D.}~\bibnamefont {{Shoemaker}}}, \ and\
  \bibinfo {author} {\bibfnamefont {P.}~\bibnamefont {{Laguna}}},\ }\href
  {\doibase 10.1103/PhysRevD.87.084008} {\bibfield  {journal} {\bibinfo
  {journal} {\prd}\ }\textbf {\bibinfo {volume} {87}},\ \bibinfo {eid} {084008}
  (\bibinfo {year} {2013})},\ \Eprint {http://arxiv.org/abs/1210.1891}
  {arXiv:1210.1891 [gr-qc]} \BibitemShut {NoStop}%
\bibitem [{\citenamefont {{Berger}}(2013)}]{Edo:2013}%
  \BibitemOpen
  \bibfield  {author} {\bibinfo {author} {\bibfnamefont {E.}~\bibnamefont
  {{Berger}}},\ }\href@noop {} {\bibfield  {journal} {\bibinfo  {journal}
  {ArXiv e-prints}\ } (\bibinfo {year} {2013})},\ \Eprint
  {http://arxiv.org/abs/1311.2603} {arXiv:1311.2603 [astro-ph.HE]} \BibitemShut
  {NoStop}%
\bibitem [{\citenamefont {{Clausen}}\ \emph {et~al.}(2013)\citenamefont
  {{Clausen}}, \citenamefont {{Sigurdsson}},\ and\ \citenamefont
  {{Chernoff}}}]{Clausen:2013}%
  \BibitemOpen
  \bibfield  {author} {\bibinfo {author} {\bibfnamefont {D.}~\bibnamefont
  {{Clausen}}}, \bibinfo {author} {\bibfnamefont {S.}~\bibnamefont
  {{Sigurdsson}}}, \ and\ \bibinfo {author} {\bibfnamefont {D.~F.}\
  \bibnamefont {{Chernoff}}},\ }\href {\doibase 10.1093/mnras/sts295}
  {\bibfield  {journal} {\bibinfo  {journal} {\mnras}\ }\textbf {\bibinfo
  {volume} {428}},\ \bibinfo {pages} {3618} (\bibinfo {year} {2013})},\ \Eprint
  {http://arxiv.org/abs/1210.8153} {arXiv:1210.8153 [astro-ph.HE]} \BibitemShut
  {NoStop}%
\bibitem [{\citenamefont {{Schutz}}(2011)}]{Schutz:2011}%
  \BibitemOpen
  \bibfield  {author} {\bibinfo {author} {\bibfnamefont {B.~F.}\ \bibnamefont
  {{Schutz}}},\ }\href {\doibase 10.1088/0264-9381/28/12/125023} {\bibfield
  {journal} {\bibinfo  {journal} {Classical and Quantum Gravity}\ }\textbf
  {\bibinfo {volume} {28}},\ \bibinfo {eid} {125023} (\bibinfo {year}
  {2011})},\ \Eprint {http://arxiv.org/abs/1102.5421} {arXiv:1102.5421
  [astro-ph.IM]} \BibitemShut {NoStop}%
\bibitem [{\citenamefont {{Dominik}}\ \emph {et~al.}(2014)\citenamefont
  {{Dominik}}, \citenamefont {{Berti}}, \citenamefont {{O'Shaughnessy}},
  \citenamefont {{Mandel}}, \citenamefont {{Belczynski}}, \citenamefont
  {{Fryer}}, \citenamefont {{Holz}}, \citenamefont {{Bulik}},\ and\
  \citenamefont {{Pannarale}}}]{Dominik:2014}%
  \BibitemOpen
  \bibfield  {author} {\bibinfo {author} {\bibfnamefont {M.}~\bibnamefont
  {{Dominik}}}, \bibinfo {author} {\bibfnamefont {E.}~\bibnamefont {{Berti}}},
  \bibinfo {author} {\bibfnamefont {R.}~\bibnamefont {{O'Shaughnessy}}},
  \bibinfo {author} {\bibfnamefont {I.}~\bibnamefont {{Mandel}}}, \bibinfo
  {author} {\bibfnamefont {K.}~\bibnamefont {{Belczynski}}}, \bibinfo {author}
  {\bibfnamefont {C.}~\bibnamefont {{Fryer}}}, \bibinfo {author} {\bibfnamefont
  {D.}~\bibnamefont {{Holz}}}, \bibinfo {author} {\bibfnamefont
  {T.}~\bibnamefont {{Bulik}}}, \ and\ \bibinfo {author} {\bibfnamefont
  {F.}~\bibnamefont {{Pannarale}}},\ }\href@noop {} {\bibfield  {journal}
  {\bibinfo  {journal} {ArXiv e-prints}\ } (\bibinfo {year} {2014})},\ \Eprint
  {http://arxiv.org/abs/1405.7016} {arXiv:1405.7016 [astro-ph.HE]} \BibitemShut
  {NoStop}%
\bibitem [{\citenamefont {{Tai}}\ \emph {et~al.}(2014)\citenamefont {{Tai}},
  \citenamefont {{McWilliams}},\ and\ \citenamefont {{Pretorius}}}]{Tai:2014}%
  \BibitemOpen
  \bibfield  {author} {\bibinfo {author} {\bibfnamefont {K.~S.}\ \bibnamefont
  {{Tai}}}, \bibinfo {author} {\bibfnamefont {S.~T.}\ \bibnamefont
  {{McWilliams}}}, \ and\ \bibinfo {author} {\bibfnamefont {F.}~\bibnamefont
  {{Pretorius}}},\ }\href@noop {} {\bibfield  {journal} {\bibinfo  {journal}
  {ArXiv e-prints}\ } (\bibinfo {year} {2014})},\ \Eprint
  {http://arxiv.org/abs/1403.7754} {arXiv:1403.7754 [gr-qc]} \BibitemShut
  {NoStop}%
\bibitem [{\citenamefont {{McLaughlin}}(2013)}]{NANOMaura:2013}%
  \BibitemOpen
  \bibfield  {author} {\bibinfo {author} {\bibfnamefont {M.~A.}\ \bibnamefont
  {{McLaughlin}}},\ }\href {\doibase 10.1088/0264-9381/30/22/224008} {\bibfield
   {journal} {\bibinfo  {journal} {Classical and Quantum Gravity}\ }\textbf
  {\bibinfo {volume} {30}},\ \bibinfo {eid} {224008} (\bibinfo {year}
  {2013})},\ \Eprint {http://arxiv.org/abs/1310.0758} {arXiv:1310.0758
  [astro-ph.IM]} \BibitemShut {NoStop}%
\bibitem [{\citenamefont {{Sesana}}(2013)}]{Sesana:2013CQG}%
  \BibitemOpen
  \bibfield  {author} {\bibinfo {author} {\bibfnamefont {A.}~\bibnamefont
  {{Sesana}}},\ }\href {\doibase 10.1088/0264-9381/30/22/224014} {\bibfield
  {journal} {\bibinfo  {journal} {Classical and Quantum Gravity}\ }\textbf
  {\bibinfo {volume} {30}},\ \bibinfo {eid} {224014} (\bibinfo {year}
  {2013})},\ \Eprint {http://arxiv.org/abs/1307.2600} {arXiv:1307.2600
  [astro-ph.CO]} \BibitemShut {NoStop}%
\bibitem [{\citenamefont {{Seoane}}\ \emph {et~al.}(2013)\citenamefont
  {{Seoane}}, \citenamefont {{Aoudia}}, \citenamefont {{Audley}}, \citenamefont
  {{Auger}}, \citenamefont {{Babak}}, \citenamefont {{Baker}} \emph
  {et~al.}}]{TGU:2013}%
  \BibitemOpen
  \bibfield  {author} {\bibinfo {author} {\bibfnamefont {P.~A.}\ \bibnamefont
  {{Seoane}}}, \bibinfo {author} {\bibfnamefont {S.}~\bibnamefont {{Aoudia}}},
  \bibinfo {author} {\bibfnamefont {H.}~\bibnamefont {{Audley}}}, \bibinfo
  {author} {\bibfnamefont {G.}~\bibnamefont {{Auger}}}, \bibinfo {author}
  {\bibfnamefont {S.}~\bibnamefont {{Babak}}}, \bibinfo {author} {\bibfnamefont
  {J.}~\bibnamefont {{Baker}}},  \emph {et~al.},\ }\href@noop {} {\bibfield
  {journal} {\bibinfo  {journal} {ArXiv e-prints}\ } (\bibinfo {year}
  {2013})},\ \Eprint {http://arxiv.org/abs/1305.5720} {arXiv:1305.5720
  [astro-ph.CO]} \BibitemShut {NoStop}%
\bibitem [{\citenamefont {{Gair}}\ \emph {et~al.}(2013)\citenamefont {{Gair}},
  \citenamefont {{Vallisneri}}, \citenamefont {{Larson}},\ and\ \citenamefont
  {{Baker}}}]{GairL:2013}%
  \BibitemOpen
  \bibfield  {author} {\bibinfo {author} {\bibfnamefont {J.~R.}\ \bibnamefont
  {{Gair}}}, \bibinfo {author} {\bibfnamefont {M.}~\bibnamefont
  {{Vallisneri}}}, \bibinfo {author} {\bibfnamefont {S.~L.}\ \bibnamefont
  {{Larson}}}, \ and\ \bibinfo {author} {\bibfnamefont {J.~G.}\ \bibnamefont
  {{Baker}}},\ }\href@noop {} {\bibfield  {journal} {\bibinfo  {journal}
  {Living Reviews in Relativity}\ }\textbf {\bibinfo {volume} {16}},\ \bibinfo
  {pages} {7} (\bibinfo {year} {2013})},\ \Eprint
  {http://arxiv.org/abs/1212.5575} {arXiv:1212.5575 [gr-qc]} \BibitemShut
  {NoStop}%
\bibitem [{\citenamefont {{The LIGO Scientific Collaboration}}\ \emph
  {et~al.}(2014{\natexlab{a}})\citenamefont {{The LIGO Scientific
  Collaboration}}, \citenamefont {{the Virgo Collaboration}}, \citenamefont
  {{Aasi}}, \citenamefont {{Abbott}}, \citenamefont {{Abbott}}, \citenamefont
  {{Abbott}}, \citenamefont {{Abernathy}}, \citenamefont {{Acernese}},
  \citenamefont {{Ackley}}, \citenamefont {{Adams}},\ and\ \citenamefont
  {et~al.}}]{Inter:2014}%
  \BibitemOpen
  \bibfield  {author} {\bibinfo {author} {\bibnamefont {{The LIGO Scientific
  Collaboration}}}, \bibinfo {author} {\bibnamefont {{the Virgo
  Collaboration}}}, \bibinfo {author} {\bibfnamefont {J.}~\bibnamefont
  {{Aasi}}}, \bibinfo {author} {\bibfnamefont {B.~P.}\ \bibnamefont
  {{Abbott}}}, \bibinfo {author} {\bibfnamefont {R.}~\bibnamefont {{Abbott}}},
  \bibinfo {author} {\bibfnamefont {T.}~\bibnamefont {{Abbott}}}, \bibinfo
  {author} {\bibfnamefont {M.~R.}\ \bibnamefont {{Abernathy}}}, \bibinfo
  {author} {\bibfnamefont {F.}~\bibnamefont {{Acernese}}}, \bibinfo {author}
  {\bibfnamefont {K.}~\bibnamefont {{Ackley}}}, \bibinfo {author}
  {\bibfnamefont {C.}~\bibnamefont {{Adams}}}, \ and\ \bibinfo {author}
  {\bibnamefont {et~al.}},\ }\href@noop {} {\bibfield  {journal} {\bibinfo
  {journal} {ArXiv e-prints}\ } (\bibinfo {year} {2014}{\natexlab{a}})},\
  \Eprint {http://arxiv.org/abs/1403.6639} {arXiv:1403.6639 [astro-ph.HE]}
  \BibitemShut {NoStop}%
\bibitem [{\citenamefont {{The LIGO Scientific Collaboration}}\ \emph
  {et~al.}(2014{\natexlab{b}})\citenamefont {{The LIGO Scientific
  Collaboration}}, \citenamefont {{the Virgo Collaboration}}, \citenamefont
  {{Aasi}}, \citenamefont {{Abbott}}, \citenamefont {{Abbott}}, \citenamefont
  {{Abbott}}, \citenamefont {{Abernathy}}, \citenamefont {{Acernese}},
  \citenamefont {{Ackley}}, \citenamefont {{Adams}},\ and\ \citenamefont
  {et~al.}}]{SGRBaLIGO:2014}%
  \BibitemOpen
  \bibfield  {author} {\bibinfo {author} {\bibnamefont {{The LIGO Scientific
  Collaboration}}}, \bibinfo {author} {\bibnamefont {{the Virgo
  Collaboration}}}, \bibinfo {author} {\bibfnamefont {J.}~\bibnamefont
  {{Aasi}}}, \bibinfo {author} {\bibfnamefont {B.~P.}\ \bibnamefont
  {{Abbott}}}, \bibinfo {author} {\bibfnamefont {R.}~\bibnamefont {{Abbott}}},
  \bibinfo {author} {\bibfnamefont {T.}~\bibnamefont {{Abbott}}}, \bibinfo
  {author} {\bibfnamefont {M.~R.}\ \bibnamefont {{Abernathy}}}, \bibinfo
  {author} {\bibfnamefont {F.}~\bibnamefont {{Acernese}}}, \bibinfo {author}
  {\bibfnamefont {K.}~\bibnamefont {{Ackley}}}, \bibinfo {author}
  {\bibfnamefont {C.}~\bibnamefont {{Adams}}}, \ and\ \bibinfo {author}
  {\bibnamefont {et~al.}},\ }\href@noop {} {\bibfield  {journal} {\bibinfo
  {journal} {ArXiv e-prints}\ } (\bibinfo {year} {2014}{\natexlab{b}})},\
  \Eprint {http://arxiv.org/abs/1405.1053} {arXiv:1405.1053 [astro-ph.HE]}
  \BibitemShut {NoStop}%
\bibitem [{\citenamefont {{Huerta}}\ \emph {et~al.}(2014)\citenamefont
  {{Huerta}}, \citenamefont {{Kumar}}, \citenamefont {{Gair}},\ and\
  \citenamefont {{McWilliams}}}]{Huerta:2014a}%
  \BibitemOpen
  \bibfield  {author} {\bibinfo {author} {\bibfnamefont {E.~A.}\ \bibnamefont
  {{Huerta}}}, \bibinfo {author} {\bibfnamefont {P.}~\bibnamefont {{Kumar}}},
  \bibinfo {author} {\bibfnamefont {J.~R.}\ \bibnamefont {{Gair}}}, \ and\
  \bibinfo {author} {\bibfnamefont {S.~T.}\ \bibnamefont {{McWilliams}}},\
  }\href@noop {} {\bibfield  {journal} {\bibinfo  {journal} {ArXiv e-prints}\ }
  (\bibinfo {year} {2014})},\ \Eprint {http://arxiv.org/abs/1403.0561}
  {arXiv:1403.0561 [gr-qc]} \BibitemShut {NoStop}%
\bibitem [{\citenamefont {{Huerta}}\ \emph {et~al.}(2012)\citenamefont
  {{Huerta}}, \citenamefont {{Kumar}},\ and\ \citenamefont
  {{Brown}}}]{Huerta:2012}%
  \BibitemOpen
  \bibfield  {author} {\bibinfo {author} {\bibfnamefont {E.~A.}\ \bibnamefont
  {{Huerta}}}, \bibinfo {author} {\bibfnamefont {P.}~\bibnamefont {{Kumar}}}, \
  and\ \bibinfo {author} {\bibfnamefont {D.~A.}\ \bibnamefont {{Brown}}},\
  }\href {\doibase 10.1103/PhysRevD.86.024024} {\bibfield  {journal} {\bibinfo
  {journal} {\prd}\ }\textbf {\bibinfo {volume} {86}},\ \bibinfo {eid} {024024}
  (\bibinfo {year} {2012})},\ \Eprint {http://arxiv.org/abs/1205.5562}
  {arXiv:1205.5562 [gr-qc]} \BibitemShut {NoStop}%
\bibitem [{\citenamefont {{Huerta}}\ and\ \citenamefont
  {{Gair}}(2011{\natexlab{a}})}]{Huerta:2011a}%
  \BibitemOpen
  \bibfield  {author} {\bibinfo {author} {\bibfnamefont {E.~A.}\ \bibnamefont
  {{Huerta}}}\ and\ \bibinfo {author} {\bibfnamefont {J.~R.}\ \bibnamefont
  {{Gair}}},\ }\href {\doibase 10.1103/PhysRevD.83.044020} {\bibfield
  {journal} {\bibinfo  {journal} {\PRD}\ }\textbf {\bibinfo {volume} {83}},\
  \bibinfo {pages} {044020} (\bibinfo {year} {2011}{\natexlab{a}})},\ \Eprint
  {http://arxiv.org/abs/1009.1985} {arXiv:1009.1985 [gr-qc]} \BibitemShut
  {NoStop}%
\bibitem [{\citenamefont {{Huerta}}\ and\ \citenamefont
  {{Gair}}(2011{\natexlab{b}})}]{Huerta:2011b}%
  \BibitemOpen
  \bibfield  {author} {\bibinfo {author} {\bibfnamefont {E.~A.}\ \bibnamefont
  {{Huerta}}}\ and\ \bibinfo {author} {\bibfnamefont {J.~R.}\ \bibnamefont
  {{Gair}}},\ }\href {\doibase 10.1103/PhysRevD.83.044021} {\bibfield
  {journal} {\bibinfo  {journal} {\PRD}\ }\textbf {\bibinfo {volume} {83}},\
  \bibinfo {pages} {044021} (\bibinfo {year} {2011}{\natexlab{b}})},\ \Eprint
  {http://arxiv.org/abs/1011.0421} {arXiv:1011.0421 [gr-qc]} \BibitemShut
  {NoStop}%
\bibitem [{\citenamefont {{Huerta}}\ and\ \citenamefont
  {{Gair}}(2011{\natexlab{c}})}]{smallbody}%
  \BibitemOpen
  \bibfield  {author} {\bibinfo {author} {\bibfnamefont {E.~A.}\ \bibnamefont
  {{Huerta}}}\ and\ \bibinfo {author} {\bibfnamefont {J.~R.}\ \bibnamefont
  {{Gair}}},\ }\href {\doibase 10.1103/PhysRevD.84.064023} {\bibfield
  {journal} {\bibinfo  {journal} {\prd}\ }\textbf {\bibinfo {volume} {84}},\
  \bibinfo {pages} {064023} (\bibinfo {year} {2011}{\natexlab{c}})},\ \Eprint
  {http://arxiv.org/abs/1105.3567} {arXiv:1105.3567 [gr-qc]} \BibitemShut
  {NoStop}%
\end{thebibliography}%

\end{document}